\documentclass[superscriptaddress,floatfix,nofootinbib,longbibliography,prx,twocolumn]{revtex4-1}
\usepackage{graphicx,amsfonts,amssymb,amsmath,hyperref,hypcap,enumerate}
\usepackage{slashed}
\usepackage{amsmath,amssymb,bm,graphicx,dsfont}
\usepackage[latin9]{inputenc}
\usepackage{amsmath}
\usepackage{slashed}
\usepackage{dsfont}
\usepackage{amstext}
\usepackage{amssymb}
\usepackage{amsbsy}
\usepackage{amsthm}
\usepackage{epsfig}
\usepackage{graphicx}
\usepackage{bbm}
\usepackage{hyperref}
\usepackage{color}
\usepackage{slashed}
\newcommand{\<}{\langle}

\newcommand{\up}{\uparrow}
\newcommand{\down}{\downarrow}
\renewcommand{\>}{\rangle}

\renewcommand{\v}[1]{\mathbf{#1}} 

\renewcommand{\d}{\partial}

\newcommand{\be}{\begin{equation}}
\newcommand{\ba}{\begin{align}}
\newcommand{\ee}{\end{equation}}
\newcommand{\bea}{\begin{eqnarray}}
\newcommand{\eea}{\end{eqnarray}}
\newcommand{\beq}{\begin{equation}}
\newcommand{\eeq}{\end{equation}}
\newcommand{\beqn}{\begin{eqnarray}}
\newcommand{\eeqn}{\end{eqnarray}}

\newcommand{\la}{\langle}
\newcommand{\ra}{\rangle}

\newcommand{\lp}{\left(}
\newcommand{\rp}{\right)}

\newcommand{\Z}{\mathbb{Z}}

\newcommand{\nccp}{NCCP}
\newcommand{\dd}{\mathrm{d}}
\newcommand{\f}{\frac}

\renewcommand{\vec}[1]{{\bf #1}}
\renewcommand{\hat}[1]{{\widehat #1}}

\def\nn{\nonumber\\}

\newcommand{\red}{}
\newcommand{\blue}{}

\begin{document}
\title{Deconfined quantum critical points: symmetries and dualities}
\author{Chong Wang}
\affiliation{Department of Physics,  Harvard University,
Cambridge, MA 02138, USA}
\affiliation{Kavli Institute for Theoretical Physics, University of California, Santa Barbara, CA 93106, USA}
\author{Adam Nahum}
\affiliation{Department of Physics, Massachusetts Institute of
Technology, Cambridge, MA 02139, USA}
\affiliation{Theoretical Physics, Oxford University, 1 Keble Road, Oxford OX1 3NP, United Kingdom}
\author{Max A. Metlitski}
\affiliation{Department of Physics, Massachusetts Institute of
Technology, Cambridge, MA 02139, USA}
\affiliation{Perimeter Institute for Theoretical Physics, Waterloo, ON N2L 2Y5, Canada}
\affiliation{Kavli Institute for Theoretical Physics, University of California, Santa Barbara, CA 93106, USA}
\author{Cenke Xu}
\affiliation{Department of Physics, University of California,
Santa Barbara, CA 93106, USA}
\affiliation{Kavli Institute for Theoretical Physics, University of California, Santa Barbara, CA 93106, USA}
\author{T. Senthil}
\affiliation{Department of Physics, Massachusetts Institute of
Technology, Cambridge, MA 02139, USA}
\date{\today}

\begin{abstract}

\noindent
The deconfined quantum critical point (QCP), separating  the N\'eel and
 valence bond solid phases  in a 2D antiferromagnet, was proposed as an example of $2+1$D criticality  fundamentally different
from standard Landau-Ginzburg-Wilson-Fisher {criticality}. In this
work we present multiple  equivalent descriptions of deconfined QCPs, and use these to  address the possibility of enlarged emergent symmetries in the low energy limit. The easy-plane
deconfined QCP, besides its previously discussed self-duality, is
 dual to $N_f = 2$ fermionic quantum electrodynamics
(QED), which has its own self-duality and hence may have an
O(4)$\times Z_2^T$ symmetry. We propose several dualities for the deconfined QCP with
${\mathrm{SU}(2)}$ spin symmetry which together make natural the emergence of a previously suggested  $SO(5)$ symmetry rotating the N\'eel and VBS orders. 
 These emergent symmetries are implemented anomalously. {The  associated infra-red theories can also be viewed as surface descriptions of 3+1D topological paramagnets, giving further insight into the dualities. We describe a number of numerical tests of these dualities.  We also discuss the possibility of `pseudocritical' behavior for deconfined critical points, and the meaning of the dualities and emergent symmetries in such a scenario.}

\end{abstract}

\maketitle

\tableofcontents

\section{Introduction}

Zeus, the ruler of the Olympian gods, often conceals his
identity by changing himself into different forms. Strongly
interacting conformal field theories (CFTs), which underlie many
different states of matter, can sometimes also
be described by Lagrangians with very different forms. In other
words, two seemingly different `dual' Lagrangians may
correspond to the same CFT. The classic example of such a duality
is the equivalence between the 3D $O(2)$ Wilson-Fisher fixed point
and the Higgs transition of bosonic quantum electrodynamics
(QED) with one flavor of complex
boson~{\cite{Peskin,bosonvortexdh,bosonvortexfl}}. {\red Either theory
describes interacting lattice bosons at the quantum phase transition between a superfluid phase, in which  $U(1)$ symmetry is spontaneously broken, and a Mott 
insulating phase, in which it is not.}

This paper studies dualities for quantum phase transitions in
two spatial dimensions that lie outside the Landau paradigm. Our focus will be on non-Landau
transitions between two conventional phases, each of which is well
described by a Landau order parameter. The paradigmatic example of
such a phase transition occurs in two dimensional quantum magnets.
Square lattice spin-$1/2$ magnets allow (as the interactions are
changed) a conventional N\'eel antiferromagnetic phase which
breaks spin rotation symmetry, and a `Valence Bond Solid' (VBS)
phase: a crystal of spin singlets which preserves the spin
rotation symmetry while breaking lattice symmetries. A field
theory for a putative continuous phase transition between the
N\'eel and VBS phases was described in
Refs.~\cite{deccp,deccplong}. The theory
--- known as the noncompact CP$^1$ model (NCCP$^1$) --- is
formulated in terms of fractionalized `spinon' degrees of freedom
(a bosonic field $z_\alpha$, with $\alpha = 1,2$ an
 ${\mathrm{SU}(2)}$ flavor index) coupled to a non-compact $U(1)$ gauge
field $b$. Neither the spinon nor the gauge photon however exist
as deconfined quasiparticles in either phase. The phase transition
has hence been dubbed a `deconfined quantum critical point'.
Numerical work on specific quantum magnets and related systems
\cite{SandvikJQ,melkokaulfan,lousandvikkawashima,Banerjeeetal,Sandviklogs,Kawashimadeconfinedcriticality,Jiangetal,deconfinedcriticalityflowJQ,DCPscalingviolations,emergentso5,MotrunichVishwanath2,kuklovetalDCPSU(2),Bartosch,CharrierAletPujol, Chenetal,Aletextendeddimer,powellmonopole} shows a
striking (apparently\footnote{It is not yet clear whether the transition is truly second order or whether it displays only `quasiuniversal' behavior up to a very large but finite length scale \cite{Jiangetal, deconfinedcriticalityflowJQ,
Kawashimadeconfinedcriticality, Sandviklogs, kuklovetalDCPSU(2),sandvik2parameter,DCPscalingviolations,SimmonsDuffinSO(5),Nakayama}. We will discuss this in detail in the text.}) continuous phase transition with properties
broadly consistent with field theoretic expectations.   The deconfined criticality
scenario also generalizes to $SU(N)$ magnets with large $N$, where
there is a second order phase transition that is under good
theoretical control.

Analytic progress on the field theories for ${\mathrm{SU}(2)}$ deconfined
quantum critical points\footnote{The continuous global symmetry
of the theory is actually $SO(3) \times O(2)$ and not ${\mathrm{SU}(2)}
\times U(1)$ (see Sec.~\ref{symm_su2}). By a slight but standard abuse of terminology we
will nevertheless refer to it as the ${\mathrm{SU}(2)}$ $\nccp^1$ model.}
has been challenging. Refs.~\cite{tanakahu} and \cite{tsmpaf06}
showed that a formulation directly in terms of the Landau order
parameters for the two phases was possible using a non-linear
sigma model, but required the addition of a `topological' term to
correctly capture their competition/intertwinement. This term
endows the topological defects of each order parameter with
nontrivial symmetry properties, enabling the Landau forbidden
phase transition. This sigma model formulation gave rise to the
possibility that the phase transition may have a large emergent
symmetry, which rotates the two Landau order parameters (N\'eel
and VBS) into each other.

Remarkably, recent numerical work finds evidence for the
emergence of such a higher symmetry. Specifically a model described by
$\nccp^1$ (a field theory which
naively has only ${SO(3) \times O(2)}$ symmetry) was seen to have an
emergent ${\mathrm{SO}(5)}$ symmetry at the critical point, at the length
scales accessible in the calculations \cite{emergentso5}.  A good understanding of this
emergent extra symmetry is currently not available.

In a different direction, many fascinating new dualities for field
theories with $U(1)$ gauge fields have been found very
recently \cite{wangsenthil15b,MaxAshvin15,dualdrMAM,qeddual,seiberg1,karchtong,murugan,seiberg2}.
These dualities originated from studies of the surface of three
dimensional symmetry protected topological (SPT)
phases \cite{wangsenthil15b,MaxAshvin15,qeddual}, their relation to
three dimensional quantum spin liquids with an emergent $U(1)$
gauge field \cite{wangsenthil15a,MaxAshvin15,Max15}, and the
physics of the half-filled Landau level of two dimensional
electrons \cite{sonphcfl,wangsenthil15c,geraedtsnum}. {\red Other related dualities were discussed in a high energy context (see Ref.~\cite{Aharony2016} and references therein). }

From a modern point of view, the classic infra-red duality between the $O(2)$ Wilson-Fisher theory and bosonic QED is natural because the two field theories have the same global symmetry, namely $U(1)$, the same allowed quantum numbers for gauge-invariant local operators (here bosons of integer charge) and the same anomalies (none in this case). The two ultra-violet Lagrangians can therefore be viewed as descriptions of the same physical system with different bare interactions, making it possible that their long distance behaviour is the same. The dualities mentioned above are extensions of this idea to situations in which the  operator content and anomalies are nontrivial. In this paper we apply this philosophy to the field theories for deconfined critical points.

We {propose and analyze} dualities involving these field theories, paying
special attention to the realization of symmetries. These dual descriptions give a new way of understanding emergent symmetries relating the Landau order
parameters of the N\'eel and VBS phases.

For the easy-plane version of the $NCCP^1$ model (describing the
N\'eel-VBS transition in magnets with $XY$ spin symmetry) several
dual descriptions have been discussed in the old and recent
literature, as we review below.  Some of these dual theories are formulated in terms of bosonic fields while others involve fermionic fields.  These boson/fermion fields are coupled to a dynamical $U(1)$ gauge field. 
Here we unify these different dual
descriptions into a common duality web, and clarify the emergent
symmetries of the putative critical fixed point.

For the ${\mathrm{SU}(2)}$--symmetric $NCCP^1$ model, we propose a dual fermionic
description as massless $QED_3$ coupled to a critical real scalar
field. We refer to this theory as QED$_3$-Gross-Neveu. We show that
this duality implies the emergent ${\mathrm{SO}(5)}$ symmetry at this
deconfined critical point (as observed in numerical simulations).
We are then led to propose a duality web for the  ${\mathrm{SU}(2)}$
symmetric $NCCP^1$ theory as well.  The existence of this duality
web provides an alternate point of view on the emergence of the
${\mathrm{SO}(5)}$ symmetry at the deconfined critical point.

Remarkably the duality web implies that the ${\mathrm{SU}(2)}$-symmetric $NCCP^1$ model is itself self-dual.  Indeed, if we assume this self-duality, the ${\mathrm{SO}(5)}$ symmetry follows as an inescapable consequence. Conversely the existing evidence for the emergent ${\mathrm{SO}(5)}$ symmetry strongly supports the conjectured self-duality of the ${\mathrm{SU}(2)}$-symmetric $NCCP^1$
model.

We will show that useful insight into these field theories is obtained by realizing them at the boundary
of $3+1$D bosonic Symmetry Protected Topological (SPT) phases.
This allows the theories to be regularized in a way which preserves the full internal symmetry of the putative IR fixed point. By contrast, the full internal symmetry of the IR theory cannot be incorporated  microscopically in a strictly 2D quantum magnet: it can only be emergent. In field theoretic parlance the symmetry of the IR theory is anomalous, and the anomaly is cancelled when the theory  resides at the boundary of a $3+1$D SPT phase.
Furthermore, in the easy plane
case, we show how the `bulk' $3+1$D description provides a very
simple explanation for the existence of the duality web and the
symmetry realizations of the various theories contained therein.
For the ${\mathrm{SU}(2)}$ invariant case,  with its putative emergent ${\mathrm{SO}(5)}$ symmetry, we describe a manifestly ${\mathrm{SO}(5)}$ invariant formulation in terms of massless fermions coupled to an ${\mathrm{SU}(2)}$ gauge field, a theory we denote $N_f = 2$ QCD$_3$. This 2+1D theory is shown to have the same anomaly as the proposed ${\mathrm{SO}(5)}$ invariant fixed point associated with the $\nccp^1$ theory. This allows us to show that there is a corresponding bulk SPT phase of bosons with global ${\mathrm{SO}(5)}$ symmetry (i.e. an ${\mathrm{SO}(5)}$ `topological paramagnet'). This boson SPT  is characterized in the bulk by its response to an external background ${\mathrm{SO}(5)}$ gauge field.  This response includes a non-trivial discrete theta angle, introduced in Ref.~\cite{ahseta}, which  distinguishes it from a  trivial gapped phase of ${\mathrm{SO}(5)}$--symmetric bosons. The $2+1$D theories with anomalous ${\mathrm{SO}(5)}$ symmetry are alternative descriptions of the surface of this $3+1$D boson SPT phase.

It is important to distinguish {two different} versions of
statements about duality of quantum field theories that are
conflated in the literature. First, there are `weak' duality
statements. These assert that the two theories in question have
the same local operators, the same symmetries, and the same
anomalies (if any). In condensed matter parlance, this means that
the two theories `live in the same Hilbert space' and can be
viewed as descriptions of the same microscopic system in different
limits. These weak dualities are nontrivial statements that can be
unambiguously derived. In the context of the present paper their
main interest is that they open up the possibility of `strong'
dualities. The strong dualities will hold if the putatively dual
theories flow, without fine-tuning, to the same nontrivial IR fixed point.
For the theories discussed here, it is these strong dualities that
would imply the emergence of large, exact symmetries in the infrared.  We will
not derive the strong dualities in this paper but rather view them
as plausible conjectures suggested by the weak dualities.

In fact, the strong dualities can be relevant to the physics up to a very
large length scale even in the absence of a true fixed point,  if
the system shows quasi-universal `pseudocritical' behavior up to a
large length scale. We emphasize that it is not yet clear whether the theories we discuss do flow to nontrivial IR fixed points: this is an ongoing question for numerical work.  But for the ${\mathrm{SU}(2)}$ symmetric $\nccp^1$ model, simulations  show that there is at least apparent critical
behavior up to a remarkably large length scale. Numerical evidence
for ${\mathrm{SO}(5)}$ in this regime supports the applicability of the
${\mathrm{SO}(5)}$ web of dualities. For QED$_3$, very recent simulations argued
for a flow to a conformal fixed point~\cite{qedcft,kevinQSH,so4qsh},
while earlier studies argued for (very weak) chiral symmetry
breaking~\cite{kogut2}. For the easy-plane $\nccp^1$ model the current numerical
consensus is that the transition is weakly first order. The
duality to QED$_3$ suggests that it may be worth revisiting the
N\'eel-VBS transition in easy plane magnets and related models to
look for a second order transition.

We describe consequences of the strong duality conjectures that
may be tested in future numerical work. Our proposed duality web for
${\mathrm{SU}(2)}$-invariant $\nccp^1$ and QED$_3$--Gross-Neveu involves an
emergent ${\mathrm{SO}(5)}$ symmetry, and leads to clear and testable
predictions for the behavior of 2-flavor QED$_3$ when it is
coupled to a critical real scalar field. The web of dualities
involving easy-plane $\nccp^1$ and two-flavour QED$_3$ are
naturally thought of in terms of a `mother' theory with an $O(4)$
symmetry which rotates the N\'eel and VBS order parameters. For
QED$_3$ this emergent symmetry should have striking numerically
accessible consequences. {Our results also show how numerical and analytical studies of QED$_3$ and QED$_3$--GN will provide new information about deconfined criticality.}

The duality transformations we employ involve global symmetries with a $U(1)$ subgroup.  For a $2+1$D CFT with a global $U(1)$ symmetry there are two basic formal transformations  --- denoted $S$ and $T$ --- which map the theory to other inequivalent theories with a global $U(1)$ symmetry, assumed  also to be CFTs \cite{kapstr,witten03}. Our duality transformations can be viewed within this framework. However there are a  number of caveats about the standard use of the  $S$ and $T$ transformations which we discuss in Appendix~\ref{sandt}. Making standard assumptions about the effect of $S$ and $T$ on CFTs allows  stronger assertions about deconfined critical points and their symmetries than those discussed above. However it is not clear at this point whether these standard assumptions can be trusted far from the context in which they were originally discussed, i.e. in non-supersymmetric theories that are far from any large $N$ limit.

\section{Preliminaries and Summary of Results}
\label{prelim}

\subsection{Deconfined quantum criticality: $NCCP^1$ and related
models}
\label{deccprev}

We first briefly recall the theory of deconfined quantum critical
points in quantum magnets. For a spin-$1/2$ quantum
antiferromagnet on a two dimensional square lattice, the
transition between the N\'eel ordered magnet and the Valence Bond
Solid (VBS) paramagnet is potentially second order and is
described by the $NCCP^1$ field theory
\begin{equation}
\label{nccp1su2} {\cal L}_0 = \sum_{\alpha = 1,2} |D_b z_\alpha|^2
- \left(|z_1|^2 + |z_2|^2\right)^2.
\end{equation}
Here $z_\alpha$ ($\alpha = 1,2$) are bosonic spinons coupled to a
dynamical $U(1)$ gauge field $b$, and $D_{b,\mu}=\partial_{\mu}-ib_{\mu}$ is the covariant derivative. (This action and  subsequent
similar actions are short-hands for the appropriate strongly
coupled  Wilson-Fisher critical theory where a background gauge
field has been promoted to a dynamical field; unless otherwise
specified they are written in Minkowski signature.)   The model
has a global $SO(3)$ symmetry under which $z_\alpha$ transforms
as a spinor.\footnote{{Though $z_\alpha$ transforms as a spinor
under $SU(2)$, rotations in the center of $SU(2)$ can be
compensated by a $U(1)$ gauge transformation so that the spin
rotation symmetry of the model is $\frac{SU(2)}{Z_2} = SO(3)$. }}
 In the microsopic lattice spin model, this corresponds to the
$SO(3)$ spin rotation. It also has a global $U(1)$ symmetry
associated with the conservation of the flux\footnote{It is
customary in condensed matter literature to call a $U(1)$ gauge
field ``non-compact" if its flux is conserved. In high energy
literature a $U(1)$ gauge field is called ``compact" if monopole
(instanton) operators are local. In this paper we use the
condensed matter notation, and call our gauge fields
``non-compact". Of course they are also ``compact" in the high
energy sense, {\em i.e} monopole operators can, in principle, be
added to the action.} of $b$. In the microsopic lattice spin
model, this is not an exact symmetry. Consequently monopole
operators (which pick up a phase under a $U(1)$ rotation)
 must be added to the Lagrangian.
However, it is known that lattice symmetries ensure that the
minimal  allowed monopole operator (with continuum angular momentum $\ell = 0$) has
strength 4. Analytic arguments \cite{deccp,deccplong} and numerical
calculations \cite{SandvikJQ,lousandvikkawashima,DCPscalingviolations} strongly support the
possibility that these monopoles are irrelevant at the critical
fixed point of Eq.~\eqref{nccp1su2}. The N\'eel phase is obtained
when $z_\alpha$ is condensed, and the VBS phase when $z_\alpha$ is
gapped. The N\'eel phase breaks $SO(3)$ to a $U(1)$ subgroup while
the VBS phase breaks the $U(1)$ flux conservation symmetry. The
N\'eel order parameter is simply $\v{N} = z^\dagger \v{\sigma} z$
($\v{\sigma}$ are Pauli matrices), and the VBS order parameter is
the strength 1 monopole operator ${\cal M}_b$ which creates $2\pi$
flux of $b$.

If the underlying spin model has only O(2) ($XY$) spin symmetry
--- corresponding to conservation of the $z$-component of spin,
together with a discrete $\pi$-rotation of the spins around the
$x$-axis which we denote $\mathcal{S}$ ---  then the N\'eel-VBS
phase transition is described by the theory \be \label{epnccp1}
{\cal L}_{ep-cp1} = \sum_{\alpha = 1,2} |D_b z_\alpha|^2 -
\left(|z_1|^4 + |z_2|^4\right)+... . \ee This is known as the easy
plane $NCCP^1$ model. In this model the $XY$ N\'eel order parameter
is $N_x + i N_y = 2 z_1^*z_2$, while the VBS order parameter is
the monopole operator $\mathcal{M}_b$.  Note that under the $Z_2$
spin flip symmetry $\mathcal{S}$ we have 
\be \label{calSdef}
{\mathcal S}: \quad z \to \sigma_x z, \quad b \to b. 
\ee
Then under ${\mathcal S}$ the $XY$
N\'eel order parameter transforms as $N_x + i N_y \to N_x - i N_y$  and the  VBS
order parameter ${\cal M}_b$ is invariant, as
expected microscopically.  Later in the paper we will describe the
action of time reversal and lattice symmetries for square lattice antiferromagnets (Sec.~\ref{symm_ep}).

The easy-plane theory is known to be self-dual~\cite{lesikav04},
in the sense that it is dual to another easy-plane $NCCP^1$ theory
\be \label{epcp1selfdual} {\cal L}_{ep-cp1-dual} = \sum_{\alpha =
1,2} |D_{\tilde{b}} w_\alpha|^2 - \left(|w_1|^4 + |w_2|^4\right)+... ,
\ee in which the roles of the two order parameters are switched: $w_1^*w_2$
is the VBS order parameter, while $\mathcal{M}_{\tilde{b}}$ is the
$XY$ order parameter. This self-duality is obtained by applying the
particle-vortex duality to both spinons: $z_1\to w_2$, $z_2\to
w_1^*$. Since the boson mass term is odd under the particle-vortex
duality, the self-duality sends $|z_1|^2\to -|w_2|^2$ and
$|z_2|^2\to -|w_1|^2$.

The IR fates of the two $NCCP^1$ models, and their generalizations
with an $N$-component spinon field $z_\alpha$, have been discussed
extensively. They flow to conformal field theories within a $1/N$
expansion. Directly at $N = 2$, numerical calculations see an
apparently continuous transition in the $SU(2)$ invariant $NCCP^1$
model, but with drifts in some critical properties {(which we will discuss in Sec.~\ref{pseudocritical}).}
Further recent studies show the
emergence of an $SO(5)$ symmetry that rotates the N\'eel and VBS
order parameters into one another. For the easy plane case the
current wisdom is that the N\'eel-VBS transition is weakly first
order. However, as we will discuss at length, the potential duality
with $QED_3$ may make it interesting to examine this further.

These gauge theories give a natural route to a second order
transition between two distinct symmetry broken phases, despite
the fact that such a transition is naively forbidden by  the Landau
theory. In contrast to the standard Landau-Ginzburg-Wilson
description, the critical theory is expressed in terms of
`deconfined' degrees of freedom (the spinons and the gauge field)
which do not describe sharp quasiparticles in either phase.
Physically the breakdown of the Landau paradigm occurs because the
topological defects of either order parameter carry non-trivial
quantum numbers{: the skyrmion defect of the N\'eel
phase carries quantum numbers under lattice symetries \cite{HaldaneBerry,ReSaSUN,deccp,deccplong}, and
the vortex defect of the VBS phase carries
spin-$1/2$ \cite{mlts04}.}

There is an alternative formulation \cite{tanakahu,tsmpaf06} for
the competition between the two order parameters directly in terms
of a non-linear sigma model. In the $SU(2)$ invariant case, we
define a real $5$-component unit vector $n^a$ ($a = 1,...., 5$)
such that $n^{3,4,5}$ correspond to the three components of the
N\'eel vector, and $n^{1,2}$ to the two real components of the VBS
order parameter. The intertwined fluctuations of the two order
parameters are then described by an $SO(5)$ action with a
Wess-Zumino-Witten (WZW) term at level $1$:
\begin{equation}
S =  \frac{1}{2g}  \int d^3x \, (\partial  n^a)^2 + 2\pi
\Gamma\left[ n^a \right].
\end{equation}
The WZW term $\Gamma$ is defined in the standard way: the field
$n^a$ defines a map from spacetime $S^3$ to the target space
$S^4$, and $\Gamma$ is the ratio of the volume in $S^4$ traced out
by $n_a$ to the total volume of $S^4$. If $n^a(x,u)$ is any smooth
extension of $n^a(x)$ such that $n^a(x,0) = (0,0,0,0,1)$ and
$n^a(x,1) = n^a(x)$, then
\begin{equation}
\Gamma  =  \frac{\epsilon_{abcde}}{\textrm{Area}(S^4)}
\int_{0}^{1} d\,u \int d^3 x \, n^a\partial_x n^b \partial_y n^c
\partial_t n^d \partial_u n^e.
\end{equation}
In order to share the symmetry of the $\nccp^1$ model, the above
action must also be supplemented with anisotropy terms that break
$SO(5)$ to $SO(3) \times U(1)$. The WZW term correctly captures
the non-trivial quantum numbers of the topological defects and is
responsible for the non-Landau physics.  For example,
if the $U(1)$ symmetry is spontaneously broken, a vortex in the
$U(1)$ order parameter will carry spin-$1/2$ under the unbroken
$SO(3)$.

The easy plane case can be obtained from this theory by setting
$n^5 = 0$. This then leads to an $O(4)$ non-linear sigma model in
$2+1$ spacetime dimensions supplemented with a $\theta$ term at
$\theta = \pi$:

\be
S=\int d^3x\left[\frac{1}{2g}(\partial n^a)^2+\frac{\theta \epsilon_{abcd}}{\textrm{Area}(S^3)}n^a\partial_tn^b\partial_xn^c\partial_yn^d  \right].
\label{eq:introsigmapi}
\ee

 The value $\theta=\pi$ is robust as a result of
the $\mathbb{Z}_2$ spin-flip symmetry ${\mathcal S}$ of the
easy-plane $NCCP^1$ model, which changes the sign of $n_5$ and
therefore acts as $\theta\to-\theta$. This topological term is
once again responsible for the non-trivial structure of the
topological defects.

The sigma model formulation raises the possibility that the phase
transition described by Eq.~\eqref{nccp1su2} may have an emergent
$SO(5)$ symmetry ($O(4)$ in the easy plane case). However, we
should emphasize that the sigma model is well-defined as a continuum field theory only in the weak coupling limit, where it is
ordered. Here there is a clear semiclassical picture for the
effect of the WZW term ($\theta$ term in the $O(4)$ case) on the
topological defects in the ordered state.
For the transition itself --- driven by anisotropy for the $SO(5)$
or $O(4)$ vector --- this ordered state corresponds to a first
order phase transition. To study second order Landau-forbidden
transitions we need to extend the model to strong coupling, and
look for a disordered but power-law correlated  $SO(5)$ invariant
fixed point\footnote{Specifically we need to look for a
power-law correlated fixed point which has only one relevant
perturbation that breaks $SO(5)$ to  $SO(3) \times O(2)$ (or rather
the symmetry of the lattice model, which is slightly smaller).
This perturbation will correspond to the operator whose
coefficient is tuned to place the lattice model at its critical
point.}. At strong coupling the sigma model theory is
non-renormalizable and requires an alternative formulation as a
continuum field theory. Physically, disordered phases of the sigma
model (defined with an explicit UV cutoff) correspond to phases
where topological defects of the order parameter have
proliferated. Thus a modification of the topological defects leads
to modifications of the corresponding disordered phases. The sigma
model formulation thus exposes the seed, in the ordered phase, of
the impending non-Landau physics of the disordered critical
regime.

  Yet another formulation \cite{tanakahu,tsmpaf06} of the intertwinement of the N\'eel and VBS orders that maintains manifest $SO(5)$ symmetry may be  obtained by starting with a fermionic spinon description of the square lattice spin-$1/2$ magnet. This naturally leads to a low energy theory of two flavors of massless Dirac fermions coupled to a dynamical $SU(2)$ gauge field --- a theory we denote $N_f = 2$ $QCD_3$.  This theory will be useful for some purposes: we discuss it further in  Secs.~\ref{manifest SO(5)}, \ref{sec:mathy}.
  
{Finally, deconfined critical field theories also  arise in the context of phase transitions between trivial and SPT phases \cite{ashvinsenthil,kevinQSH,so4qsh}. We will review this connection as needed later in the paper.}
 
\subsection{Fermionic $N_f = 2$ QED$_3$ and related models}
\label{qedrev}

We now turn our attention to fermionic massless QED$_3$ models
with $N_f = 2$ flavors of two-component fermions:\footnote{Here we use a more
``traditional" procedure of defining a Dirac fermion action.
Namely a single Dirac fermion should come together with a
Chern-Simons term at level $k=\pm1/2$ to avoid a gauge anomaly. A more
precise way to define this theory is to use the procedure in
Ref.~\cite{wittenreview,seiberg1} where the partition function of
a massless Dirac fermion is written as $ Z_\psi =
|Z_\psi| e^{-i\pi \eta[A,g]/2}$, where $A$ is the gauge field,
either dynamical or background, and $g$ is the space-time metric.
$\eta$ is defined in terms of eigenvalues of the Dirac
operator \cite{wittenreview}, see Eq.~(\ref{eq:etadef}). {This form corresponds to UV-completing the massless Dirac theory by adding two extra `heavy' Dirac fermions with the same sign of Dirac mass. This enables retaining flavor $SU(2)$ rotations as an exact symmetry. If we do not care about this $SU(2)$ symmetry then we can choose the two heavy fermions to have opposite masses in which case the partition function is real. } Further strictly speaking $A$ should
be regarded as a spin$_c$ connection and not an ordinary $U(1)$
gauge field, which means that fields with odd charge are
fermions \cite{Max15}.   The more precise form of the Lagrangian will be presented in detail in Appendix~\ref{PreciseL}.}
\beqn \label{qed3} {\cal L}_{qed} = \sum_{j = 1}^2 i\bar{\psi}_j
\slashed{D}_a \psi_j + \cdots \eeqn where $\slashed{D}_a=\gamma^{\mu}D_{a,\mu}$
is the
gauge covariant Dirac operator that involves a dynamical
noncompact U(1) gauge field $a_\mu$ (we choose $\gamma^{0,1,2}=\{\sigma^y, i\sigma_z, i\sigma_x \}$ and $\bar{\psi}=\psi^{\dagger}\gamma^0$). The flavor symmetry of the model will play an important role in our discussion. We will often, but not always,  restrict attention to the case with symmetry under $SU(2)$ rotations between the two flavors.  In addition there is a global $U(1)$ symmetry associated with the conservation of the flux of the gauge field $a$.  The  theory then has
 manifest global $\frac{SU(2) \times
U(1)}{Z_2}$ symmetry.\footnote{Note that  the fermions themselves transform as spinors under the flavor $SU(2)$ but rotations by the element of the center $Z_2$ can be compensated by a $U(1)$ gauge rotation.  This might naively lead to the expectation that the global symmetry,  excluding charge conjugation, is $SO(3) \times U(1)$.   However the theory has monopole operators which are local and which transform as spinors
under the $SU(2)$. Rotations by the center of $SU(2)$ on these can now be compensated by $\pi$ rotations under the flux conservation {\em global} $U(1)$, and therefore the global symmetry apparent in the Lagrangian, excluding charge conjugation, is $\frac{SU(2) \times U(1)}{Z_2}$.} 
(The full manifest symmetry of the field theory is larger once charge conjugation is included.\footnote{Including charge conjugation gives $[\mathrm{SU}(2)\times \operatorname{Pin}(2)_-]/Z_2$. The notation $\operatorname{Pin}(2)_-$ means that the charge conjugation operation, which reverses the global U(1) charge, squares to the $-1$ element of U(1).}) It will sometimes however be convenient to consider a more general class of QED$_3$ theories where the two fermion species are not related by $SU(2)$ rotations {but only by a discrete exchange, so that SU(2) is reduced to $\operatorname{Pin}(2)_-$. Below we often neglect discrete symmetry generators, and will refer to this case as having $\mathrm{U}(1)\times \mathrm{U}(1)$ symmetry.}\footnote{The full manifest symmetry in this case is $[\operatorname{Pin}(2)_-\times \operatorname{Pin}(2)_-]/Z_2$.}

By applying the fermion-fermion duality of a single species of Dirac fermion to each of the two fermion species, Refs.~\cite{qeddual,karchtong,seiberg2} demonstrated
that, similar to the bosonic easy-plane $CP^1$ model, this theory
is self-dual, $i.e.$ it is dual to another $N_f = 2$ QED \be
\label{qeddual} {\cal L}_{qed-dual} = \sum_{j = 1}^2
i\bar{\chi}_j\slashed{D}_{\tilde{a}} \chi_j + \cdots. \ee
Given that a particular basis in flavour space had to be selected to perform this duality, we are, strictly speaking, restricting to theories with just  $U(1) \times U(1)$ {continuous} symmetry. The dual theory in Eqn. \ref{qeddual} then should also only be taken to have $U(1) \times U(1)$  {continuous} symmetry. However, we will later discuss the possibility that with full $SU(2)$ flavor symmetry this duality survives.
As in the
easy-plane $NCCP^1$ model, the roles of the gauge-flux conservation
symmetry and the relative phase rotation symmetry between the two
Dirac fermions are exchanged in the dual QED theory. The
self-duality is obtained by applying the fermionic particle-vortex
duality~\cite{sonphcfl,wangsenthil15a,wangsenthil15b,MaxAshvin15}
to both flavors of fermions: $\psi_1\to \chi_2$, $\psi_2\to
\chi_1$. Since the Dirac mass term is odd under the
particle-vortex duality, the self-duality sends
$\bar{\psi}_1\psi_1\to -\bar{\chi}_2\chi_2$ and
$\bar{\psi}_2\psi_2\to -\bar{\chi}_1\chi_1$.

The IR fate of QED$_3$ at $N_f=2$ is controversial at this stage.
It is not clear whether at low energy the Dirac fermions will
spontaneously break the flavor symmetry and gain a mass of the
form $m\bar{\psi}\sigma^z\psi$ --- a long-standing issue known as
chiral symmetry breaking. Recent numerics~\cite{qedcft}, however,
suggests the possibility that this theory may be stable in the IR
(although an earlier study suggests spontaneous
chiral symmetry breaking~\cite{kogut2}).

We will also be interested in the phases and phase transition of
this model when a coupling to an extra scalar $\phi$ is allowed.
The resulting model has the Lagrangian \be \label{lqed3gn} {\cal
L}_{qed-gn} =  \sum_{j = 1}^2 i\bar{\psi}_j \slashed{D}_a \psi_j +
\phi \bar{\psi} \psi + V(\phi) \ee Here we have included a
potential $V(\phi) = V(-\phi)$ for the scalar field $\phi$ (we
have suppressed its kinetic term for notational simplicity).  The
theory is time-reversal symmetric if under  time reversal $\phi
\rightarrow - \phi$. As the potential $V(\phi)$ is tuned, we expect
a phase transition between a time-reversal symmetric phase where
$\langle \phi \rangle = 0$ and a time reversal broken one where
$\langle \phi \rangle \neq 0$. We will usually refer to
Eq.~\eqref{lqed3gn} when tuned to this transition as the
QED$_3$-Gross-Neveu model (QED$_3$--GN for short).

Interestingly, with some assumptions, Ref.~\cite{tsmpaf06} showed
that the low energy behavior of $N_f = 2$ QED$_3$ was described by
the $O(4)$ sigma model at $\theta = \pi$ discussed in the previous
subsection, again with the proviso that the sigma model needs to
extended to strong coupling. This suggests a connection between
the bosonic $NCCP^1$ theories and the fermionic QED$_3$ theories.
Below we will sharpen this connection through precise duality
statements. This will also enable us to understand the emergent IR
symmetries of these theories at their critical point.

\subsection{Summary of results}

We now summarize the key results in this paper. We also point out the sections that discuss these statements (and their subtleties) in detail. This section can be viewed as a map of the paper.

\begin{enumerate}

\item 

Both the easy plane and the $SU(2)$ symmetric $NCCP^1$ models are
part of a web of dualities.

We begin with the easy plane model. It turns out that the easy-plane $NCCP^1$ model is dual to fermionic QED$_3$ with $N_f=2$ Dirac fermions. This duality was first mentioned in Ref.~\cite{karchtong}, and will be discussed in detail in Sec.~\ref{easy_plane_duality}.  As discussed in the literature and reviewed earlier in this section, the two theories also possess their own self-dualities. This leads to a web of four theories, mutually dual to each other, as summarized below:
\begin{widetext}
\bea \label{epweb} &&|D_{b + B} z_1|^2+|D_{b + B'}z_2|^2-
|z_1|^4 - |z_2|^4 - \frac{1}{2\pi}bd(B + B')-\frac{1}{2\pi}BdB'-\frac{1}{2\pi}B'dB'
\nn \Longleftrightarrow  \hspace{20pt} && |D_{\tilde{b} - B}
w_1|^2+|D_{\tilde{b} + B'}w_2|^2- |w_1|^4 - |w_2|^4 -
\frac{1}{2\pi}\tilde{b}d(B' - B)+\frac{1}{2\pi}BdB'-\frac{1}{2\pi}B'dB' \nn
\Longleftrightarrow  \hspace{20pt} && \bar{\psi}_1i\slashed{D}_{a
- B}\psi_1+\bar{\psi}_2i\slashed{D}_{a + B}\psi_2
+\frac{1}{2\pi}adB'+\frac{1}{4\pi}(BdB-B'dB') \nn
\Longleftrightarrow  \hspace{20pt} &&
\bar{\chi}_1i\slashed{D}_{\tilde{a} -
B'}\chi_1+\bar{\chi}_2i\slashed{D}_{\tilde{a} + B'}\chi_2
+\frac{1}{2\pi}\tilde{a}dB+\frac{1}{4\pi}(BdB-B'dB'). \eea
\end{widetext}
Here $b$ and $\tilde{b}$ are ordinary dynamical
$U(1)$ gauge fields,  $a$ and $\tilde{a}$ are dynamical gauge fields whose charge-$1$ fields are fermions (they are formally known as spin$_c$
connections). We have also included \textit{background} $U(1)$
gauge fields  $B$ and $B'$.\footnote{Strictly speaking, we will take $B + B'$ and $B-B'$ to be properly quantized $U(1)$ gauge fields.} Various background Chern-Simons terms
are included to ensure that all the theories have the
same response.  Despite the profusion of background terms, the dynamical content of the theories in Eq.~\ref{epweb} is simple. Sec.~\ref{easy_plane_duality} will also discuss various possibilities for the IR fates of the theories in the duality web, paying careful attention to symmetries.

There is a similar web of dualities for the $SU(2)$-invariant
$NCCP^1$ model, which we discuss in Sec.~\ref{su2cp1-qedgn}. The structure of this duality web is very similar to that of the easy-plane case: the $SU(2)$-invariant $NCCP^1$ model is dual to the QED$_3$-Gross-Neveu (QED-GN) theory with $N_f=2$, and the two theories both admit their own self-dualities. We summarize the mutual dualities of the four theories in the web below: 
\bea \label{so5dualweb} &&\sum_{\alpha
=1,2}|D_bz_\alpha|^2-\lp  |z_1|^2+ |z_2|^2 \rp^2 \nn
\Longleftrightarrow \hspace{20pt} &&\sum_{\alpha
=1,2}|D_{\tilde{b}}w_\alpha|^2-\lp |w_1|^2+|w_2|^2 \rp^2 \nn
\Longleftrightarrow \hspace{20pt}
&&\sum_{j=1,2}\bar{\psi}_ji\slashed{D}_a\psi_j + \phi\sum_{j =
1,2}\bar{\psi}_j \psi_j +V(\phi)  \nn \Longleftrightarrow
\hspace{20pt}
&&\sum_{j=1,2}\bar{\chi}_ji\slashed{D}_{\tilde{a}}\chi_j
-\phi\sum_{j = 1,2}\bar{\chi}_j \chi_j +V(\phi), \eea
where the potential of Ising scalar field $V(\phi)$ is tuned to the critical point.

\item

Understanding the duality web allows some powerful statements about emergent symmetries of possible IR fixed points for the two $NCCP^1$ models which is one of the main goals of this paper. 
In the context of deconfined quantum criticality we will show that in both the easy plane and $SU(2)$ symmetric cases the emergent symmetry enables rotating the Landau order parameters of the two
phases on either side of the transition.

In the easy plane case, the duality web in its strongest form implies an emergent $O(4)$ symmetry. The most basic local (gauge-invariant) operators are the order parameters $(n_1, n_2, n_3, n_4)$, which form a vector representation of the $O(4)$ symmetry. Since $SO(4)\sim (SU(2)\times SU(2))/\Z_2$, the vector operators can rearranged into $SU(2)$ spinors. The complex doublet
\be
{(\Phi_1, \Phi_2) \sim (\overbrace{n_3+in_4}^{\text{N\'eel}}, \overbrace{n_1+in_2 }^{\text{VBS}} )}
\ee 
is a fundamental under the first $SU(2)$, and ${(\Phi_1^*,-\Phi_2)}$ is a fundamental under the second. The improper $O(4)$ reflection is represented as complex conjugation on one of the components of $\Phi$. The two complex operators are represented in each theory as:
\bea
(\Phi_1,\Phi_2) &\sim &( z_1^{\dag}z_2,    \mathcal{M}_b)  \nn
&\sim & ( \mathcal{M}^{\dagger}_{\tilde{b}},  w_2^{\dag}w_1) \nn
&\sim & ( \psi_2^{\dagger}\mathcal{M}_a,  (\psi_1^{\dagger}\mathcal{M}_a)^{\dagger}) \nn
&\sim & \left( (\chi_2^{\dagger}\mathcal{M}_{\tilde{a}})^{\dagger}, (\chi_1^{\dagger}\mathcal{M}_{\tilde{a}})^{\dagger} \right),
\eea
 where $\mathcal{M}_b$ is the monopole (instanton) operator for the gauge field $b$. (In QED the monopole configuration induces one zero mode for each Dirac fermion, and gauge invariance requires exactly one of the two zero modes to be filled {\cite{kapustinqed}}. The notation $\psi_1^{\dagger}\mathcal{M}_a$ denotes a monopole in $a$ with the zero mode from the Dirac fermion $\psi_1$ filled.) The $O(4)$ symmetry is discussed in detail in Sec.~\ref{easy_plane_duality}. Its implications for numerical simulations are discussed in Sec.~\ref{implications}.

In the $SU(2)$-invariant case, the duality web implies an emergent $\mathrm{SO}(5)$ symmetry. The most basic local operators form a vector representation of this $\mathrm{SO}(5)$: $(n_1,n_2,n_3,n_4,n_5)$. They are represented in $NCCP^1$ and QED-GN theories as
\begin{widetext}
 \bea
(n_1, n_2, n_3, n_4, n_5)&
 \sim & (2\,{\rm Re} \mathcal{M}_b,2\, {\rm Im}\mathcal{M}_b, z^\dag {\sigma_x} z, z^\dag {\sigma_y} z,z^\dag {\sigma_z} z)  \nn 
&\sim & (w^\dag {\sigma_x} w, -w^\dag {\sigma_y} w,2\, {\rm Re} \mathcal{M}_{\tilde{b}}, - 2\,{\rm Im}\mathcal{M}_{\tilde{b}}, w^\dag {\sigma_z}w) \nn  
&\sim& ({\rm Re} (\psi_1^{\dagger}\mathcal{M}_a), -{\rm Im}  (\psi_1^{\dagger}\mathcal{M}_a), {\rm Re}(\psi_2^{\dagger}\mathcal{M}_a), {\rm Im} (\psi_2^{\dagger}\mathcal{M}_a), \phi) \nn
&\sim& ({\rm Re} (\chi_1^{\dagger}\mathcal{M}_{\tilde{a}}), -{\rm Im}  (\chi_1^{\dagger}\mathcal{M}_{\tilde{a}}), {\rm Re}(\chi_2^{\dagger}\mathcal{M}_{\tilde{a}}), -{\rm Im} (\chi_2^{\dagger}\mathcal{M}_{\tilde{a}}), \phi)
\eea
\end{widetext}

This $\mathrm{SO}(5)$ symmetry is discussed in detail in Sec.~\ref{su2cp1-qedgn}. It has been numerically observed in Ref.~\cite{emergentso5}, providing a strong support to our duality web. Its further implications are discussed in Sec.~\ref{PhaseD} and \ref{implications}.

\item
The easy-plane theory has several $\Z_2$ (or $\Z_2$-like) symmetries which are anomalous. In the context of lattice quantum magnetism these symmetries include the $\Z_2$ spin-flip, time-reversal and lattice translation symmetries (see Sec.~\ref{discrete_symm_section}). Under these symmetries the Lagrangian is invariant only up to a background term:
\be
\mathcal{L}\to \mathcal{L}-\frac{1}{2\pi}B_1dB_2,
\ee
where $B_1=B-B'$ and $B_2=B+B'$ are the properly quantized background $U(1)$ gauge fields. On the lattice this anomaly is harmless since one of the $U(1)$ symmetries is really a discrete lattice rotation symmetry. However, if one wants to formulate the theory with all these symmetries realized in an on-site manner, the theory can only exist on the surface of a three-dimensional bulk. The symmetry anomaly can be canceled by a bulk mutual $\Theta$-term
\be
-\frac{1}{4\pi}\int_{Bulk}dB_1\wedge dB_2.
\ee
All the dualities on the surface are then related to the electric-magnetic dualities in the bulk. Many symmetry actions that appear to be complicated on the surface (in certain pictures) become obvious once the bulk view is taken. We discuss this in Sec.~\ref{bulk} for the easy plane theory.

\item
None of the field theories in the duality web, Eq.~\eqref{so5dualweb}, posesses the full $\mathrm{SO}(5)$ symmetry explicitly --- the symmetry is at best emergent in the IR.    Further, just as in the easy-plane case,  the $\mathrm{SO}(5)$ symmetry is anomalously realized. We also discuss two (renormalizable) field theories with explicit $\mathrm{SO}(5)$ symmetry in Sec.~\ref{manifest SO(5)}. The first one is QCD$_3$ with $N_f=2$:
\be
\mathcal{L} =\sum_{v=1,2} i \bar{\psi}_v \gamma^{\mu} (\d_{\mu} - i a_{\mu})\psi_v, \label{eq:QCD3sum}
\ee
where $a$ is an $SU(2)$ gauge field, and $\psi_{1,2}$ are two $SU(2)$-fundamental fermions. This theory can be obtained from the square lattice spin-$1/2$ model through a standard parton construction with a $\pi$-flux mean field ansatz, {and it has an $\mathrm{SO}(5)$ symmetry which becomes manifest when (\ref{eq:QCD3sum}) is written in terms of Majorana fermions}. The second theory is a Higgs descendent of QCD$_3$, where the $SU(2)$ gauge symmetry is Higgsed down to $U(1)$:
\be
\mathcal{L}=\sum_{i=1}^{4}i\bar{\psi}_i\gamma^{\mu}(\partial_{\mu}-ia_{\mu})\psi_i+(\lambda\mathcal{M}_a+h.c.), \label{eq:QEDN4sum}
\ee where $a_{\mu}$ is now a $U(1)$ gauge field, and the term $\mathcal{M}_a$ represents (schematically) monopole tunneling (instanton) events. In both theories the Dirac fermions transform in the spinor representation of $\mathrm{SO}(5)$. The $\mathrm{SO}(5)$-vector operators are simply the mass operators that are time-reversal invariant.

While the IR fates of the theories (\ref{eq:QCD3sum}) and (\ref{eq:QEDN4sum}) are unknown, both theories have the same symmetry anomaly as the deconfined critical point. Therefore, one possible scenario, among others, is that one of these theories will flow to the deconfined critical point in the IR.

\item
The full $\mathrm{SO}(5)$ symmetry of the deconfined critical point is anomalous, as revealed already by the sigma model approach. The manifestly $\mathrm{SO}(5)$-invariant QCD theory makes it possible to analyze the anomaly in more detail. We show in Secs.~\ref{sec:QCD3SPT} and \ref{sec:mathy} that QCD$_3$  with $N_f=2$,  with the full $SO(5) \times Z_2^T$ symmetry, can only be realized on the surface of a three-dimensional bosonic symmetry-protected topological (SPT) state. To characterize this SPT state, we can introduce a background $\mathrm{SO}(5)$ gauge {\red bundle  $A^5$ in the $3+1$D bulk. The topological response to $A^5$ is given by a discrete theta-angle (in contrast to the more familiar theta-angle in $3+1$D which can be continuously varied). Sec.~\ref{sec:QCD3SPT} provides a physical derivation of these results which are then rederived by more formal methods in Sec.~\ref{sec:mathy}.\footnote{A brief review of some math concepts relevant to this Sec. ~\ref{sec:mathy} is given in Appendix \ref{mathapp}.}. We show that  
the partition function of the $3+1$D SPT for a given $\mathrm{SO}(5)$ gauge field configuration is
\begin{equation}
Z[A^5] = |Z[A^5]| e^{i\pi  \int w_4[A^5]},
\end{equation}
where $w_4[A^5]$ is known as the $4$th Stiefel-Whitney class of the $SO(5)$ gauge bundle.  Though the IR fate of   QCD$_3$ with $\mathrm{SO}(5) \times Z_2^T$ symmetry is not known, we show that it must either break this symmetry spontaneously, or flow to a CFT. Gapped symmetry preserving phases (even with topological order) are prohibited. 

\item
We also discuss the implications of these dualities (for example, for numerical simulations) extensively in Sec.~\ref{implications}.   We outline a variety of  numerical tests of many aspects of the dualities. We also show how numerical calculations on fermionic QED$_3$ and QED$_3$-GN may provide a new handle on issues associated with deconfined critical points.

\item
In Sec.~\ref{pseudocritical} we discuss what is currently known about the deconfined critical points from simulations. In particular we discuss the possibility of `pseudocritical' behavior for deconfined critical points \cite{DCPscalingviolations}. It is possible that the theories discussed in this paper do not flow to stable CFTs, which in the context of deconfined critical point means that the transition is ultimately first order. But the flow to instability could be very slow, giving rise to a very large correlation length, and scaling behavior can still hold up to this very large scale (with exponents drifting as the scale increases). This pseudocritical scenario could potentially reconcile various seemingly conflicting results from numerical simulations: the observance of the emergent $SO(5)$ symmetry, the drifting of the scaling behavior, and the constraints on exponents of  $SO(5)$ invariant CFTs from conformal bootstrap \cite{SimmonsDuffinSO(5),Nakayama}. In this scenario, the dualities and emergent symmetries discussed in this paper can still hold below the (very large) correlation length.

We should point out that the pseudocritical scenario may be  more broadly relevant to many quantum phase transitions: in such a scenario the system shows quantum critical behavior above a very low temperature scale $T^*$, below which the criticality eventually disappears.

}

\end{enumerate}

{\red Related field theoretic work on these dualities has recently appeared in Ref.~\cite{bhse2017} while this paper was in preparation. }

\section{Easy plane $NCCP^1$ and fermionic $N_f = 2$ QED$_3$}
\label{easy_plane_duality}

We propose a duality of the easy plane $NCCP^1$ model to  fermionic $N_f = 2$
QED$_3$: 
\begin{widetext}
\ba &|D_{b + B} z_1|^2+|D_{b + B'}z_2|^2- |z_1|^4 -
|z_2|^4 - \frac{1}{2\pi}bd(B + B')-\frac{1}{2\pi}BdB'-\frac{1}{2\pi}B'dB' \\
\Longleftrightarrow \quad & \bar{\psi}_1i\slashed{D}_{a -
B}\psi_1+\bar{\psi}_2i\slashed{D}_{a + B}\psi_2
+\frac{1}{2\pi}adB'+\frac{1}{4\pi}(BdB-B'dB').
\label{epduality} \end{align} 
\end{widetext}
Here $b$ and $a$ are the dynamical
$U(1)$ gauge fields,\footnote{Strictly speaking $a$ is a spin$_c$
connection.} and we have also included \textit{background} $U(1)$
gauge fields  $B$ and $B'$. Various background Chern-Simons terms
are included to ensure that the theories on the two sides have the
same response (we will elaborate on this below).  In Appendix~\ref{PreciseL}, we present the above duality in a more precise and compact, but less physically intuitive, form.  The identification of the N\'eel and VBS order parameters in the $z$-theory was reviewed in Sec. \ref{deccprev} above.  We will identify these order parameters in the fermionic description in Sec. \ref{justify} below.

Both theories have been examined in independent numerical studies.
This duality  implies that two topical issues (fixed points of $QED_3$ at
$N_f=2$ and the easy-plane $NCCP^1$) are closely related. A `weak'
form of the duality is the assertion that easy-plane $\nccp^1$ is
equivalent to $QED_3$ perturbed by interactions which break the
flavor $SU(2)$ symmetry to $U(1)$. These perturbations are
formally irrelevant at the free UV fixed point of QED$_3$ where
the gauge field $a_\mu$ is decoupled from the fermions. If these interactions
are not important for the IR behavior, as one would naively guess,
then the IR fate of easy plane $NCCP^1$ will be the same as the the fate of 
QED$_3$ with full flavor $SU(2)$ symmetry, and  the IR behavior
will have an enlarged emergent symmetry  if QED$_3$ flows to a nontrivial fixed point.

In the deconfined criticality context, the easy-plane models that
have been studied show first order transitions {\cite{kukloveasyplane, kragseteasyplane, kauleasyplane1, kauleasyplane2,
Chenetal}}. However, on the
fermionic side, a recent numerical study~\cite{qedcft} found
evidence for an IR CFT in Eq.~\eqref{qed3} with $SU(2)$ flavor
symmetry (in contrast to earlier calculations~\cite{kogut2}).
Recent numerics on the quantum phase transition between a
particular $2D$ bosonic Symmetry Protected Topological (SPT) state
and the trivial state, which presumably is described by
Eq.~\eqref{qed3} with tuning parameter $\sum_{j=1}^2 m
\bar{\psi}_j\psi_j$, also show evidence of a continuous phase
transition~\cite{kevinQSH,so4qsh}.\footnote{In
Ref.~\cite{kevinQSH,so4qsh} the quantum phase transition was
discussed with the O(4) nonlinear sigma model with a topological
term, which according to Ref.~\cite{tsmpaf06} is the low energy
effective field theory of the $N_f = 2$ QED$_3$.} In light of the
duality it may be interesting to revisit easy plane deconfined
criticality.

\subsection{Justification for the duality}
\label{justify}

We first show how the fermionic dual of the easy
plane $NCCP^1$ model may be justified. This duality has already been mentioned in Ref.~\cite{karchtong}.

First recall the `basic' boson-fermion duality \cite{seiberg1}
relating the free Dirac fermion to a Wilson-Fisher boson coupled
to $U(1)_1$. There are two closely related versions of this
duality (for early works on related dualities see Ref.~\cite{ChenFisherWu,BarkeshliMcGreevy}). The Dirac side is (with our conventions)
 \begin{equation}
 \label{freeD}
 {\cal L}_{f1} = i\bar{\psi}\slashed{D}_{A}\psi.
 \end{equation}
 One version of the dual boson theory is
 \begin{equation}
 \label{dualb1}
{\cal L}_{b1} = |D_b \phi|^2 - |\phi|^4 + \frac{1}{2\pi} bdA +
\frac{1}{4\pi}
 bdb + \frac{1}{8\pi} A d A.
 \end{equation}
The other dual boson theory is
 \be
 \label{dualb2}
{\cal L}_{b2} = |D_{\hat{b}} \hat{\phi}|^2 - |\hat{\phi}|^4 - \frac{1}{2\pi} \hat{b} dA -  \frac{1}{4\pi} \hat{b} d \hat{b} - \frac{1}{8\pi} A d A  
 .
 \ee
The two boson theories are simply related:
$\hat{\phi}$ is the vortex dual of $\phi$. Let us also recall the
mapping of the relevant `mass' operators. On the Dirac side, a
mass term $m \bar{\psi} \psi$ with $m > 0$ maps to $r|\phi|^2$
with $ r > 0$ in Eq.~\eqref{dualb1} while it maps to $- \hat{r}
|\hat{\phi}|^2$ with $\hat{r} > 0$ in Eq.~\eqref{dualb2}. This
change of sign of the boson mass between the two bosonic theories
is exactly what we expect given that $\hat{\phi}$ is the vortex
dual of ${\phi}$.

Now, starting with the interacting fermionic theory in
Eq.~\eqref{epduality}, we use the dual theory in Eq.~\eqref{dualb1}
for the first flavor of fermion and the dual theory in
Eq.~\eqref{dualb2} for the second flavor.  The resulting dual
theory, in terms of bosons $\phi_I$ and dynamical gauge fields
$b_I$ ($I=1,2$) is 
\begin{widetext}\beqn {\cal L}'_{b12} = \sum_{I=1,2} {\cal
L}[\phi_I, b_I] + \frac{1}{2\pi} b_1 d(a - B)  + \frac{1}{4\pi}
b_1db_1 + \frac{1}{8\pi} (a - B) d (a - B) - \frac{1}{2\pi} b_2
d(a + B)  \cr\cr-  \frac{1}{4\pi}b_2 d b_2 -
\frac{1}{8\pi} (a + B) d (a + B) + \frac{1}{2\pi} a d B'+\frac{1}{4\pi}(BdB-B'dB') 
,\label{dual_int} \eeqn \end{widetext} where $\mathcal{L}[\phi_I, b_I]$ contains
the kinetic and potential terms for $\phi_I$ and $b_I$.
Integrating out the dynamical gauge field $a$ will impose the
following constraint: \beqn b_1 - b_2 - B + B' = 0. \eeqn This
implies that we can define a new dynamical gauge field $b$ such
that $b_1 = b + B$, and $b_2 = b + B'$. Then Eq.~\eqref{dual_int}
becomes exactly the first line of Eq.~\eqref{epduality} (after
identifying $\phi$ with $z_1$ and $\hat{\phi}$ with $z_2$).

In addition to the formal derivation above, in the following we
perform various consistency checks of the duality and make a few
further comments. We will defer until Sec. \ref{symm_ep} a detailed
discussion of the matching of symmetries (explicit or emergent) on
the two sides, and their implications.

\begin{enumerate}

 \item
Clearly both sides have (at least) $U(1) \times U(1)$ symmetry,
probed by the two background gauge fields $B$ and $B'$. On the
boson side the gauge-invariant operator $z_1^* z_2$ has charges
$q_B = + 1$, $q_{B'} = - 1$, and the monopole operator ${\cal
M}_b$ has $q_B = q_{B'} = +1$. It is actually more natural to
define $B_1 = B - B'$ and $B_2 = B + B'$ and to define
$\Phi_{B_1}$ and $\Phi_{B_2}$ as the order parameters charged
under the corresponding global $U(1)$s. That is, if we let $(q_1,
q_2)$ denote the charges under $U_{B_1}(1)$ and $U_{B_2}(1)$, then
$\Phi_{B_1}$ carries charges $(1,0)$ and $\Phi_{B_2}$ carries
$(0,1)$. We have \ba\label{Phi_operators} \Phi_{B_1} &= z_1^\ast
z_2, & \Phi_{B_2} &= {\cal M}_b.
\end{align}
These are the two Landau order parameters (N\'eel and VBS
respectively in the quantum magnet realization), one of which
orders on each side of the putative deconfined QCP.

On the fermion side, a monopole operator ${\cal M}_a$ is
associated with two complex fermion zero modes $f_{1,2}$ from the two Dirac
fermions, and gauge invariance requires filling one of the zero
modes \cite{kapustinqed}. Therefore the operators  $f^\dagger_j
{\cal M}_a$ are gauge invariant bosons with charges $q_B = - 1$,
$q_{B'} = - 1$ (for $j = 1$) and $q_B = + 1$, $q_{B'} = - 1$ (for
$j = 2$). Clearly they can be identified with the corresponding
Landau order parameters on the bosonic side.

The bosonic side of Eq.~\eqref{epduality}, after a redefinition
$b\to b-B'$, can be written as simply \be\label{epdualitysimpler}
|D_{b+B_1}z_1|^2+|D_bz_2|^2-|z_1|^4-|z_2|^4-\frac{1}{2\pi}bdB_2.
\ee Below we will use both forms of the easy-plane $NCCP^1$
Lagrangian.

%

\item The derivation above makes clear that the correspondence for
mass operators is \be\label{mass_operator_correspondence} {m_1
\bar{\psi}_1 \psi_1 + m_2 \bar{\psi_2}\psi_2} \quad
\Longleftrightarrow \quad {m_1 |z_1|^2 - m_2 |z_2|^2}. \ee This is
because we used the first boson-fermion duality (\ref{dualb1}) for
the first fermion, and the second boson-fermion duality
(\ref{dualb2}) for the second fermion. It is easy to check that
the phases obtained by adding such mass operators match in the
bosonic and fermionic descriptions:

On the bosonic side turning on $r (|z_1|^2 + |z_2|^2)$ will drive
the system into an   ordered phase with   $\langle \Phi_{B_1}
\rangle \neq 0$ or one with  $\langle \Phi_{B_2} \rangle \neq 0$
depending on the sign of $r$.\footnote{{In the quantum magnetism
realization these are simply the N\'eel and VBS ordered states.}} If
$r>0$ we have $\<z_1\>,\<z_2\>\neq0$, a Higgsed gauge field
$b_{\mu}$, and a nonzero  expectation value $\<z_1^*z_2\>\neq0$
which leads to a `Meissner effect' for $B_1$. If $r<0$ we have a
free dynamical Maxwell photon at low energy which leads to the
Meissner effect for $B_2$, via the mutual Chern Simons term in
(\ref{epdualitysimpler}). The mass term corresponds on the
fermionic side to $r \bar{\psi} \sigma^z \psi$, which gaps out the
fermions. Integrating them out produces the term $\frac{1}{2\pi}
{\rm sgn}(m) adB$. Together with the $\frac{1}{2\pi}adB'$ term in
(\ref{epduality}), this again leads to the Meissner effect of
either $B_1$ or $B_2$ depending on the sign of $m$. This is
consistent with the operator identification discussed above.

\item We can also turn on an anti-symmetric mass ${\mu (|z_1|^2 -
|z_2|^2)}$ on the bosonic side. For $\mu>0$ we have $\<z_1\>\neq0$
but $\<z_2\>=0$, which gaps out $b$ and sets $b=-B$. So we get a
gapped phase with response
$\frac{1}{2\pi}(BdB-B'dB')=\frac{1}{2\pi}B_1dB_2$. If we
reinterpret $B$ as a `charge' probe and $B'$ as a `spin' probe,
this corresponds to the response of the bosonic integer quantum
hall state \cite{luav12,tsml13}.   For $\mu<0$ we get a gapped
phase with a trivial response. (In the context of deconfined
criticality in quantum magnets, where $U(1)_{B_2}$ is only
emergent, both of these are {trivial phases in which the N\'eel order parameter is 
polarized along the $z$ axis.})

On the fermionic side these phases can be reproduced by a
symmetric mass term
${\mu(\bar{\psi}_1\psi_1+\bar{\psi_2}\psi_2)}$: integrating out
the fermions leaves the action \be \label{trivialphases}
\f{1}{4\pi} {\rm sgn}(\mu) \left(ada+BdB \right) +
\frac{1}{2\pi}adB'+\frac{1}{4\pi}(BdB-B'dB'). \ee Since the term
$\pm\frac{1}{4\pi}ada$ is a trivial topological field theory, it
can be integrated out
safely 
and we produce the same response theory as on the bosonic side. It is
also easy to see that no thermal hall conductance (i.e. no chiral
edge state) is generated for either phase (in either picture), in
agreement with known results \cite{luav12,tsml13} for the trivial
insulator and the bosonic integer quantum hall state.

The responses above indicate that the massless theories of primary
interest to us can  also be viewed as describing the phase
transition between the trivial and bosonic integer quantum Hall
insulators. The fermionic picture for this transition was
discussed in Refs.~\cite{tarunashvin,LuLee} using parton constructions (see also Appendix \ref{altdl} for a different perspective) and more recently in Ref.~\cite{hewire} using a coupled-wire construction.  This bosonic IQH
transition has been seen numerically, and the results show
evidence of a continuous phase transition~\cite{kevinQSH,so4qsh}.   


{
The mean field phase diagram implied by the above is summarized in Fig.~\ref{epphasediagram}, both
in the deconfined criticality context (where $U(1)_\text{N\'eel}$
is an exact microscopic spin symmetry and  $U(1)_\text{VBS}$ is
emergent at the critical point) and for the boson integer quantum
Hall transition. This mean field picture captures only the topology of the phase diagram adjacent to the putative critical point: in general the phase boundaries will meet at a cusp (since ${|z_1|^2 + |z_2|^2}$ and ${|z_1|^2 - |z_2|^2}$ will have different scaling dimensions) and will not lie along the axes.
}

\begin{figure}
\begin{center}
\includegraphics[width=2.5in]{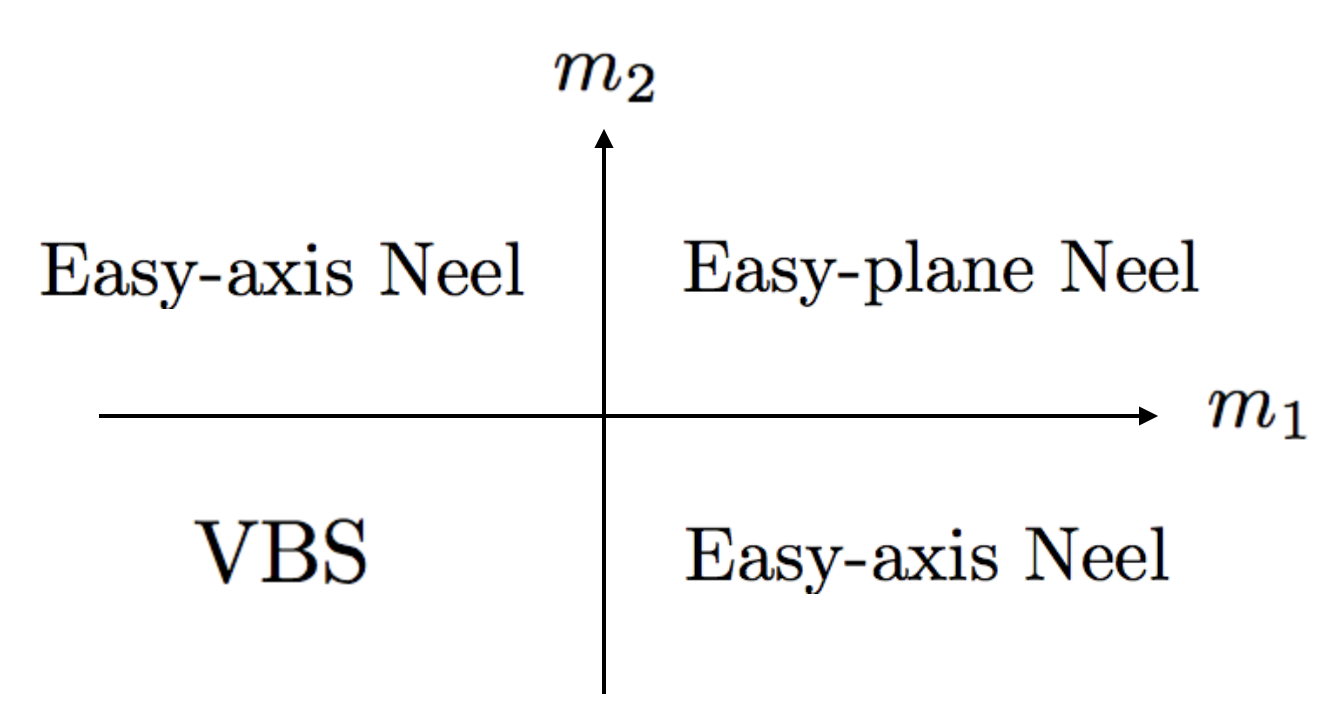} \hspace{20pt}
\includegraphics[width=2.5in]{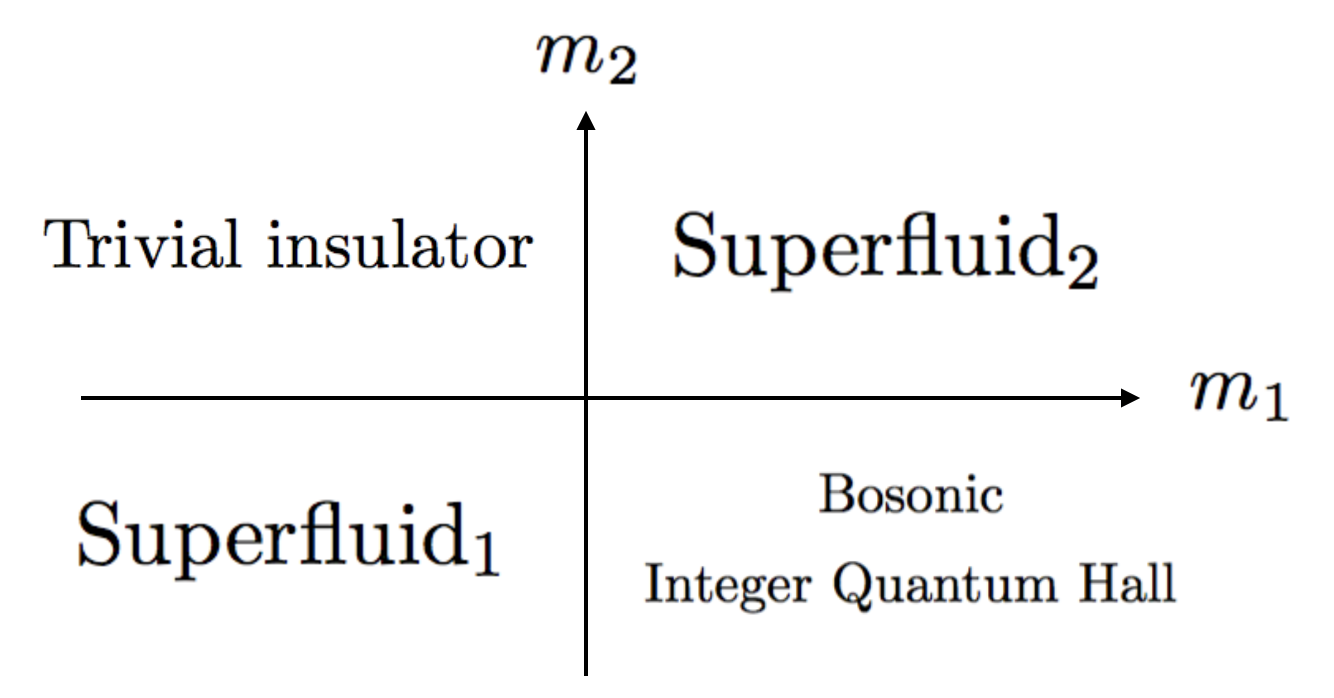}
\end{center}

\caption{Mean-field phase diagram for the mass term
$m_1|z_1|^2+m_2|z_2|^2$ in the easy-plane $NCCP^1$ model. The same
phase diagram is obtained from the QED theory with mass term
$m_1\bar{\psi}_1\psi_1-m_2\bar{\psi}_2\psi_2$. The upper panel is
realized in the context of the quantum magnet, where the $U(1)_{B_2}$
symmetry is only emergent. The lower panel is realized in the
context of the integer quantum hall transition of bosons, where the
two superfluid phases correspond to superfluids of the first or
second layer, i.e. up or down components of the pseudospin. (Going beyond mean field will move the phase boundaries away from the axes.)}
\label{epphasediagram}
\end{figure}

\item There is potentially a dual formulation of $N = 2$ QED$_3$
directly in terms of the bosonic monopoles $ f_j^\dagger{\cal
M}_a$ which are precisely the $\Phi_{B_1}, \Phi_{B_2}$ defined in
Eq.~\eqref{Phi_operators}. The most we know about this theory if we want to
formulate it directly in terms of the physical bosons is that it
has the structure of an (anisotropic) $O(4)$ sigma model at
$\theta = \pi$ \cite{tsmpaf06} discussed in Sec. \ref{deccprev}.  This is
consistent with the various other known connections between this
sigma model and the QED$_3$ theory. This effective field theory is
a convenient language for discussing the emergent $O(4)$ symmetry
that would be required if the strong duality conjecture holds for
these theories.

\end{enumerate}

\subsection{Symmetries}
\label{symm_ep}

We now study the symmetries of the
various dual actions, and the implications of the dualities for
emergent symmetries of the easy-plane deconfined quantum critical points. Here we will discuss the symmetries from a physical point
of view natural in condensed matter physics. This will make
plausible the statements about non-trivial emergent symmetries.

\subsubsection{Continuous symmetries}
\label{easy_plane_cts_symmetries}

In this section we will see how the duality web leads to the
possibility that the easy plane $NCCP^1$ theory could have an
emergent  $O(4) \times Z_2^T$ symmetry.\footnote{Here $Z_2^T$
refers to time reversal which is antiunitary.} In
Sec.~\ref{allowed1} we will express the
requirements for this symmetry enhancement more formally in terms
of the properties of the putative  `mother' $O(4)$ fixed point. Here
we discuss how the dualities, if they hold in a `strong' form,
lead to this emergent symmetry.

In the duality between easy-plane $\nccp^1$ and QED$_3$ we naively
only expect that the  continuous symmetry of the resulting fermion
theory is  $U(1) \times U(1)$.  As explained in Sec. \ref{qedrev}, the fermionic action we have
written down apparently however has manifest $\frac{SU(2) \times
U(1)}{Z_2}$ symmetry with the
$SU(2)$ corresponding to rotations between the two fermion
flavors. Thus, we should allow for terms that break this apparent
flavor $SU(2)$ down to $U(1)$. The mass term $\bar{\psi}\sigma^z
\psi$ will accomplish that but this is precisely the operator
whose coefficient is tuned to zero at the transition. The minimal
operators with no derivatives that break the flavor $SU(2)$
symmetry are thus four fermion terms, {\em e.g.} $\left(\bar{\psi}
\sigma^z \psi \right)^2$.
At the free Dirac fermion fixed point
these operators are strongly irrelevant.
So if the $SU(2)$ symmetric QED theory has a nontrivial IR fixed
point, it is plausible that perturbations breaking $SU(2)$ to
$U(1)$ are also irrelevant here, and that the theory with
microscopic $U(1)$ flavor symmetry flows to this point and has
emergent $SU(2)$ flavor symmetry in addition to the other
symmetries already present.

Now consider the self-duality of QED$_3$. In the derivation of
Ref.~\cite{qeddual} of this self-duality a priori we only
know that the $U(1) \times U(1)$ symmetry of one side maps to the
$U(1) \times U(1)$ symmetry of the other side, with the role of
the two $U(1)$ symmetries being exchanged by duality (the flavor
conservation symmetry on one side becomes the flux conservation
symmetry on the other side). Again it is naively plausible that
the $SU(2)$ flavor symmetry emerges in the infrared on both sides
of the duality (see Sec.~\ref{allowed1} for
a more careful discussion).
The two flavor $SU(2)$ symmetries on the two sides are distinct
from each other, implying that the full continuous symmetry of the
QED$_3$ theory is then $(SU(2) \times SU(2))/Z_2= SO(4)$. A
version of this argument was previously made in
Ref.~\cite{seiberg2,xucheng}\footnote{To be completely
precise Ref.~\cite{seiberg2} stated that the enhanced
symmetry is $Spin(4)$ while according to our discussion it is
$SO(4)$. In particular we will not find local operators
transforming under the $(\frac{1}{2},0)$ or $(0, \frac{1}{2})$
representations.}. We will clarify the conditions under which this
symmetry enhancement actually happens.



On the fermionic side, the two flavors of Dirac fermions $\psi_j$
form a spin-$1/2$ representation under one of the two $SU(2)$ subgroups in $SO(4)$, and
the dual Dirac fermions $\chi_j$ form a spin-$1/2$ representation under the other $SU(2)$ subgroup. These fermions do not transform in a simple way under the whole $SO(4)$ group, but this is not problematic since they are not gauge invariant. The gauge invariant operators ${\cal
M}_a^\dagger f_j$ transform as a spinor under the flavor
$SU(2)$ of the $\psi$ theory. As these are identified with the
boson operators $z_1^* z_2$ and ${\cal M}_b$ it follows that these
two operators are a spinor under this $SU(2)$. Thus this $SU(2)$
rotates 
\be \left(
\begin{array}{ccc}
\Phi_{B_1}^* \\
- \Phi_{B_2}
\end{array}
\right)\sim
\left(
\begin{array}{ccc}
n_3 - i n_4 \\
-n_1 - i n_2
\end{array}
\right) \ee
 as a spinor. It is easy to see that under the flavor
$SU(2)$ of the dual  QED$_3$ theory, 
\be
\label{o4vec}
 \left(
\begin{array}{ccc}
\Phi_{B_1}  \\
\Phi_{B_2}
\end{array}
\right)\sim
\left(
\begin{array}{ccc}
n_3 + i n_4 \\
n_1 + i n_2
\end{array}
\right) \ee
is rotated as a spinor. This means that the $SO(4)$
simply rotates the four real components of $\Phi_{B_1},
\Phi_{B_2}$, i.e. $(n_1, n_2, n_3, n_4)$, into one another. In the
quantum magnetism realization these are precisely the N\'eel and VBS
order parameters.

\subsubsection{Discrete symmetries}\label{discrete_symm_section}
Let us now turn to discrete symmetries.  We have already mentioned
the $Z_2$ spin flip symmetry $\mathcal S$. For the quantum
magnetism realization in spin-$1/2$ square lattice magnets, we
must also discuss lattice translation, lattice rotation, lattice reflection, and time
reversal symmetries.
\begin{enumerate}
\item
{\em $Z_2$ spin flip symmetry}

The $Z_2$ spin flip $\mathcal{S}$ corresponds in the microscopic
spin model to a rotation of the spin at each site by $\pi$ around
the {$x$-axis}. This is a  subgroup of spin SO(3) symmetry that is
presumed to  be retained in the easy plane model.  In the
context of easy-plane deconfined criticality, this symmetry
ensures that the only tuning parameter across the transition is
the \textit{symmetric} mass term $r(|z_1|^2+|z_2|^2)$.

The full action of $\mathcal S$ in the easy-plane $NCCP^1$ theory
in the presence of background fields is: \be z_{1} \leftrightarrow
z_2, \ \ \ B \leftrightarrow B', \ \ \ b \rightarrow b. \ee
Equivalently, $\mathcal S$ takes $B_1 \leftrightarrow - B_1$, $B_2
\leftrightarrow B_2$.  As emphasized in Sec. \ref{deccprev}, the
corresponding action on the N\'eel and VBS  order parameters is
\be {\mathcal S}(\Phi_{B_1} ) = \Phi_{B_1}^*, ~~~{\mathcal
S}(\Phi_{B_2}) = \Phi_{B_2}. \ee Thus this $Z_2$ acts as an
improper $O(4)$ rotation on the vector $(n_1, n_2, n_3, n_4)$
formed from the four real components of these fields.

In QED$_3$, $\mathcal S$ is a transformation between $\psi_j$ and
its dual fermion $\chi_j$. Since the mass terms transform under the fermion
self-duality as
$\bar{\psi}_1\psi_1\leftrightarrow-\bar{\chi}_2\chi_2$, the
antisymmetric fermion mass term $\bar{\psi}\sigma^z\psi$ is
invariant, which is consistent with the operator identification
described above.

We should point out that in the continuum field theory this
symmetry is actually anomalous. In both the boson and fermion
pictures the Lagrangian picks up an extra term under this symmetry
operation: \be \label{z2anomaly}
\mathcal{L}\longrightarrow\mathcal{L}+\frac{1}{2\pi}(B'dB'-BdB)=\mathcal{L}-\frac{1}{2\pi}B_1dB_2. 
\ee For deconfined criticality realized in a lattice spin system,
this anomaly is harmless because the $U(1)$ symmetry probed by
$B_2$ is really a discrete lattice rotation symmetry. However, if the symmetries are
on-site, this theory can only be regularized on the surface of a three
dimensional bulk. We shall discuss this in more detail in
Sec.~\ref{bulk}.

Including time reversal which we discuss below, the full symmetry
of the easy plane $NCCP^1$ fixed point   {may thus  be} $O(4) \times
Z_2^T$.  Note that the enlargement of $SO(4)$ to $O(4)$ is also
expected from the standpoint of the non-linear sigma model with a
theta term, Eq.~(\ref{eq:introsigmapi}).  If $O(4)$ is broken to $SO(4)$ the value of $\theta$
can be varied away from $\pi$: this is plausibly a relevant
perturbation.

\item
{\em Bosonic self-duality symmetry}

There is also a $Z_2$ subgroup, which we denote ${\mathcal
S}_{\psi}$, of the $SU(2)$ flavor symmetry of the QED$_3$: \beqn
{\mathcal S}_\psi: \psi_1 \leftrightarrow \psi_2, \ \ \ B
\rightarrow - B, \ \ \ a \rightarrow a, \ \ \ B' \rightarrow B'.
\eeqn This ${\mathcal S}_\psi$ symmetry is not a microscopic
symmetry for the quantum magnet. In NCCP$^1$ it becomes the
bosonic self-duality, $z_\alpha \to w_\alpha$: \beqn \Phi_{B_1} \leftrightarrow
\Phi^\ast_{B_2}. \eeqn This also shows that the $SU(2)$ flavor
group of the QED$_3$ theory must act in a highly nonlocal fashion
in the $NCCP^1$ theory.

On the QED side, imposing ${\mathcal S}_\psi$ symmetry forces the
mass term to be symmetric, $m \bar{\psi}_j\psi_j$, which gives a
transition between two distinct (SPT) gapped phases. On the
bosonic side, this symmetry allows the antisymmetric mass term
$\mu(|z_1|^2-|z_2|^2)$, which also gives the SPT transition.

\item
{\em Time reversal}

We now specialize to realizations of these deconfined critical
points at the N\'eel-VBS transition in square lattice spin-$1/2$
magnets.  Microscopic symmetries then include --- in addition to
the spin rotation and spin flip $\mathcal S$ symmetries ---  time
reversal  and lattice symmetries.

Time reversal $T$ is anti-unitary and acts on the $NCCP^1$ fields as
\beqn
T(z_\alpha)  & =  & \epsilon_{\alpha\beta} z_\beta \\
T(b) & =  & -b - B - B' \\
T(B) & = & B, ~~~ T(B') = B'. \eeqn (For the gauge fields we only
indicate the time reversal action on the spatial  components; the
time component will transform with the opposite sign.)  Here
$\epsilon = i\sigma_y$ is antisymmetric with $\epsilon_{12} = 1$.
This is consistent with $\Phi_{B_1} \rightarrow - \Phi_{B1}^*$ and
$\Phi_{B2} \rightarrow \Phi_{B2}^*$, as befits the N\'eel and VBS
order parameters respectively (written in complex form).  Note
that, as with the $Z_2$ spin flip symmetry, the bosonic
Lagrangian is only invariant up to an anomaly: \be
\label{Tanomaly}
\mathcal{L}\to\mathcal{L}+\frac{1}{2\pi}(B'dB'-BdB)=\mathcal{L}-\frac{1}{2\pi}B_1dB_2.
\ee

On the dual QED$_3$ side, time-reversal acts as the product
${\mathcal S}_\psi \mathcal{T}$ transformation   under
which
\ba T(\psi) &=
{\mathcal S}_\psi\gamma_0\psi, &  T(a) &= - a. \end{align}

Note that the QED$_3$ theory also has the same anomaly in
Eq.~\eqref{Tanomaly}.

\item
{\em Translation symmetry}

It suffices to discuss unit lattice translations along one
direction, say the $y$-direction (${x \to x}$, ${y \to y+a}$ where
$a$ is the lattice spacing) which we dub $T_y$.  The N\'eel  and VBS
orders clearly transform as 
\ba \label{Typhys} T_y(\Phi_{B_1} ) &=
- \Phi_{B_1} & T_y(\Phi_{B_2}) &= \Phi_{B_2}^*.
\end{align}

In the $NCCP^1$ theory, this is implemented as
\beqn
\label{Tycp1}
T_y(z_\alpha) & =  & \epsilon_{\alpha\beta} z_\beta^* , \\
T_y(b) & = & -b \\
T_y(B) & = & -B',~~~ T_y(B') = - B.
\eeqn

Just like with the $Z_2$ spin flip, or time reversal symmetry, the
$NCCP^1$ Lagrangian is invariant under $T_y$ only up to an anomaly
(shift by $\frac{1}{2\pi} B_1 dB_2$). On the QED$_3$ side,  $T_y$
takes the fermions $\psi$ to their fermionic duals $\chi$ though
the detailed transformation is different from that of the  $Z_2$
spin flip symmetry. Specifically we have
\ba \label{Tyqed}
\psi_1 &\leftrightarrow -\chi_1^\dagger &
 \psi_2 &\leftrightarrow \chi_2^\dagger &
a &\leftrightarrow -\tilde{a}. 
\end{align}
Note that the symmetric mass
$\bar{\psi}\psi$ is odd under $T_y$ while the anti-symmetric mass
$\bar{\psi}\sigma^z\psi$ is even, exactly as expected from the
corresponding operators on the bosonic side.

A simple way to understand the action of $T_y$ is as follows.
From  Eq.~\eqref{Typhys} we see that the action of $T_y$ on the
physical order parameters is similar to that of $\mathcal S$ if we
exchange the N\'eel and VBS orders.  To make this precise consider
the modified translation $\tilde{T}_y = U_{B_1}(\pi) T_y$ which
combines translation with a $\pi$ rotation of the easy plane N\'eel
vector. Then the physical order parameters transform as 
\ba
\tilde{T}_y(\Phi_{B_1} ) &= \Phi_{B_1}, &
\tilde{T}_y(\Phi_{B_2}) &=
\Phi_{B_2}^*.
\end{align}
Thus we see that it is precisely the $\mathcal S$
operation performed after a charge-conjugation transformation $\mathcal{C}$ that takes all the fields to their charge-conjugate. This
identification is nicely consistent  with the action of $T_y$ in
the QED$_3$ theory in  Eq.~\eqref{Tyqed}. The extra
$(-)$ sign in Eq.~\eqref{Tyqed} is simply due to the additional
$U_{B_1}(\pi)$ in the definition of $\tilde{T}_y$.

\item
{\em Rotation and reflection symmetries}

Lattice rotations by $\frac{\pi}{2}$ about a square lattice site
act very simply on the N\'eel and VBS vectors.  We have 
\ba
\label{Rphys} R_{\frac{\pi}{2}}(\Phi_{B_1})& = \Phi_{B_1},  &
R_{\frac{\pi}{2}}(\Phi_{B_2}) &=  i \Phi_{B_2}.
\end{align}
This is part of
a $U_{B_2}(1)$ rotation whose role in the various dualities we
have already discussed. A site-centered lattice reflection $R_y$, say about the
$x$-axis ($x \rightarrow x, y \rightarrow -y$) acts as 
\ba
\label{refphys} R_y(\Phi_{B_1}) &= \Phi_{B_1},   &    R_y(\Phi_{B_2}) &=
\Phi_{B_2}^*.
\end{align}
To define its action simply let us denote, for
any gauge field $A = (A_0, A_x, A_y)$ the reflected version  by
$RA = (A_0, A_x, - A_y)$.  Then  $R_y$ acts in the $z$-formulation
as 
{
\ba
R_y(z_\alpha) &=  z_\alpha, & R_y(b) &= Rb + RB + RB',  \\
R_y(B) &= - RB', & R_y(B')& = - RB.
\end{align}
It is readily checked that the $NCCP^1$ Lagrangian is invariant under this transformation and there is no anomaly.} On the fermion side,
$R_y$ again involves a duality transformation between $\psi$ and
$\chi$: \be \label{Ryqed} \psi_1 \to \gamma_y \chi_2, ~~\psi_2 \to
\gamma_y \chi_1, ~~ a \to R\tilde{a}. \ee

\end{enumerate}

\subsection{Allowed symmetry breaking terms}
\label{allowed1}

The strong forms of the dualities discussed here involve the
emergence of higher symmetries than are present in the UV
Lagrangians. In order for the dualities to hold in the IR without
fine-tuning, the hypothetical higher-symmetry fixed point must
exist and must be stable to all perturbations allowed by the
symmetry of the UV theory. Here we clarify these stability
requirements. As will be discussed in Sec.~\ref{pseudocritical}, it is also
possible that there is no fixed point with the higher symmetry,
but that there is a pseudocritical regime up to a large but finite
lengthscale $\xi$; in this case the requirements should be
interpreted in terms of the effective scaling dimensions in this
regime.

We consider perturbations of the hypothetical
$\mathrm{O}(4)$--symmetric point relevant to $N_f = 2$ QED$_3$ and
the easy-plane $\nccp^1$ model.  Here we will take  QED$_3$ to be
defined with full $SU(2)$ flavor symmetry, as done for instance in
the lattice calculations of Ref.~\cite{qedcft}.  We will see that the
conditions for the emergence of $\mathrm{O}(4)$ are more stringent
for easy-plane $\nccp^1$ than for QED. Therefore in principle it
is possible that the self-duality of QED$_3$ could hold, with
emergent $\mathrm{O}(4)$, but that easy-plane $\nccp^1$ could fail
to flow to this fixed point. (By contrast, the
requirements for the emergence of $\mathrm{SO}(5)$ are similar
for the bosonic and fermionic theories, as we will see in Sec.~\ref{allowed2}.)

 To begin, the hypothetical fixed point must be stable to $O(4)$-singlet scalar perturbations. We certainly expect a relevant perturbation which is invariant under $\mathrm{SO}(4)=[SU(2)\times SU(2)]/Z_2$ but not under improper
$\mathrm{O(4)}$ transformations: in the language of the sigma
model  for the field $(n_1, n_2,
n_3, n_4)$ (Secs.~\ref{deccprev},~\ref{easy_plane_cts_symmetries}) this corresponds
to varying the coefficient of the $\theta$ term away from $\pi$.
But this perturbation is harmless as it is forbidden by time reversal,
and in  the easy-plane $\nccp^1$ model also by the $Z_2$ spin-flip
symmetry $\mathcal{S}$ (Sec.~\ref{discrete_symm_section}).

Apart from the $\mathrm{O}(4)$ vector order parameters
$n_a$ defined in Eq.~\ref{o4vec} it is natural to expect the next leading scalar operators to be those in the two and
four-index symmetric tensor representations of $\mathrm{O}(4)$.
We denote these $X^{(2)}_{ab}$ and $X^{(4)}_{abcd}$. At the level
of symmetry, \ba X^{(2)}_{ab} &\sim n_a n_b - \delta_{ab} n^2/4, &
X^{(4)}_{abcd} & \sim n_a n_b n_c n_d - (\ldots),
\end{align}
where the subtraction $(\ldots)$ makes the operator traceless.
$X^{(2)}$ is certainly relevant.

The two-index symmetric tensor $X^{(2)}_{ab}$ corresponds to the $(1,1)$ representation of
$\mathrm{SU}(2)\times \mathrm{SU}(2)$. All components of this
operator are therefore forbidden by the explicit $\mathrm{SU}(2)$
of QED. In the easy plane model one component, $\sum_{a=1}^2X^{(2)}_{aa}$, is allowed but is precisely the
tuning parameter for the N\'eel-VBS transition: i.e. $\f{1}{2}
(|\Phi_{B_2}|^2 - |\Phi_{B_1}|^2)$ in terms of the complex N\'eel
and VBS order parameters. The four-index symmetric tensor
$X^{(4)}$ is the $(2,2)$ representation of
$\mathrm{SU}(2)\times\mathrm{SU}(2)$, so again all components are
forbidden by the explicit $\mathrm{SU}(2)$ of QED. However in the
easy-plane $\nccp^1$ model the N\'eel-VBS anisotropy
$\sum_{a=1}^2\sum_{b=3}^4 X^{(4)}_{aabb}$ is allowed, and
microscopic models on the square lattice will also allow
$\sum_{a=1}^2 X^{(4)}_{aaaa}$.\footnote{{In fact for the lattice antiferromagnet another symmetry- breaking term is allowed: $(\partial_x n_1)^2+(\partial_y n_2)^2+...$ (see also note in Sec.~\ref{allowed2}). (This operator lives in a spin-2 representation of spatial rotations; for a $3D$ CFT, unitarity bounds indicate that spin-$2$ operators have RG eigenvalue $\leq 0$.) This perturbation is absent for the continuum easy-plane $NCCP^1$ field theory, where the VBS $U(1)$ is an exact internal symmetry.}}

Since the easy plane model allows an $\mathrm{O}(4)$--breaking
perturbation that is forbidden for QED, it is conceivable that the
QED self-duality holds, with emergent $\mathrm{O(4)}$, but that
the {strong} 
duality with easy-plane $\nccp^1$ does not hold. This scenario
would apply if there was an $\mathrm{O}(4)$ (pseudo)critical
regime in which $X^{(4)}$ was relevant, but  the perturbations
allowed in the QED theories were irrelevant.

 The explicit
$\mathrm{SU}(2)$ of QED restricts such perturbations to
representations of the form $(0, \text{integer})$. In the sigma
model language, the simplest such terms involve two
derivatives and four powers of $n$, so are plausibly irrelevant,
as argued in Ref.~\cite{tsmpaf06}. In other words, if there
is an $\mathrm{O(4)}$ fixed point which is stable to
$\mathrm{O}(4)$--singlet perturbations, it is  very likely that
QED flows to it.

From the point of view of the fermionic Lagrangians, both types of
perturbation (those allowed in the easy-plane model and those
allowed in QED) can be cast as four-fermion terms, giving a
plausibility argument for their irrelevance. The breaking of the
{symmetry} of QED to {that of} the
easy plane model {allows} a four-fermion operator, as
discussed in Sec.~\ref{easy_plane_cts_symmetries}. For the
possible emergence of $\mathrm{SU}(2)\times \mathrm{SU}(2)$,
consider the following `weak' form of the fermion-fermion duality.
We expect that in principle there is an exact UV duality between a
cutoff version of QED with $\mathrm{SU}(2)_B\times
\mathrm{U}(1)_{B'}$ symmetry and an effective field theory version
of QED (with a highly fine-tuned Lagrangian) in which the
{explicit} {continuous} symmetry is $\mathrm{U}{(1)}_B\times
\mathrm{U}(1)_{B'}$. (For convenience we label the groups by the
probe field for the U(1) part.) Since the simplest terms that
break $\mathrm{SU}(2)_{B'}$ are four-fermion terms in this dual
description, it is plausible that $\mathrm{SU}(2)_{B'}$ will
emerge in the continuum, giving full $\mathrm{SU}(2)_B\times
\mathrm{SU}(2)_{B'}$.

\section{$SU(2)$ symmetric $NCCP^1$ and fermionic QED$_3$--GN}
\label{su2cp1-qedgn}

We now turn to the deconfined critical point with full $SU(2)$ symmetry. We propose a duality between the $SU(2)$ symmetric $NCCP^1$ model and the $N_f = 2$ QED$_3$--GN  theory,
\ba
\label{so5duality}
&\sum_{\alpha = 1,2} |D_b z_\alpha|^2
- \left(|z_1|^2 + |z_2|^2\right)^2\\
\Longleftrightarrow \quad & \sum_{j=1,2}\bar{\psi}_j
i\slashed{D}_a\psi_j + \phi\sum_{j = 1,2}\bar{\psi}_j \psi_j +
V(\phi), \label{so5dualityl2}
\end{align}
where $\phi$ is a critical Ising field (real scalar), with the
Ising terms $(\partial\phi)^2-\phi^4$ suppressed for notational convenience. The QED$_3$-GN model has not been studied
numerically as far as we know. The duality suggests the
possibility of critical behavior with emergent symmetry.

To justify this duality, we consider the phases of the QED$_3$--GN Lagrangian in Eq.~\eqref{so5dualityl2}, making use of the results for the pure QED in Sec~\ref{easy_plane_duality}.  First, the
phase with a positive mass term for $\phi$ and $\langle \phi
\rangle = 0$ is expected to be equivalent to QED$_3$, and dual to
the critical easy plane NCCP$^1$ theory. What does the coupling to
$\phi$ mean in the bosonic theory? The mass term  ${\bar{\psi}
\psi}$  is identified with $\left(|z_1|^2 - |z_2|^2\right)$ in
Eq.~\ref{mass_operator_correspondence}, so
 the coupling to the scalar field becomes
\be \label{phiz} \phi \left(|z_1|^2 - |z_2|^2\right). \ee Now
consider the QED$_3$-GN theory in the phase where $\langle \phi
\rangle \neq 0$, induced by turning on a negative mass term for
$\phi$. This gives the symmetric mass term for QED$_3$ and the
antisymmetric mass for easy-plane $\nccp^1$, as discussed under
Eq.~\eqref{trivialphases}. The theory becomes trivially gapped
(except for response terms).  Finally, what does the phase
transition associated with the onset of $\langle \phi \rangle$ of
the QED$_3$-GN model correspond to on the boson side? We propose
that it is the critical point of the $SU(2)$ invariant $NCCP^1$
model.

The discussion of the phase diagram above gives a basic
consistency check on this duality. The operator identification
goes as follows: the Ising field $\phi$ on the fermionic side is
dual to ${|z_1|^2-|z_2|^2}$ on the bosonic side. The Ising mass
$-\lambda\phi^2$ is dual to the anisotropy $\lambda|z_1|^2|z_2|^2$ on the bosonic side.\footnote{Strictly speaking the anisotropy term dual to the Ising mass should be $\lambda[|z_1|^2|z_2|^2-\alpha(|z_1|^2+|z_2|^2)]$, where the second term is added to keep the theory critical when $\lambda>0$. We discuss this further in Sec.~\ref{PhaseD}.} The phases with $\<\phi\>=0$ and
$\<\phi\>\neq 0$ correspond respectively to the easy-plane
critical theory and to a gapped state with easy-axis N\'eel order.

To be more precise, the duality with the critical $SU(2)$
invariant $\nccp^1$ model requires an emergent $SO(5)$ symmetry:
the basic assumption underlying the duality is that allowed terms
in each theory which break this symmetry are irrelevant.  We will
make this explicit in Sec.~\ref{allowed2}.
Once again we postpone a detailed discussion of the matching of
the symmetries of the two sides and their implications until Sec.
\ref{symm_su2}.



\subsection{Duality from the sigma model}

We now provide an alternative understanding of the proposed
duality, from the standpoint of the non-linear sigma model
description of the $SU(2)$ invariant $NCCP^1$ model. We propose
the equivalence between the 2+1D $\mathrm{SO}(5)$ nonlinear sigma model
with a WZW term at level 1 (extended to a strong-coupling fixed
point) and $N=2$ QED$_3$ deformed with a quartic interaction
term to a critical point at $\lambda = \lambda_c$: \beqn &&
\mathcal{L} = \frac{1}{g} (\partial_\mu \vec{n})^2 +
\frac{2\pi}{\Omega_4} \int_0^1 du \ n^a
\partial_u n^b \partial_x n^c \partial_y n^d \partial_t n^e \cr\cr
&\Longleftrightarrow & \mathcal{L} = \sum_{j = 1}^2 \bar{\psi}_j
i\slashed{D}_a \psi_j + \frac{\lambda}{2} \left( \sum_{j=1}^2
\bar{\psi}_j\psi_j \right)^2. \eeqn We also identify a relevant
perturbation on both sides of the duality: \beqn u (n_5)^2 \sim
 u \left( \sum_{j=1}^2 \bar{\psi}_j\psi_j \right)^2. \eeqn

When $u > 0$ we expect that $n_5$ can be treated as effectively
zero in the $\mathrm{SO}(5)$ sigma model, so that this theory reduces to an O(4) nonlinear sigma model
with a $\Theta-$term at $\Theta=\pi$: \beqn \mathcal{L} =
\frac{1}{g} (\partial_\mu \vec{n})^2 + \frac{\Theta}{\Omega_3} n^a
\partial_t n^b \partial_x n^c \partial_y n^d, \quad \Theta = \pi. \eeqn This theory
has been shown \cite{tsmpaf06}, with some assumptions, to be the
low energy effective theory of $N=2$ QED$_3$. {Consistent with this,} a positive $u$  in the
QED$_3$--Gross-Neveu model will drive the
system back to QED$_3$.

When $u < 0$, the $\mathrm{SO}(5)$ sigma model develops a nonzero
expectation value $\langle n_5 \rangle$. This spontaneously breaks
the $Z_2$ subgroup of the $\mathrm{SO}(5)$. Depending on the sign of $\langle
n_5 \rangle$, {the effective O(4) sigma model for the remaining components (which we may imagine deriving by integrating out fluctuations in $n_5$) will  flow to either}
$\Theta = 2\pi $ or $\Theta = 0$. The QED$_3$--Gross-Neveu model with $u < 0$ spontaneously
condenses $\langle \sum_j \bar{\psi}_j\psi_j \rangle$. This precisely yields the two phases with $\Theta = 0$ and $2\pi$. A condensate of $\langle \sum_j \bar{\psi}_j\psi_j
\rangle$ spontaneously breaks the $Z_2$ subgroup of O(4) (the
$Z_2$ transformation that exchanges the two SU(2) subgroups),
which is also the $Z_2$ subgroup of SO(5): $Z_2$ takes $(n_1, n_2,
n_3, n_4, n_5)$ to $(-n_1, -n_2, -n_3, \ n_4, -n_5)$.

\subsection{Self-dualities}
\label{selfdual}


We showed that the easy-plane duality Eq.~\eqref{epduality}
naturally motivates the duality  of the $SU(2)$ symmetric $NCCP^1$
theory in Eq.~\eqref{so5duality}. Following the same logic, the
self-dualities of easy-plane $NCCP^1$ and the QED$_3$ theories
also motivate further self-dualities with higher symmetries. For
the $NCCP^1$ model, this self-duality reads
 \ba
\label{cp1selfdual} &\sum_{\alpha
=1,2}|D_bz_\alpha|^2-\lp |z_1|^2+|z_2|^2 \rp^2 \nn
\Longleftrightarrow \quad & \sum_{\alpha
=1,2}|D_{\tilde{b}}w_\alpha|^2-\lp   |w_1|^2+|w_2|^2 \rp^2.
\end{align}
In the dual theory, the $U(1)$ phase rotation symmetry of the
local operator $w_1^*w_2$ corresponds to the flux conservation of
the $b$ gauge field in the original theory; likewise the flux
conservation of $\tilde{b}$ in the dual theory corresponds to the
$U(1)$ phase rotation symmetry of $z_1^*z_2$ in the original
theory. One can check the consistency of this duality by turning
on an anisotropy term \be
\lambda|z_1|^2|z_2|^2\sim\lambda|w_1|^2|w_2|^2. \ee When
$\lambda>0$, both theories flow to easy-plane $NCCP^1$, where
the self-duality holds.\footnote{Strictly speaking, when $\lambda>0$, we need to compensate the anisotropy term with a mass term $-\alpha\lambda(|z_1|^2+|z_2|^2)$ (likewise for the $w$-theory), to keep the theories on both sides critical. We discuss this further in Sec.~\ref{PhaseD}.} When $\lambda<0$, both theories flow to
the easy-axis limit, where the system becomes trivially gapped.

Similarly for the QED$_3$-GN model we have the self-duality \bea
\label{qedgnselfdual} \sum_{j=1,2}\bar{\psi}_ji\slashed{D}_a\psi_j
+ \phi\sum_{j = 1,2}\bar{\psi}_j \psi_j +V(\phi)  \nn
\Longleftrightarrow \hspace{20pt}
\sum_{j=1,2}\bar{\chi}_ji\slashed{D}_{\tilde{a}}\chi_j
-\phi\sum_{j = 1,2}\bar{\chi}_j \chi_j +V(\phi). \eea The
switching of the two global $U(1)$ symmetries in the two pictures
is similar to that in the self-duality of $NCCP^1$. The consistency
of this duality is checked by turning on a mass term for the Ising
scalar fields on both sides: when $\langle\phi\rangle\neq0$ both
sides are trivially gapped, and when $\phi$ is gapped the duality
follows from the self-duality of $QED_3$.

\subsection{Symmetries}
\label{symm_su2}


We now study the symmetries of the $NCCP^1$ theory and its duals. We
simply assume the correctness of the proposed dualities.

The bosonic side has a manifest {\red $SO(3)\times O(2)$ symmetry, where
the $z$ bosons are $SO(3)$ spinors, and the global $U(1)$
symmetry is simply the flux symmetry of the gauge field $b$. The
fermionic side also has a (different) {manifest $[{\mathrm{SU}(2)\times \mathrm{U}(1)}]/{Z_2}$
symmetry (more precisely $[{{\mathrm{SU}}(2)\times \operatorname{Pin}(2)_-}]/{Z_2}$ when charge conjugation is accounted for).

How do these symmetries act on physical operators? Collecting the various gauge-invariant  order parameters, we have  (recall that $\mathcal{M}_b$ is the monopole operator in the $NCCP^1$ model)
\begin{align}\notag
&(  2  {\rm Re} \mathcal{M}_b,
2{\rm Im}(\mathcal{M}_b),
 z^\dag \sigma_x z, z^\dag \sigma_y z,z^\dag \sigma_z z
 )\\ 
 &\sim (n_1, n_2, n_3, n_4, n_5).\label{SO(5)_vector}
\end{align}
The operators
$(n_3, n_4, n_5)\sim (z^\dag \sigma_x z, z^\dag \sigma_y z,z^\dag \sigma_z z)$
form a fundamental representation of the $\mathrm{SO}(3)$
symmetry in the
$\nccp^1$ model, while ${(n_1+in_2, n_3 - i n_4)\sim 2(\mathcal{M}_b, z_2^*z_1)}$ transforms
as a spin-$1/2$ representation of the flavor $\mathrm{SU}(2)$ symmetry of
the QED$_3$ theory. It is easy to see that the vector in Eq.~\ref{SO(5)_vector} then forms  a
vector representation
of an enlarged $SO(5)$ symmetry. 

The $O(2)$ symmetry of the $NCCP^1$ theory acts as follows: proper rotations act only on $(n_1, n_2)$, while improper rotations also multiply $(n_3, n_4, n_5)$ by a minus sign. [Therefore the $\mathrm{SO}(3)\times \mathrm{O}(2)$ of $NCCP^1$ is indeed a subgroup of $\mathrm{SO}(5)$.] The $U(1)$ flux symmetry of the QED$_3$ theory acts as a common phase rotation of both $n_1 + i n_2$ and $n_3 - i n_4$. Finally the charge conjugation symmetry of QED$_3$, which reverses the charge under this $U(1)$, acts as $(n_1, n_2, n_3, n_4)\rightarrow (-n_3, -n_4, n_1, n_2)$. 
}

A different and interesting perspective is provided  by taking the proposed strong self-duality of the $SU(2)$ invariant $NCCP^1$ model as our logical starting point.
  In the original model,
the operators
$(z^\dag \sigma_x z, z^\dag \sigma_y z,z^\dag \sigma_z z)$
form a vector under a global
$SO(3)$ symmetry. In the dual model, the operators
$(w^\dag \sigma_x w, w^\dag \sigma_y w,z^\dag \sigma_z w)$
form a vector
under another global $SO(3)$. Since
$z^\dag \sigma_z z \sim w^\dag \sigma_z w$,
it is easy to see that the
operators
${(w^\dag \sigma_x w, w^\dag \sigma_y w,
z^\dag \sigma_x w, z^\dag \sigma_y z, z^\dag \sigma_z z)
}$
form a vector under an enlarged $SO(5)$ symmetry.  Thus,  the {strong} self-duality of the $SU(2)$ invariant $NCCP^1$ model implies the presence of an enlarged $SO(5)$ symmetry. {Conversely,} the numerical evidence for the emergence of $SO(5)$ symmetry at the N\'eel-VBS transition may be taken as support for the proposed self-duality of the $SU(2)$--invariant $NCCP^1$ model.

Further support comes from considering a deformation  of the model to reach the easy plane model. In the $z$-theory, this can be accomplished by perturbing with the operator
$(z^\dagger \sigma_z z)^2$. In the $w$-theory the same operator is represented as $(w^\dagger \sigma_z w)^2$.  We expect that this is a relevant perturbation (see Sec. \ref{allowed2}). The resulting flow leads directly to the self-duality of the easy plane $NCCP^1$ model which has been independently derived (in its weak form).  This is a good consistency check on the self duality of the $SU(2)$ invariant model. 

The action of discrete symmetries is similar to the easy plane
case discussed above.  Note, however, that the ${\mathcal S}$
symmetry is a subgroup of the $SU(2)$ flavor symmetry of the
$NCCP^1$ theory, and that ${\mathcal S}_\psi$ is a subgroup of the
$SU(2)$ flavor symmetry of QED$_3$. We, thus, do not need to
consider them separately.  For square lattice spin-$1/2$ magnets,
the action of time reversal and lattice symmetries may be readily
inferred from the easy plane case once we recognize that the extra
field $\phi$ transforms identically to $z^\dag \sigma_z z$.

\subsection{Allowed symmetry breaking terms}
\label{allowed2}

We now study the $SO(5)$-breaking operators that are allowed by microscopic symmetries. These operators must be irrelevant in order for the dualities and the emergent symmetries to hold in the IR without fine-tuning. Again we point out that in the situation where there is no fixed point with the higher symmetry,
there could still be a pseudocritical regime up to a large but finite lengthscale $\xi$ (see Sec.~\ref{pseudocritical}); in this case the requirements should be interpreted in terms of the effective scaling dimensions in this regime.

The putative emergent symmetry for the $\nccp^1$ model and
QED-Gross-Neveu is $\mathrm{SO}(5)$. A natural guess is that the
leading scalar operators, apart from the $\mathrm{SO}(5)$ vector
$n_a$ defined in Eq.~\eqref{SO(5)_vector}, are those in the two and
four-index symmetric tensor representations of $\mathrm{SO}(5)$.
We denote these $X^{(2)}_{ab}$ and $X^{(4)}_{abcd}$. At the level
of symmetry, \ba X^{(2)}_{ab} &\sim n_a n_b - \delta_{ab} n^2/5, &
X^{(4)}_{abcd} & \sim n_a n_b n_c n_d - (\ldots),
\end{align}
where the subtraction $(\ldots)$ makes the operator traceless.
$X^{(2)}$ is certainly relevant.\footnote{See \cite{emergentso5} for numerical
results for correlation functions of $X^{(2)}$ components.} The
microscopic symmetries of $\nccp^1$ allow the perturbation
$\sum_{a=3}^5 X^{(2)}_{aa}$, which is an anisotropy between N\'eel
and VBS ($\sim \f{2}{5}[n_3^2+n_4^2+n_5^2]- \f{3}{5} [n_1^2
+n_2^2]$) and QED-GN allows $X^{(2)}_{55}$,  corresponding to the
mass term for the scalar field $\phi$.  Since these are the
perturbations that are tuned away to reach the critical point,
they do not pose a problem for stability. However stability does
require the irrelevance of $X^{(4)}$, as this gives rise to
further symmetry--allowed perturbations. The symmetries of
$\nccp^1$ allow the higher N\'eel-VBS anisotropy
$\sum_{a=1}^2\sum_{b=3}^5 X^{(4)}_{aabb}$. For a quantum
antiferromagnet on the square lattice,\footnote{For the quantum
antiferromagnet we are also allowed the term $(\partial_x n_1)^2 +
(\partial_y n_2)^2+...$. The emergence of $U(1)$ symmetry for the VBS
in JQ model simulations \cite{SandvikJQ, lousandvikkawashima} implies that this term is also
irrelevant. This term is
absent for the model of  \cite{DCPscalingviolations,emergentso5} since this model is isotropic in
space-time.}  the anisotropy $\sum_{a=1}^2 X^{(4)}_{aaaa}$, which
breaks the $U(1)$ symmetry for the VBS down to  $Z_4$,  is also
allowed. In QED-GN the anisotropy $X^{(4)}_{5555}$ is allowed.

Stability also requires the irrelevance of all
$\mathrm{SO}(5)$-singlet scalar operators. As discussed in
Sec.~\ref{pseudocritical}, this requirement is in tension with
conformal bootstrap results \cite{SimmonsDuffinSO(5), Nakayama}. However, the
numerical evidence for $\mathrm{SO}(5)$ suggests that there is at
least a pseudocritical regime where allowed $SO(5)$--breaking
perturbations, including $X^{(4)}$, are effectively irrelevant.

\subsection{Phase diagram}
\label{PhaseD}

We now discuss the phase diagram of the quantum magnet near the $SU(2)$-invariant deconfined critical point, allowing for a perturbation that breaks the spin symmetry to easy-plane. We assume the emergence of $SO(5)$ symmetry at the $SU(2)$ critical point.

It is useful to organize perturbations into representations of the $SO(5)$ symmetry.  The two perturbations that we must consider live in the symmetric tensor representation $X^{(2)}$  of $SO(5)$ discussed above (we drop the superscript),
\be
X_{ab}\sim n_an_b-\frac{\delta_{ab}}{5}{n}^2,
\ee
and we denote them
\begin{align}
\mathcal{O}_1 &= X_{11} + X_{22},
&
\mathcal{O}_2 & = X_{55}.
\end{align}
First, the leading perturbation allowed in an $SU(2)$-symmetric spin model is ${\delta \mathcal{L} = \lambda_1\mathcal{O}_1}$, which drives the system into the N\'eel ordered phase for $\lambda_1 <0$ and into the VBS phase for $\lambda_1 > 0$.  Second, breaking the spin symmetry down to that of the easy plane model allows the anisotropy ${\delta \mathcal{L} = \lambda_2 \mathcal{O}_2}$. Again we already know the effect of this operator on its own: it drives the system into an easy-axis-ordered gapped phase for $\lambda_2>0$,  and to the N\'eel-VBS phase transition of the easy plane model, potentially with $O(4)$ symmetry, for $\lambda_2<0$.

The full phase diagram for small $(\lambda_1$, $\lambda_2)$  follows by $SO(5)$ symmetry, assuming that the only ordered phases in the vicinity of the $SU(2)$ critical point are those mentioned above. Essentially, each of the three ordered phases is determined by which components of $n_a$ are favored by the potential ${\lambda_1(X_{11}+X_{22})+ \lambda_2 X_{55}}$.  More formally, the transition  line $\lambda_1 = \lambda_2 > 0$ between the Ising and VBS ordered phases may be fixed by noting that it corresponds to the perturbation
\be
\lambda_1 (X_{11}+ X_{22} + X_{55} ) = - \lambda_1 (X_{33} + X_{44} ),
\ee
where we have used the tracelessness of $X$. This is related to the N\'eel ordered line $\lambda_1 < 0$, $\lambda_2 = 0$ by the $SO(5)$ rotation  $n_1\leftrightarrow n_3$, $n_2\leftrightarrow n_4$, which is precisely the  self-duality of the $NCCP^1$ theory. The phase diagram in the $(\lambda_1,\lambda_2)$ basis is shown in Fig.~\ref{so5pd1}.

\begin{figure}[h]
\begin{center}
\includegraphics[width=3.5in]{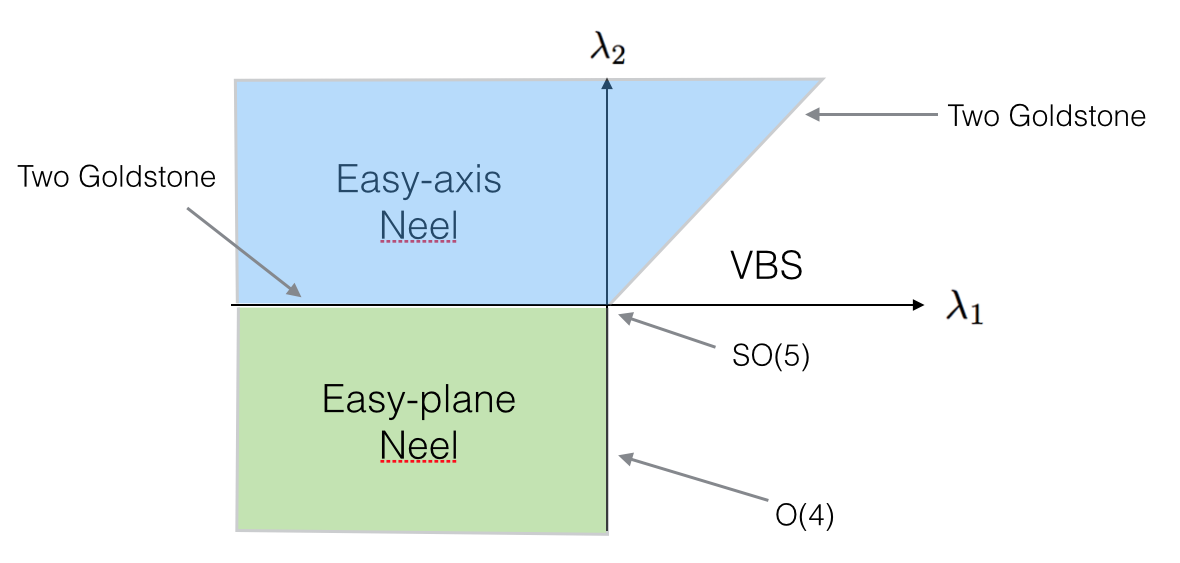}
\end{center}
\caption{Phase diagram near the $SO(5)$-invariant fixed point with perturbation of the form $\lambda_1(X_{11}+X_{22})+ \lambda_2 X_{55}$.} \label{so5pd1}
\end{figure}

If we neglect perturbations that are irrelevant at the $SO(5)$ fixed point, the transition between VBS  and  easy-axis order  is governed by a Lagrangian with an emergent $O(3)$ symmetry rotating $(n_1, n_2, n_5)$. This is spontaneously broken to $SO(2)$ once $\lambda_1=\lambda_2$ flows to an order one value, yielding a pair of Goldstone modes.  In reality, these Goldstone modes are only approximate if the bare $\lambda_1=\lambda_2$ is finite;  the emergent $O(3)$ is  explicitly broken by  dangerously irrelevant higher anisotropies\footnote{Components of $X^{(4)}$ in the notation of Sec.~\ref{allowed2}. These are presumed to be irrelevant at the $SU(2)$ critical point, but are relevant in the $O(3)$-breaking phase.} which are allowed by the symmetries of the lattice model. However these anisotropies will appear only at a parametrically large lengthscale when the bare $\lambda_1 = \lambda_2$ is small.

Similarly, the line $\lambda_1=0$, $\lambda_2<0$ which leads to the easy-plane deconfined transition has an emergent $O(4)$ symmetry when higher anisotropies are neglected. Here however it is possible that the $O(4)$ symmetry survives to asymptotically long lengthscales: this depends on the ultimate fate of the easy-plane theory.

The structure of the phase diagram above could be tested numerically. The most basic test is that the phase boundaries all meet at nonzero angles, showing that the distinct components of $X$ have the same scaling dimension.\footnote{See Ref.~\cite{emergentso5} for a direct numerical test of this using correlation functions.} There is also universal information in the slopes of the phase boundaries. In the microscopic model a more natural basis for perturbations is ${\delta \mathcal{L} = \tilde \lambda_1 \mathcal{O}_1 + \tilde \lambda_2 {\widetilde{\mathcal{O}}}_2 }$, where $\mathcal{O}_1$ is the lattice operator which drives the N\'eel-VBS transition and ${\widetilde{\mathcal{O}}}_2 \sim X_{55} + \f{1}{3}(X_{11} + X_{22})$ is a modified easy-plane anisotropy. The numerical coefficient in the latter is fixed by demanding that it belongs to the traceless symmetric tensor representation of spin $SO(3)$: ${\mathcal{A}_{ij} \sim X_{ij} + \f{1}{3}(X_{11} + X_{22})}$, where  $i,j=3,4,5$. In $NCCP^1$,  $\mathcal{O}_1 \sim - |z|^2$ and $\widetilde{\mathcal{O}}_2 \sim [|z_1|^2-|z_2|^2]^2-\frac{1}{3}[|z_1|^2+|z_2|^2]^2$. When we draw the phase diagram  in the $(\tilde \lambda_1, \tilde \lambda_2)$ plane, the easy-axis N\'eel/VBS transition line is at $\tilde \lambda_1 = 2 c \tilde \lambda_2$, and the easy-plane N\'eel/VBS transition line is at $\tilde \lambda_1 = - c \tilde \lambda_2$. The constant $c>0$ is arbitrary since the normalization of the lattice operators is arbitrary, but the ratio of the slopes of the two lines is a fixed constant which could be checked numerically. The phase diagram in this $(\tilde{\lambda}_1,\tilde{\lambda}_2)$ basis is shown schematically in Fig.~\ref{so5pd2}.

\begin{figure}[h]
\begin{center}
\includegraphics[width=3.5in]{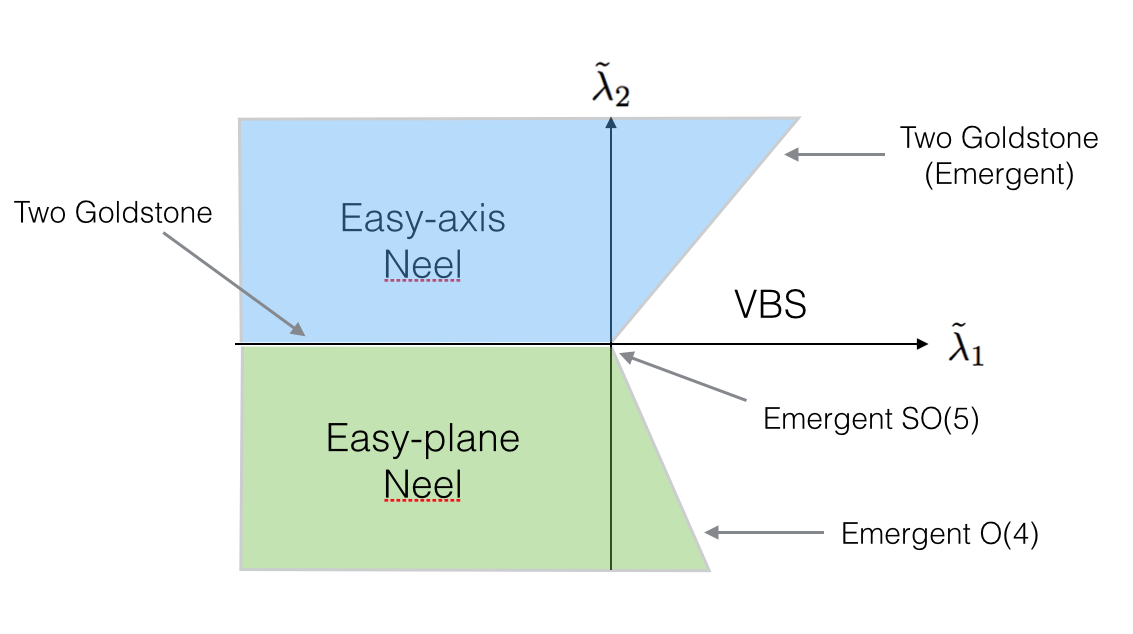}
\end{center}
\caption{Phase diagram near the $SO(5)$-invariant fixed point with perturbation of the form $-\tilde{\lambda}_1(|z_1|^2+|z_2|^2)+ \tilde{\lambda}_2 [(|z_1|^2-|z_2|^2)^2-\frac{1}{3}(|z_1|^2+|z_2|^2)^2]$. This is the natural perturbation to consider in the context of deconfined criticality in quantum magnets. The emergent $SO(5)$ symmetry requires that the slope of the lower transition line is twice that of the upper transition line.} \label{so5pd2}
\end{figure}

Alternately   we may check universal amplitudes using correlation functions, once the location of one of the nontrivial transition lines is determined. Let us normalize $X_{ab}$ so that
\be
\label{so5normal}
\langle X_{ab}(x) X_{cd}(0)\rangle = \frac{1}{2}\left(\delta_{ac} \delta_{bd} + \delta_{ad}\delta_{bc} - \frac{2}{5} \delta_{ab}\delta_{cd} \right) \frac{1}{x^{2\Delta_2}},
\ee
where $\Delta_2$ is the scaling dimension of  $X_{ab}$.  Assume that we can identify (numerically) either the perturbation $\mathcal{O}_2 \sim - X_{55}$ which drives the system along the VBS/easy-plane N\'eel phase boundary, or the perturbation ${\mathcal{O}_3\sim X_{11}+X_{22}+X_{55}}$ that drives the system to the first-order transition between the VBS and the easy-axis N\'eel state. Then by Eq.~\eqref{so5normal} the following statements, independent of normalization, should be true:
\begin{eqnarray}
\frac{\langle\mathcal{O}_1(x)\mathcal{O}_2(0)\rangle^2}{\langle\mathcal{O}_1(x)\mathcal{O}_1(0)\rangle\langle\mathcal{O}_2(x)\mathcal{O}_2(0)\rangle}&=&\frac{1}{6},
\nonumber \\
\frac{\langle\mathcal{O}_1(x)\mathcal{O}_3(0)\rangle^2}{\langle\mathcal{O}_1(x)\mathcal{O}_1(0)\rangle\langle\mathcal{O}_3(x)\mathcal{O}_3(0)\rangle}&=&\frac{4}{9}.
\end{eqnarray}

Similar tests are possible in the QED-Gross-Neveu theory, if the fixed point is found. There ${\mathcal{O}_2 \sim \phi^2}$ is the Ising mass operator  that drives the system through the Gross-Neveu transition between $QED_3$ and the gapped phase in which $\phi$ has condensed. The fermion chiral mass ${\bar{\psi}_1\psi_1-\bar{\psi}_2\psi_2}$ is a mixture of $\mathcal{O}_1$ and $\mathcal{O}_2$. The $SU(2)$ flavour symmetry of QED requires the chiral mass to be orthogonal to $\mathcal{O}_2$, so  by Eq.~\eqref{so5normal}
\be
\bar{\psi}_1\psi_1-\bar{\psi}_2\psi_2\sim \mathcal{O}_1+\frac{1}{2}\mathcal{O}_2.
\ee

Now if we consider a perturbation of the form $m_{\phi}\phi^2+m_{\psi}(\bar{\psi}_1\psi_1-\bar{\psi}_2\psi_2)$, the phase diagram will look like Fig.~\ref{so5pd3}. The phase diagram is symmetric under the reflection across the $m_{\phi}$ axis simply because of the fermion flavor symmetry. The two transitions lines near the gapped phase are given by $m_{\phi}=\pm\frac{1}{2c'}m_{\psi}>0$ with $c'>0$ being a normalization-dependent constant. So $c'$ alone does not provide nontrivial information. However, it enters into the ratio of correlation functions:
\be
\frac{\langle\phi^2(x)\phi^2(0)\rangle}{\langle(\bar{\psi}_1\psi_1-\bar{\psi}_2\psi_2)(x)(\bar{\psi}_1\psi_1-\bar{\psi}_2\psi_2)(0)\rangle}=\frac{4}{5}(c')^2, \label{eq:cpsq}
\ee
which is in principle testable.

\begin{figure}[h]
\begin{center}
\includegraphics[width=3.5in]{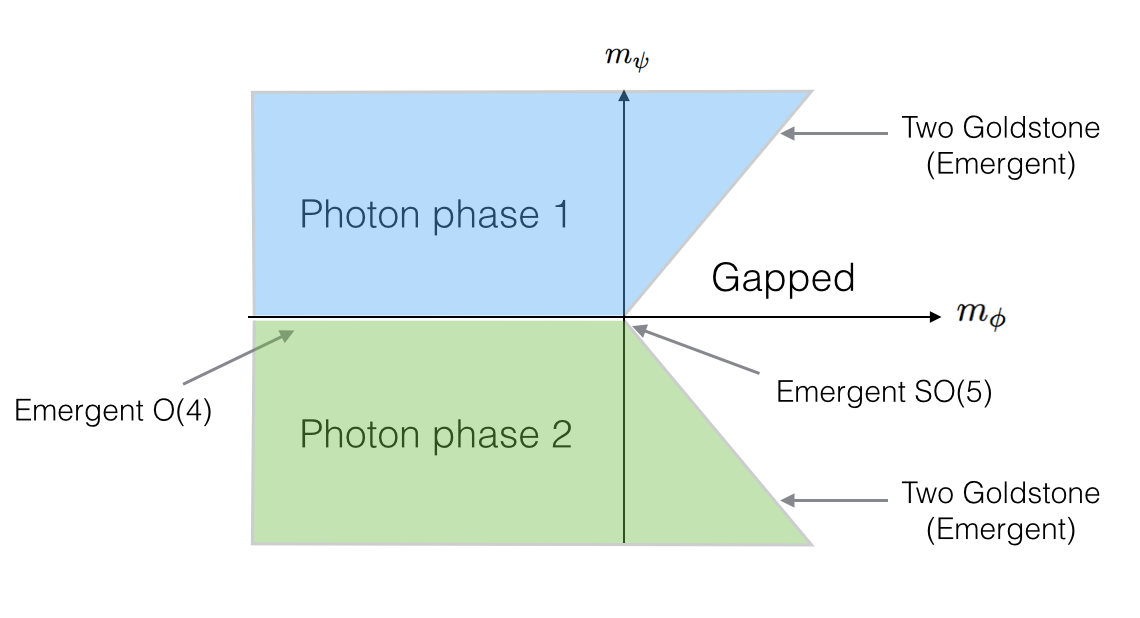}
\end{center}
\caption{Phase diagram near the $SO(5)$-invariant fixed point with perturbation of the form $m_{\phi}\phi^2+m_{\psi}(\bar{\psi}_1\psi_1-\bar{\psi}_2\psi_2)$. This is the natural perturbation to consider in the context of QED$_3$--Gross-Neveu theory. The emergent $SO(5)$ symmetry predicts that the slope of the transition lines is related to the relative amplitude of the correlation functions of the two operators, Eq.~(\ref{eq:cpsq})} \label{so5pd3}
\end{figure}

We end this section with a discussion on the nature of the transitions at  $m_{\phi}=\pm\frac{1}{2c'}m_{\psi}>0$ in Fig.~\ref{so5pd3}. As discussed already, we expect the two transitions to be direct first order, instead of broadening into coexistence phases. How do we understand this from a fermion mean-field point of view? We can calculate the mean field free energy with respect to $m_{\psi}$ and $\la\phi\ra$, treating the fermions as almost non-interacting, which is valid when the fermion flavor number $N_f\to\infty$. The result is
proportional to \be E\sim |m_\psi+\la\phi\ra|^3+|m_\psi-\la\phi\ra|^3. \ee
Interestingly, this function gives no preference to either
scenario (first order or coexistence). Presumably fluctuations
beyond mean-field will break this degeneracy and lead to a direct
transition.

\section{Bulk interpretation: $U(1) \times U(1)$ theory}
\label{bulk}

From the sigma model point of view, it is known that the
symmetries of the field theories discussed so far have to be
anomalous \cite{ashvinsenthil,BiNLSM}. Deconfined criticality can
nevertheless be realized in quantum magnets because lattice
rotation symmetries are not on-site, and therefore can be
implemented in a seemingly anomalous fashion in the continuum theory. If we want all the
symmetries to be on-site, the theories must be regularized on the
boundary of a $3+1$D bulk. In this section we discuss the bulk
physics corresponding to the easy-plane deconfined critical point.
This provides considerable insight into the duality web and the
unconventional symmetry actions of the theory.

In Eq.~\eqref{z2anomaly} we see that the spin-flip    ${\mathcal
S}$, time reversal, and other symmetries  are anomalous: \be
\mathcal{L}\to\mathcal{L}-\frac{1}{2\pi}B_1dB_2. \ee We will
initially focus on the $\mathcal S$ symmetry. This anomaly is
natural from the sigma model approach: the $\mathcal{S}$ symmetry
is an improper $O(4)$ rotation $(n_1,n_2,n_3,n_4)\to (n_1,-
n_2,n_3,n_4)$ which fixes $\theta=\pi$, and such a symmetry is
typically expected to be anomalous.

The anomaly can be cured by placing the $2+1$D theory at the
boundary of a $3+1$D bosonic SPT insulator with $[U(1) \rtimes
Z_2] \times U(1)$ symmetry.  Let us  couple $3+1$D background
gauge fields $B_1$ and $B_2$ to the two $U(1)$ symmetries such that
under $\mathcal S$ they transform as \be \label{Sbulk}
\mathcal{S}: B_1\to-B_1, B_2\to B_2. \ee A non-trivial SPT phase
of such a bosonic system then has a response characterized by a
mutual $\Theta$ term at $\Theta = \pi$ for the two gauge fields
$B_1$ and $B_2$ of the form
 \be
 -\frac{\Theta}{(2\pi)^2}\int_{Bulk}dB_1\wedge dB_2, \quad \Theta = \pi
 \ee
Notice that under  $\mathcal S$,  $\Theta\to-\Theta$, and
therefore $\Theta=\pi$ is protected\footnote{We have used the
periodicity of the physics under shifts $\Theta \to \Theta + 2\pi$
which holds for this system.} by the $\mathcal{S}$ symmetry.

Now consider the surface of this boson SPT phase.  The bulk
$\Theta$ term leads to a surface state with anomalous symmetry
realization. Clearly this anomaly is exactly the same as in the
$2+1$D easy plane $NCCP^1$ field theory, Eq.~(\ref{z2anomaly}). Specifically if we add a
bulk contribution, $-\frac{1}{4\pi} B_1dB_2$, to the Lagrangian, defining 
 \be
 {\cal L}_{z,SPT} = {\cal L}_z -\frac{1}{4\pi} B_1dB_2,
 \ee
 (where  the  extra term, which is not well-defined as a mutual
Chern-Simons term in pure $2+1$D, is really a short-hand for the bulk mutual $\Theta$-term at $\Theta=\pi$) it is easy to check that ${\cal L}_{z,SPT}$ is indeed invariant under the
spin-flip symmetry.  

Now imagine gauging the $U(1)_{B_1}\times U(1)_{B_2}$ symmetry in
the bulk. An important observation is that due to the mutual
$\Theta$-term the monopole of one species carries charge $\pm 1/2$
of the {\em other} species. Let us label the charge-monopole
lattice by $(q_{e1}, q_{e2}; q_{m1}, q_{m2})$. Here $(q_{e1},
q_{m1})$ are the electric and magnetic charges under $U(1)_{B_1}$,
and so on.  The mutual $\Theta$ term implies the relations 
\begin{align}
\label{qmlatbdir}
q_{e2} & =   \frac{q_{m1}}{2}  ~~(\mathrm{mod} ~Z),  &
q_{e1} & =  \frac{q_{m2}}{2} ~~ (\mathrm{mod} ~Z). \end{align}
Note that $q_{m1}, q_{m2}
\in Z$.

There is a correspondence between fields in the boundary theory and particles in the bulk theory: bulk electric charges correspond to electrically charged surface fields, and bulk magnetic charges correspond to vortices on the surface. For the `physical' bosons the correspondence is ${\Phi_{B_1} \sim (1,0;0,0)}$ and ${\Phi_{B_2} \sim (0,1;0,0)}$. The surface fields $z_{1,2}$ are vortices in $\Phi_{B_2}$, and they carry charge $q_{B_1} = \pm 1/2$. Their bulk avatars are thus the dyons $z_{1,2} \sim (\pm \f{1}{2}, 0;0,1)$.

The bosonic self-duality of the easy plane $NCCP^1$ theory leads
to a description in terms of complex fields $w_{1,2}$ which are
vortices of $\Phi_{B_1}$ and carry charges $q_{B_2} = \pm 1/2$.
Clearly their bulk avatars are $w_{1,2} \sim (0,\pm \frac{1}{2};
1,0)$. The surface self-duality is thus connected to the obvious
bulk duality between descriptions in terms of these two sets of
dyons.

Consider the bound states of these two kinds of dyons with quantum
numbers \begin{align} \label{bulkpsi}
 &(\frac{1}{2},  \frac{1}{2}; 1, -1), &&( -\frac{1}{2},
-\frac{1}{2}; 1, -1). \end{align}
These are both fermions.
We identify them
as the bulk avatars of $\psi_{1,2}$. This can be confirmed
directly from the surface theory. Consider the $QED_3$ theory with $\psi_{1,2}$ fermions in Eq.~\eqref{epduality}  with
the added `bulk' contribution $-\frac{1}{4\pi}B_1dB_2$. Notice that $B=(B_1+B_2)/2$. We see that the $\psi_{1,2}$ fermions indeed have the right charges and vorticities to
correspond to these bulk fermionic dyons.

Thus the duality between the easy plane $NCCP_1$ theory and the
QED$_3$ theory can be understood in terms of a bulk duality that
trades the bosonic $(\pm \frac{1}{2},0; 0,1)$ particles with the
fermionic particles of  Eq.~\eqref{bulkpsi}.

What about the dual fermions $\chi_{1,2}$? They correspond to \begin{align}
&(\frac{1}{2}, -\frac{1}{2}; 1, 1), &&( -\frac{1}{2}, \frac{1}{2};
1, 1). \end{align} Indeed this is exactly what is implied by the dual
fermionic surface theory.

The fermion-fermion duality of QED$_3$ can thus be related to a
corresponding bulk fermion-fermion duality of the $U(1) \times
U(1)$ gauge theory.

Notice that 
$(q_{e1},q_{e2};q_{m1},q_{m2})\to(-q_{e1},q_{e2};-q_{m1},q_{m2})$ under $\mathcal{S}$.
It is immediately clear that the two fermionic dyons corresponding
to $\psi_{1,2}$ become the two dyons corresponding to
$\chi_{1,2}$ under $\mathcal{S}$. This offers a bulk
interpretation of the non-trivial action of $\mathcal{S}$ on the
surface $QED_3$ theory, which exchanges $\psi_{1,2}$ and their
dual fermions $\chi_{1,2}$.

Likewise, under the fermion flavor exchange symmetry ${\cal S}_\psi$,
which acts as $B_1\leftrightarrow -B_2$, the dyons corresponding
to $z_{1,2}$ and $w_{1,2}$ are exchanged. This is a simple bulk
picture of the symmetry action in the boundary $NCCP^1$ model,
which is implemented through the self-duality transform.

Note that the $U(1) \times U(1)$ gauge theory has an $\mathrm{Sp}(4,Z)$
invariance corresponding to basis changes in the four dimensional
charge-monopole lattice.  This is because the basis change must
preserve the area of the unit cell of each two dimensional
subspace corresponding to each of the two $U(1)$ gauge theories.
The surface web of dualities we have discussed can be understood
as the effects of various $\mathrm{Sp}(4,Z)$ transformations of the bulk
gauge theory.

We should also emphasize that the bulk duality offers a simple
picture of the surface duality, but does not prove the surface
duality between IR fixed point theories.

Now let us turn briefly to time reversal which acts on $B_{1,2}$
as \be T(B_{1,2}) = B_{1,2}. \ee Under this the bulk $\Theta$ term
is odd, but as before $\Theta = \pi$ is time reversal symmetric.
Correspondingly when a surface is present, the contribution from
this bulk $\Theta$ term will exactly cancel the time-reversal
anomaly of the surface theories. The bulk charges transform under
time reversal as \be T: (q_{e1},q_{e2};q_{m1},q_{m2}) \to (-
q_{e1}, - q_{e2};q_{m1},q_{m2}).\ee It is readily checked that
this is precisely consistent with the time reversal action on each
of the surface theories.

We already saw that the translation $\tilde{T}_y$ can be
related to a combination of $\mathcal S$ and ${\mathcal S}_\psi$
and therefore does not need separate discussion.

\section{Theories with manifest ${\mathrm{SO}(5)}$ symmetry}
\label{manifest SO(5)}

So far none of our field-theoretic descriptions of the deconfined
critical point possessed explicit ${\mathrm{SO}(5)}$ symmetry in the UV:
this symmetry at best emerged in the IR. An exception was the
$\mathrm{SO}(5)$ NLSM with a WZW term at level $1$; however, this model is
non-renormalizable, so while one can infer symmetry information
from it, strictly speaking its dynamics in the disordered phase is
not well-defined. {Here we present two renormalizable
theories with explicit ${\mathrm{SO}(5)}$ symmetry, namely $N_f=2$ $QCD_3$ and its Higgs descendent $N_f=4$ compact $QED_3$.  While the IR
fates of these theories are unknown, they share the same anomaly
with the deconfined critical point. So there is the possibility (among others) that either of them may flow to the deconfined
critical point.}

\subsection{Parton construction of $N_f=2$ $QCD_3$}
\label{partonqcd3}

To see the connection between deconfined criticality and these theories, we now review the construction of the $\pi$-flux state on the square lattice \cite{wenbook}  and
demonstrate that its low energy theory, $QCD_3$, has an emergent ${\mathrm{SO}(5)}$ symmetry. The Neel and VBS order parameters transform, as expected,  as the 5 components of an $\mathrm{SO}(5)$ vector.

Consider the standard fermionic parton decomposition:
\begin{align}
\vec{S}_i &= \frac12 f^{\dagger}_{i \alpha} \vec{\sigma}_{\alpha
\beta} f_{i \beta}, & f^{\dagger}_{\alpha i} f_{\alpha i} &= 1,
\label{eq:fspinons}\end{align}
 where $i$ labels a lattice site and the
spin indices $\alpha, \beta$ are summed over. Let us form a matrix
\beq X_i = \left(\begin{array}{cc} f_{i\up} &
-f^{\dagger}_{i\down}\\f_{i\down} &
f^{\dagger}_{i\up}\end{array}\right).\eeq The decomposition
(\ref{eq:fspinons}) is invariant under local ${\mathrm{SU}(2)}$ gauge
rotations \beq {\mathrm{SU}(2)}_g:\quad X_i  \to  X_i (U^{g
}_i)^{\dagger}. \label{eq:SU2g} \eeq The physical ${\mathrm{SU}(2)}$ spin rotations act as \beq
{\mathrm{SU}(2)}_s:\quad X_i \to U^s X_i \eeq and we can rewrite
(\ref{eq:fspinons}) as $\vec{S}_i = \frac14
\mathrm{tr}(X^{\dagger}_i \vec{\sigma} X_i)$. It will
occasionally be convenient to write ${f_{\up} = \frac{1}{\sqrt{2}}
(\chi_0 + i \chi_3)}$, ${f_\down = \frac{1}{\sqrt{2}}(-\chi_2 + i
\chi_1)}$, i.e ${X = \frac{1}{\sqrt{2}}(\chi_0 + i \chi_a
\sigma^a)}$, where $\chi_{m}$, $m = 0, 1,2,3$, are Majorana
fermions.  $\chi_m$ transforms as an $SO(4)$ vector under the
combined action of ${\mathrm{SU}(2)}_s$ and ${\mathrm{SU}(2)}_g$.

We consider a mean-field state on the square lattice, \beq H_{MF}
= -\frac{i}{2} \sum_{ij} t_{ij}  \chi_{im} \chi_{j m}
\label{eq:HMF}\eeq where $t_{ij} = - t_{ji}$, and $t_{i+\hat{x},
i} = 1$, $t_{i + \hat{y},i} = (-1)^{i_x}$ so that there is $\pi$
flux through each plaquette. This mean-field explicitly preserves
the ${\mathrm{SU}(2)}_s$ and ${\mathrm{SU}(2)}_g$ symmetries, while lattice symmetries now
act in a projective manner (see below). Each flavor $m$ of
Majorana fermions produces two gapless Majorana cones, so the
low-energy theory becomes, \beq H_{MF} = i \sum_{v = 1, 2}
\chi_{m,v} \left(\tau^x \d_x - \tau^z \d_y\right) \chi_{m,v}\eeq
with $\tau$ acting on suppressed sublattice indices $\sigma \in A,
B$ (see below). The index $v$ runs over two valleys and lattice
fields are related to continuum ones in the following way. The
unit cell is doubled by $t_{ij}$. We  label sites with even
$i_x$ by $A$ and odd $i_x$ by $B$, and label unit cells by
$\bar{i} \in (2 \mathbb{Z} + 1/2, \mathbb{Z})$.  Letting
$\chi_{\bar{i}} = (\chi_{\bar{i} - \hat{x}/2, A},
\chi_{\bar{i}+\hat{x}/2, B})$, \beq \chi_{m, \bar{i}} \sim \tau^x
\chi_{m, v = 1}(x) + (-1)^{\bar{i}_y} \chi_{m, v = 2}(x).\eeq We
can rewrite the mean-field Lagrangian as \beq \mathcal{L}_{MF} = i
\bar{\chi}_{v,m} \gamma^{\mu} \d_{\mu} \chi_{v,m}, \eeq where
$\bar{\chi} = \chi^T \gamma^0$, $\gamma^0 = \tau^y$, $\gamma^x = i
\tau^z$, $\gamma^y = i \tau^x$. The action of lattice symmetries
is:
\begin{align} 
T_x:& \quad \chi \to \mu^x \chi,\\
T_y: &\quad \chi \to \mu^z \chi,\\
R_{\pi/2,A}:& \quad \chi \to e^{\pi i \tau^y/4} e^{-\pi i \mu^y/4} \chi(-y,x),\\
P_{x,A}:& \quad \chi \to \tau^z \mu^z \chi(-x,y),\\
T:& \quad \chi \to \tau^y \mu^y \chi, \quad i \to -i,
\label{eq:symQCD}\end{align}
 where $\mu$ acts on the valley index $v$ and
spin/color indices $m$ have been suppressed. The subscript $A$ on
$\pi/2$ rotation $R$ and reflection $P$ indicates that these are
around an $A$ site. These symmetries prohibit any quadratic term in
$\chi$ with no derivatives in $\mathcal{L}_{MF}$.

The mean-field theory has an $O(8)$ symmetry acting on $m, v$.
However this is broken by fluctuations of ${\mathrm{SU}(2)}_g$ gauge field
and four-fermi interactions. Let us first focus on the gauge field
fluctuations. For this purpose it is convenient to introduce a $4
\times 2$ matrix, $X_{\alpha, v; \beta}$ via \beq X_{\alpha, v;
\beta} = \frac{1}{\sqrt{2}} (\chi_{0,v} \delta_{\alpha \beta} + i
\chi_{a,v} \sigma^{a}_{\alpha \beta}). \label{eq:X42} \eeq The sublattice index is
suppressed above. The hermiticity of $\chi$ imposes an important
relation \beq X^{*} = \sigma^y X \sigma^y. \label{eq:real}\eeq
${\mathrm{SU}(2)}$ spin and ${\mathrm{SU}(2)}$ gauge transformations act on $X$ from the
left and right. The covariant derivative with respect to the
dynamical ${\mathrm{SU}(2)}_g$ gauge field $a$ acts on $X$ as  $D^a_{\mu} X =
\d_{\mu} X + i X a_{\mu}$, and $a$ transforms as \beq {\mathrm{SU}(2)}_g:\,\,
X \to X U^{\dagger}_g, \quad a_{\mu} \to U_g a_{\mu} U^{\dagger}_g - i
\d_{\mu} U_g U^{\dagger}_g. \eeq The Lagrangian including the
dynamical gauge field then is \beq \mathcal{L}_{QCD_{3}} = i {\rm
tr}(\bar{X} \gamma^{\mu} D^a_{\mu} X), \label{eq:QCD3}\eeq
with 
$\bar{X} = X^{\dagger} \gamma^0$. We see that (\ref{eq:QCD3}) is
invariant under a global symmetry, \beq \mathrm{Sp}(4):\,\, X \to L X \label{eq:SP4}\eeq
with $L \in \mathrm{Sp}(4)$ - a unitary matrix acting on spin/valley
indices $\alpha, v$ of $X_{\alpha, v; \beta}$. The fact that $L$
lies in $\mathrm{Sp}(4)$, i.e. $L^T \sigma^y L = \sigma^y$, instead of in
the larger  group $U(4)$ comes from the reality condition
(\ref{eq:real}). The lattice symmetries in (\ref{eq:symQCD}) are
elements of this $\mathrm{Sp}(4)$ global symmetry combined with spatial
symmetries of Dirac theory. We note that the global symmetry
$\mathrm{Sp}(4)$ and the gauge group ${\mathrm{SU}(2)}_g$ share a common non-trivial
element: the center $-1$. Thus, the physical global symmetry after
modding out by ${\mathrm{SU}(2)}_g$ is actually $\mathrm{Sp}(4)/Z_2 = {\mathrm{SO}(5)}$ (it is
useful to recall that $\mathrm{Sp}(4) = \mathrm{Spin}(5)$). An order parameter for
this ${\mathrm{SO}(5)}$ symmetry is given by a five component vector: \beq
n^a = {\rm tr} (\bar{X} \Gamma^a  X)  \label{eq:nQCD3}\eeq with $\vec{\Gamma} = \{
\mu^z, -\mu^x, \sigma^x \mu^y, \sigma^y \mu^y, \sigma^z \mu^y\}$.
The first two components, $n^1$, $n^2$ have precisely the
transformation properties of the $x$ and $y$ components of the VBS
order parameter, while the last three components ${\rm tr}
(\bar{X} \sigma^a \mu^y X)$ correspond to the N\'eel order
parameter.

We note in passing that if we want to be less explicit about the
full emergent symmetry of (\ref{eq:QCD3}), we can express the
Lagrangian in terms of two flavors of ${\mathrm{SU}(2)}_g$ charged complex
Dirac fermions, $\psi_{\alpha, v} = i \sigma^y_{\alpha, \beta}
X_{1, v; \beta}$, with $\alpha$ being the color index, and \beq
\mathcal{L} = i \bar{\psi}_v \gamma^{\mu} (\d_{\mu} - i a_{\mu})
\psi_v \label{eq:QCD3psi} \eeq with $\bar{\psi}_v = \psi^{\dagger}_v \gamma^0$. In other words, this theory is $N_f = 2$ $QCD_3$.

There are (at least) three possible scenarios for this theory.  First,  $N_f = 2$ $QCD_3$ could confine, and in the process spontaneously
break ${\mathrm{SO}(5)}$ symmetry by generating a condensate $\langle
n^a \rangle \neq 0$. In the setting of the spin system,  quartic terms in the Lagrangian will then
select either the VBS state or the N\'eel state. This is the boring
scenario.

Second, $N_f = 2$ $QCD_3$ could {in principle} flow to a stable gapless fixed point at which all perturbations (e.g. four-fermi couplings and velocity anisotropies) which preserve lattice and $SO(3)_{s}$ symmetries are
irrelevant. We would then have a completely stable
gapless spin-liquid phase with emergent ${\mathrm{SO}(5)}$ symmetry. (In principle QCD could also flow to a gapped $\mathrm{SO}(5)$--invariant spin liquid; as shown  in  Sec.~\ref{symmetry_enforced_gaplessness_section} this is only possible if time reversal symmetry is broken.)

Third, $N_f = 2$ $QCD_3$  could flow to a  gapless fixed point which is stable in the presence of $\mathrm{SO}(5)$, but which allows a single relevant perturbation when  ${\mathrm{SO}(5)}$ is broken to  the physical symmetry: 
the operator $\mathcal{O}_1$ in Sec.~\ref{PhaseD} (breaking ${\mathrm{SO}(5)}$ to $SO(3)_{s} \times SO(2)_{VBS}$). Then $N_f  =2$ $QCD_3$
tuned to an ${\mathrm{SO}(5)}$ symmetric point describes the deconfined
critical point, and perturbing it by $\mathcal{O}_1$
drives it into either the VBS phase or the N\'eel phase. This is
the scenario relevant for this paper.

\subsection{Higgs descendent: $N_f=4$ compact $QED_3$}

Starting from  $N_f=2$ $QCD_3$, we now Higgs the
gauge group from ${\mathrm{SU}(2)}$ down to $U(1)$. We introduce and condense a scalar
field $\phi$ that transforms as a spin-$1$ vector under ${\mathrm{SU}(2)}_g$
and as a scalar under ${\mathrm{SO}(5)}$ (such a field is allowed in the
theory). After a charge-conjugation redefinition of half of the Dirac fermions, the resulting theory is
\be
\label{n4qedL}
\mathcal{L}=\sum_{i=1}^{4}i\bar{\psi}_i\gamma^{\mu}(\partial_{\mu}-ia_{\mu})\psi_i+(\lambda\mathcal{M}_a+h.c.),
\ee where $a_{\mu}$ is now a $U(1)$ gauge field, and the monopole operator
$\mathcal{M}_a$ represents 
instanton events that change the flux of $a_{\mu}$ by $2\pi$. In general such a term should be expected when the $U(1)$ gauge
field comes from Higgsing of a higher gauge symmetry. In condensed
matter language\footnote{The $QED_3$ theory discussed here is not to be confused with the ``staggered flux state" known in the spin liquid literature \cite{wenbook} (which is also described by  $QED_3$ with four Dirac fermions at low energy) --- lattice symmetries act in distinct ways in these two theories.}
 such theories are called compact $QED_3$. 

The fermion fields $\psi_i$ transform as a spinor representation
under the global ${\mathrm{SO}(5)}$ -- this follows simply from the symmetry
properties of  $QCD_3$. Naively one might expect the Lagrangian
Eq.~\eqref{n4qedL} to have a larger flavor symmetry, say $SU(4)$,
respected by the Dirac term. However, it turns out that the
monopole term  breaks the symmetry down to ${\mathrm{SO}(5)}$. This can be
seen by analyzing the fermion zero-modes \cite{kapustinqed} associated with the monopole
operator $\mathcal{M}_{0a}$: each Dirac
fermion $\psi_i$ contributes a complex fermion zero mode $f_i$ in
the monopole background, and gauge invariance requires two of the
four zero modes to be filled in the ground state, so a
gauge-invariant operator should be represented as
$f^{\dagger}_if^{\dagger}_j\mathcal{M}_{0a}$. There are in total six
of them and it is straightforward to check that they split into
$6=1\oplus 5$ with respect to the ${\mathrm{SO}(5)}$ symmetry. The monopole
operator that appears in the Lagrangian in Eq.~\eqref{n4qedL} is
precisely the ${\mathrm{SO}(5)}$ scalar monopole. It transforms nontrivially
under higher flavor symmetries, and ${\mathrm{SO}(5)}$ is the maximal flavor
group that is compatible with it.

Since this $N_f=4$ compact $QED_3$ is just a Higgs descendent of $N_f=2$ $QCD_3$, they must have the same anomaly structure.
Therefore they share the same set of possible IR
behaviors, including those discussed at the end of
Sec.~\ref{partonqcd3}. Of course the two theories could pick
different choices.

\subsection{Interpretation as surface theory of $3+1$D boson SPT}
\label{sec:QCD3SPT}

Here we show that ${\mathrm{SO}(5)}$ symmetric $N_f = 2$ $QCD_3$ (and hence
$N_f=4$ compact $QED_3$) can be interpreted as a surface theory of
a bosonic ${\mathrm{SO}(5)}$-protected $3+1$D SPT phase. This statement is
independent of the dynamics of $QCD_3$: it remains true even if
the theory spontaneously breaks ${\mathrm{SO}(5)}$. To make this statement,
we have to understand how an ${\mathrm{SO}(5)}$ background gauge field ${\cal A}^5$ enters
in $QCD_3$ and show that this theory has an anomaly, which is
precisely compensated by the $3+1$D SPT bulk. Here we
establish this using a physical argument. In Sec.~\ref{sec:mathy} we
provide a precise formal proof.

Let us first  determine the anomaly of this theory  by thinking about a background gauge field that couples to the ${\mathrm{SO}(5)}$ currents.  An 
${\mathrm{SO}(5)}$ gauge field ${\cal A}^5$ admits  $Z_2$--indexed monopole configurations, since $\pi_1({\mathrm{SO}(5)}) = Z_2$.  In 2+1D these correspond to instanton events and we can ask whether there is anything non-trivial about
them. 

 {We examine a monopole background of the following form. Consider an $SO(2) \times SO(3)$ subgroup of ${\mathrm{SO}(5)}$, with $SO(2)$ acting on the first two components of vector $n^a$ in Eq.~(\ref{eq:nQCD3}) and $SO(3)$ on the last three. Place a unit magnetic monopole in the $SO(2)$ subgroup, i.e. 
\begin{equation}
\label{so5inst}
{\cal A}^5_\mu = {\cal A}^{mon}_\mu (x)  T^1
\end{equation} 
where $T^1$ is the generator of $SO(2)$ and ${\cal A}^{mon}$ is the standard potential  associated with a unit magnetic monopole.   
In the presence of such a background ${\cal A}^5$ field, only a subgroup of the global ${\mathrm{SO}(5)}$ symmetry survives: these are rotations in an $SO(3)$ subgroup (whose generators commute with $T^1$) and the $SO(2)$ rotations generated by $T^1$ itself. }

The following argument provides a hint of the properties of the instanton. Rather than considering the QCD$_3$ theory directly, suppose we add in
an extra field $\hat{n}$, transforming in the vector representation of ${\mathrm{SO}(5)}$,
that couples to fermion bilinears through a Yukawa coupling.  In
the limit that this coupling is strong, we can integrate out the
fermions and standard methods \cite{AbanovWiegmann,tanakahu,tsmpaf06} produce an ${\mathrm{SO}(5)}$ non-linear sigma model in the $\hat{n}$ field with a level-$1$ WZW term (Sec.~\ref{deccprev}). Now the ${\mathrm{SU}(2)}$ gauge field does not couple directly to any matter field, and is expected to confine at low energy, leaving behind the ${\mathrm{SO}(5)}$ WZW model as the remaining nontrivial theory. Indeed, this 
supports the idea that  QCD$_3$ correctly
describes the  N\'eel-VBS intertwinement in square lattice quantum
magnets.   Physically the ${\mathrm{SO}(5)}$ instanton in Eqn. (\ref{so5inst}) has the effect of creating a vortex in two components of the $\hat{n}$ field. We now ask how this vortex transforms under the unbroken symmetry $SO(2)\times SO(3)$. The vortex carries no charge under $SO(2)$, but we know that the vortex transforms as a \textit{spinor} under  $SO(3)$  due to the WZW term. 
We conclude that the  instanton configuration
described above transforms  as an  $SO(3)$ spinor with zero $SO(2)$ charge also.

For a conventional ${\mathrm{SO}(5)}$ sigma model ({\em i.e} without the WZW term) the ${\mathrm{SO}(5)}$ instanton will transform trivially under $SO(2) \times SO(3)$.  The projective transformation of the instanton under the $SO(3)$ subgroup tells us that in the presence of the level-$1$ WZW term the ${\mathrm{SO}(5)}$ symmetry is realized anomalously. It cannot be realized as the on-site symmetry of any strictly 2+1D model.  Clearly the same  instanton structure also characterizes the QCD$_3$ theory. 

This is the physics of the desired anomaly.    
Note also that this instanton operator is bosonic ({\em i.e} in
relativistic parlance it has spin-$0$ under spatial rotations).

It is instructive to rederive the instanton structure of $QCD_3$ directly from the UV Majorana fermion theory. We now briefly indicate how this works out.  {It is important to recognize that the fermions that enter $QCD_3$ transform as a fundamental of $\mathrm{Sp}(4)$, although the physical global symmetry is ${\mathrm{SO}(5)} = {\mathrm{Sp}(4)}/{Z_2}$. We therefore need to lift} the ${\mathrm{SO}(5)}$ gauge field ${\cal A}^5$ to an $\mathrm{Sp}(4)$  gauge field $A^5$, {which enters the theory (\ref{eq:QCD3}) via $D_{\mu} X \to (\d_{\mu} - i A^5_{\mu})X + i X a_{\mu}$.}   For instance, consider 
\begin{equation}
\label{QCDinst}
A^5 = \frac{{\cal A}^{mon}}{2} \mu^y
\end{equation}
where  $\vec \mu$ are Pauli matrices that act on the valley index.
 Naively this may seem to require creating 
$\pi$ flux through the non-zero component of $A^5$ which apparently violates Dirac quantization for the monopole.  However, we should remember that we also have a dynamical  ${\mathrm{SU}(2)}$ gauge field $a$ that the fermions are coupled to: if the lift to $\mathrm{Sp}(4)$ is accompanied  by a $\pi$ flux instanton in one of the three components of the ${\mathrm{SU}(2)}$ gauge field, then we have a sensible configuration that satisfies Dirac quantization.\footnote{A  formal but elegant formulation of this consideration is described in  section \ref{sec:mathy} in terms of matching the second Stiefel-Whitney classes of the bundles corresponding to the background and dynamical gauge fields.}   {For instance, we may give the dynamical gauge field a background value $a = a^3 \sigma^3$, with
\beq a^3 = \frac{{\cal A}^{mon}}{2}.\eeq
It is then convenient to rewrite Eq.~(\ref{eq:QCD3}) in terms of a single color component of $X$, e.g. $X_{\alpha v, \up}$:
\beq {\cal L} = i {\bar X}_{\alpha v, \up} \gamma^{\mu}( (\d_{\mu} + i a^3_{\mu}) \delta_{\alpha \alpha'} \delta_{v v'} -i A^5_{\alpha, v; \alpha', v'}) X_{\alpha' v', \up}.\eeq
Observe that 2 Dirac fermions $X_{\alpha-}$ with $\mu^y = -1$ see a $2\pi$ flux instanton, and another 2 Dirac fermions $X_{\alpha+}$ with $\mu^y = +1$  see no background flux (we drop the color index $\up$ here).
Further, each pair of Dirac fermions transforms as a spin-$1/2$ under the global $SO(3)_{s}$ subgroup left unbroken by the ${\mathrm{SO}(5)}$ monopole. 
In the language of the state-operator correspondence, in the presence of the instanton background the $X_{\alpha -}$ fermions will form two zero modes. Charge neutrality with respect to the color gauge field $a^3$ then implies that we occupy one of these zero modes. Thus, the instanton will transform as an $SO(3)$ spinor in  agreement with the arguments above. It is also easy to see that 
it has zero charge under the unbroken $SO(2)$ subgroup of the global ${\mathrm{SO}(5)}$ symmetry. 
}

Next we want to show that this non-trivial instanton structure is
consistent at the surface of a $3+1$D boson SPT with ${\mathrm{SO}(5)}$
symmetry.  In other words we can  regularize $QCD_3$ with its full ${\mathrm{SO}(5)}$ symmetry as an on-site symmetry at the boundary of a $3+1$D bosonic SPT phase. 

First, let us discuss possible ${\mathrm{SO}(5)}$ symmetric boson SPTs in $3+1$D. Consider any short range entangled gapped phase  of an
${\mathrm{SO}(5)}$ symmetric boson theory, and again couple in background ${\mathrm{SO}(5)}$
gauge fields. The bulk again admits $Z_2$-indexed monopoles in
this gauge field, which  can be chosen to break ${\mathrm{SO}(5)}$ to $SO(2) \times SO(3)$.

Now there are logically two sharply distinct
possibilities: does the monopole transform as a spinor  under the
unbroken $SO(3)$ symmetry or not?  If it is a spinor then the original gapped state is an SPT state. The other
question we may ask is what  the charge is under the unbroken $SO(2)$ symmetry. This charge can be continuously tuned by changing the ${\mathrm{SO}(5)}$ $\theta$-angle:
\beq 
\label{so5theta}
{\cal L} = \frac{\theta}{4 (2 \pi)^2} \mathrm{Tr}_{{\mathrm{SO}(5)}} ({\cal F}^5 \wedge {\cal F}^5)\eeq 
where ${\cal F}^5_{\mu \nu} = \d_{\mu} {\cal A}^5_{\nu} - \d_{\nu} {\cal A}^5_{\mu} - i[ {\cal A}^5_{\mu},  {\cal A}^5_{\nu}]$ is the ${\mathrm{SO}(5)}$ field strength.
Since $\theta$ is a continuous parameter, in the absence of additional symmetries (e.g. time-reversal) it does not label a distinct SPT phase. It is crucial to note that changing $\theta$ does not affect the $SO(3)$ transformation properties of the monopole. Thus, the ${\mathrm{SO}(5)}$ SPT where the $SO(2)$ monopole is an $SO(3)$ spinor is an SPT rather distinct from the more familiar boson and fermion topological insulators protected by $U(1)$ and time-reversal symmetries. Below, we will discuss the topological action for this SPT. Finally, we may ask whether the monopole is a boson or fermion.\footnote{For simplicity, we may define monopole statistics by making just the $SO(2)$ part of the gauge group dynamical.}
This property may be altered by shifting $\theta \to \theta +  2\pi$ \cite{MKF2013, WS2013}.

Now, let us assume the monopole is a boson that carries no $SO(2)$ charge. Then if it does not transform as a spinor under $SO(3)$, the original gapped state is totally trivial. If it is a spinor under $SO(3)$ then the original gapped state is  an SPT state
which has the exact same monopole structure to compensate for the instanton structure of QCD$_3$ with $N_f = 2$  as a potential
boundary state.

\subsection{Explicit constructions for the $\mathrm{SO}(5)$ SPT}

First let us argue that such a 3+1D  SPT state indeed exists in a
system of ${\mathrm{SO}(5)}$ symmetric bosons by a {coupled layer} construction.
Notice that though $N_f = 2$ QCD$_3$ has an anomalous ${\mathrm{SO}(5)}$ symmetry, the anomaly disappears if we
take two copies of it.\footnote{For instance, if $X_{1,2}$ are $X$ fields from the two copies of Eq.~(\ref{eq:QCD3}),  we can add an ${\mathrm{SO}(5)}$ symmetric mass term $m Tr(\bar{X}_1 X_1 - \bar{X}_2 X_2)$. Integrating $X_{1,2}$ out, we get a pure Yang-Mills theory for $a$ with no Chern-Simons term, which is expected to confine.} This is because the ${\mathrm{SO}(5)}$ monopole (instanton) then gets a spin-1/2 (under $SO(3)$) from each copy and hence can always be made
trivial.

We can now construct the required $3+1$D
bosonic SPT state by starting with a stack of 2D layers, each containing 2 copies of QCD$_3$. We take one copy from one layer and trivialise it by
pairing with another copy from the next layer. This will give a
trivial gapped bulk, but at the boundary we are left with single
copy of QCD$_3$.

We can also construct the bulk boson SPT more explicitly using fermionic partons, following a similar approach to Refs~\cite{XuParton, WangNahumSenthil}.  Consider first a $3+1$D fermionic toplogical superconductor with $SO(8) \times Z_2^T$ symmetry. A continuum model for this state consists simply of $8$ relativistic, free, massive Majorana fermions, {
\beq {\cal L} = \sum_{i=1}^8 \bar{\chi}_i (i \gamma^{\mu} \d_{\mu} +m) \chi_i \label{eq:8Maj} \eeq
with $\bar{\chi}_i = \chi^T_i \gamma^0$. For one sign of the Majorana mass $m$ we will have a trivial gapped state, while for the other sign we will have a topological superconductor. The $2+1$D surface of this free-fermion state  correspondingly has $8$ massless Majorana cones. 
Now let us couple this system to a dynamical ${\mathrm{SU}(2)}$ gauge field $a$. As in our $2+1$D discussion, we label the 8 Majoranas $\chi$ by indices $m = 0, 1, 2, 3$; $v = 1, 2$ and form the field $X_{\alpha, v; \beta}$ in Eq.~(\ref{eq:X42}). The ${\mathrm{SU}(2)}$ gauge symmetry acts on $X$ from the right, as in Eq.~(\ref{eq:SU2g}), and the $3+1$D gauge action has the form,
\beq \mathcal{L}_{QCD_{4}} =  {\rm tr}(\bar{X} (i \gamma^{\mu} D^a_{\mu} +m) X).\label{eq:QCD4}\eeq
In the bulk this gauge theory describes a bosonic system with ${\mathrm{SO}(5)} \times Z_2^T$ symmetry. ${\mathrm{SO}(5)}$ is realized projectively on the Majorana fermions, which form an $\mathrm{Sp}(4)$ fundamental (\ref{eq:SP4}). As in $2+1$D  (Eq.~\ref{eq:QCD3psi}), we can rewrite Eq.~(\ref{eq:QCD4}) as two flavors of Dirac fermions with the same mass $m$ coupled to an ${\mathrm{SU}(2)}$ gauge field.

 What state does the theory (\ref{eq:QCD4}) realize? First consider this theory on a closed manifold. Then integrating out the massive fermions produces, at long wavelengths, the standard Yang-Mills action for the dynamical ${\mathrm{SU}(2)}$ gauge field with no topological term. Indeed, each flavor of Dirac fermions with inverted mass would give rise to an ${\mathrm{SU}(2)}$ 
$\theta$-term in the action with $\theta = \pi$,
\beq {\cal L}_\theta = \frac{\theta}{2 (2\pi)^2}  tr_{{\mathrm{SU}(2)}} f \wedge f.\label{eq:thetaSU2}\eeq
Thus, two flavors of Dirac fermions with the same mass give $\theta = 2\pi$, which is equivalent to $\theta = 0$ (see section \ref{sec:mathy} for a more careful discussion).  It is expected that the pure Yang-Mills theory will confine, and so the ground state is seemingly trivial.  Now consider the theory in the presence of a boundary.  Though the bulk is confined, the boundary is precisely  the QCD$_3$ theory of interest to us with global ${\mathrm{SO}(5)} \times Z_2^T$ symmetry.  As promised, the bulk theory therefore describes the  SPT phase of bosons with ${\mathrm{SO}(5)} \times Z_2^T$ symmetry.}  In principle this construction could also be used to write a variational (Gutzwiller-projected) wavefunction for a lattice $\mathrm{SO}(5)$ topological paramagnet.\footnote{{\red Though this parton construction made use of a $Z_2^T$ invariant fermion SPT, the final boson SPT is stable even if $Z_2^T$ is broken.  As we discuss below, with full $SO(5) \times Z_2^T$ symmetry the response to a background $SO(5)$ gauge field has a discrete $\theta$ term, but not the  conventional $\theta$ term of Eqn. \ref{so5theta}. If now $Z_2^T$ is broken but $SO(5)$ is preserved, then such a conventional theta term is allowed. Regardless, the presence of the discrete theta term means that we still have a non-trivial SPT phase. More physically the $SO(5)$ monopole wihich breaks $SO(5)$ to $SO(2) \times SO(3)$ will still transform as a spinor under $SO(3)$ but in the absence of $Z_2^T$ is allowed to have non-zero $SO(2)$ charge.} }

How do we formally describe the bulk response to a background ${\mathrm{SO}(5)}$ gauge field that captures the structure of the $Z_2$-indexed monopole in these systems? In section \ref{sec:mathy} we show that the partition function of the bulk SPT phase takes the form
\begin{equation}
\label{so5bulk}
Z[{\cal A}^5] = |Z[{\cal A}^5]| e^{i\pi  \int w_4[{\cal A}^5]}.
\end{equation}
Here $w_4[{\cal A}^5]  = 0,1$ is a quantity known as the fourth Steifel-Whitney class of the ${\mathrm{SO}(5)}$ gauge bundle ${\cal A}^5$.\footnote{Strictly speaking, ${\cal A}^5$ is the connection on the ${\mathrm{SO}(5)}$ gauge bundle. Here, we abuse the notation a bit by using ${\cal A}^5$ to label both the bundle and the connection.} The phase $e^{i\pi \int w_4[{\cal A}^5]}$ is the analog of the familiar $\theta$-term response of the standard topological insulator to background $U(1)$ gauge fields. In contrast to the usual case here the $\theta$-angle is restricted to two discrete values: $0$ (corresponding to a totally trivial state) or $\pi$ (corresponding to the SPT phase of interest to us here).  Precisely such  a discrete $\theta$ term was introduced a few years ago in Ref.~\cite{ahseta} for non-abelian gauge theories.  In that work the possibility  of such  $\theta$ terms was pointed out and some of their physical consequences were discussed. {We see that  such $\theta$ terms emerge naturally} in the response of bosonic SPT phases. Our discussion above can be viewed as a construction of an ${\mathrm{SO}(5)}$ symmetric bosonic $3+1$D model whose response  includes these discrete $\theta$ terms. Indeed, in section \ref{sec:mathy}, we show explicitly that the theory (\ref{eq:QCD4}) has this discrete $\theta$-term in its response to a background ${\mathrm{SO}(5)}$ gauge field.

\subsection{Symmetry-enforced gaplessness}
\label{symmetry_enforced_gaplessness_section}

As mentioned above the IR fate of $N_f = 2$ QCD$_3$ is at present unclear. However our understanding of the anomalous symmetry realization in this theory enables us to derive some general restrictions. We will show that either the ${\mathrm{SO}(5)} \times Z_2^T$ symmetry is spontaneously broken or the theory is gapless in the IR. This result follows purely from the anomalous symmetry realization. Indeed, it is a general feature of the surface of the bulk $3+1$D boson SPT with ${\mathrm{SO}(5)} \times Z_2^T$ symmetry discussed in the previous subsection.  Such a  phenomenon was first described for some fermion SPTs in Ref.~\cite{3dfspt2} and dubbed `symmetry enforced gaplesness'. Other examples, including
some boson SPTs, are described in Refs. \cite{wangtscflll16,skws17}.

Consider a putative gapped state of the $2+1$D theory that
preserves the ${\mathrm{SO}(5)} \times Z_2^T$ symmetry. All the
quasiparticles  of this state must transform under some
representation, possibly projective.  As usual if a quasiparticle
transforms non-projectively under ${\mathrm{SO}(5)}$ we can `screen' it
using composites of the $\hat{n}$-vector to make it a singlet
under ${\mathrm{SO}(5)}$. Therefore the only non-trivial symmetry possibility
is a quasiparticle  transforming under the 4-dimensional spinor
representation (fundamental of $\mathrm{Sp}(4)$). Let us call such
quasiparticles $X_I$.

{We will think of QCD$_3$ as living on the surface of the $3+1$D SPT described
above.} Now let us tunnel in a $Z_2$-valued ${\mathrm{SO}(5)}$ monopole through the
surface. We know that the monopole breaks ${\mathrm{SO}(5)}$ to $SO(2) \times SO(3)$,
and that it transforms in the $(0, 1/2)$
representation of $SO(2) \times SO(3)$. Therefore, {in order for $SO(2) \times SO(3)$ charge to be conserved,} there must be a
quasiparticle in this putative gapped surface state with these
properties. However the only quasiparticles which transform
non-trivially under ${\mathrm{SO}(5)}$ are the ``spinors" $X_I$. They
transform with $SO(2)$ charge of $\frac{1}{2}$ and as a spinor
under the $SO(3)$, and not under the $(0, 1/2)$ representation.
It follows that the gapped state we imagined cannot have the right
anomaly, and hence is not a possible surface state.

Note, however, that if time reversal is broken then there can be a
Hall conductivity for the ${\mathrm{SO}(5)}$ currents. Then the ${\mathrm{SO}(5)}$
monopole threading will nucleate an $SO(2)$ charge determined by
the Hall conductivity. This can then combine with  the ${\mathrm{SO}(5)}$
spinor $X_I$ to produce an object with $(0,1/2)$ quantum numbers
under the $SO(2) \times SO(3)$ symmetry, as required. Thus, if time
reversal is broken, a gapped ${\mathrm{SO}(5)}$ symmetry preserving state is
no longer prohibited. Indeed it is easy to construct a
`chiral spin liquid' state explicitly: see section \ref{sec:chiralsl}. 

The conclusion therefore is that, in the IR, $N_f = 2$ QCD$_3$
with ${\mathrm{SO}(5)} \times Z_2^T$ symmetry must either spontaneously break the
symmetry or be gapless, {\em i.e} flow to a CFT.  It cannot be
fully gapped while preserving symmetries even if we allow for
non-trivial topological order.

As another application of this result, consider the fate
of the Neel-VBS transition at the longest distances.  One
interpretation of  the existing numerics is to say that the RG
flows are  attracted to a ray with ${\mathrm{SO}(5)} \times Z_2^T$  symmetry.
If so then the eventual {destination of this ray} is either a weak first
order transition, a $Z_2^T$ broken spin liquid, or  a gapless
CFT --- a gapped, symmetry-preserving, topologically ordered
state is ruled out.\footnote{Of course if the ray ends at a gapless CFT where allowed $\mathrm{SO}(5)$-breaking terms are relevant, these terms will eventually become large and the above constraints will not apply.}

\section{QCD$_3$ as the surface of an $SO(5)$ invariant $3+1$D SPT: Formal description}
\label{sec:mathy}

Here we expand on the discussion in section \ref{sec:QCD3SPT} and demonstrate more formally that $SO(5)$ symmetric $N_f = 2$ QCD$_3$, Eq.~\ref{eq:QCD3}, can be interpreted as a surface theory of a bosonic $SO(5)$-protected $3+1$D SPT phase.  {We will first develop a precise formal description of the  anomaly of  QCD$_3$ and show how it is  compensated by the $3+1$D SPT bulk. We will also sharpen  the parton construction of this SPT outlined in Sec. \ref{sec:QCD3SPT} to derive the bulk partition function in the presence of a background $SO(5)$ gauge field. We will explicitly derive the advertised discrete theta term.} {The reader is referred to Appendix \ref{mathapp} for a brief summary of math concepts that will be useful in this section and to Appendix \ref{app:APS} for a review of the field-theoretic description of topological superconductors.}

\subsection{Gauging $SO(3)_{s}$}

First, it is useful to recall how the gauge field associated with the $SO(3)_s$ spin-rotation symmetry enters in QCD$_3$. Since $SO(3)_{s}$ symmetry is a true on-site symmetry of the lattice spin system, its implementation in QCD$_3$ must be non-anomalous. Given an $SO(3)_s$ gauge bundle $E_s$ over our space-time 3 manifold $M$,\footnote{As usual, we transition to Euclidean space-time here. Furthermore, we will 
{\blue formulate the theory on an arbitrary space-time manifold $M$. The ability to do so is a useful formal consistency check on the theory. We}
assume that the manifold $M$ is oriented: i.e. unless otherwise noted, we will not discuss ``gauging" of time-reversal symmetry.} we attempt to lift it to an $SU(2)_s$ gauge bundle $\hat{E}_s$. The resulting $SU(2)_s$ transition functions $\hat{U}^s_{ij}$ might fail to satisfy the cocycle condition: $\hat{U}^s_{ik} \hat{U}^s_{kj} \hat{U}^s_{ji} = (-1)^{w_2(E_s)_{kji}}$, where $w_2(E_s) \in H^2(M, Z_2)$ is a representative of the second Stiefel-Whitney class of $E_s$ \cite{Nakahara}. In this case, we take the transition functions of our dynamical gauge field $SU(2)_g$ to violate the cocyle condition by precisely the same factor in the center of $SU(2)_g$. That is, if we project the $SU(2)_g$ transition functions $\hat{U}^g$ to $SO(3)_g$, we get an $SO(3)$ bundle $E_g$ with $w_2(E_g) = w_2(E_s)$. The fermions $\chi_{m,v}$ then see an $SO(4) = (SU(2)_s \times SU(2)_g)/Z_2$ gauge field which satisfies the cocycle condition. In the background of this $SO(4)$ gauge field $A^{s,g}$, the action (\ref{eq:QCD3}) becomes,
\beq L = i \bar{\chi}_v\gamma^{\mu} (\d_{\mu} - i A^{s,g}) \chi_v  \label{eq:QCD3M} \eeq
where $A^{s,g}$ lives in $so(4)$ Lie algebra and acts on the spin-color index $m$ of $\chi_{m,v}$ but not on the valley index $v$. We note that in addition to the $SO(4)$ gauge field, the fermions $\chi_{m,v}$ also see a spin structure (we have suppressed the spin-connection above) with transition functions $h_{ij} \in Spin(3)$.  Every three-manifold is spin, so $h_{ij}$ can always be chosen to satisfy the cocycle condition (as we assumed in the discussion above). However, it will be more useful to think of $h_{ij}$ as arbitrary lifts to $Spin(3)$ of the $SO(3)$ transition functions in the tangent bundle. The fermions only see a combination of $SU(2)_s$, $SU(2)_g$ and $Spin(3)_{TM}$ transition functions, so we only require this combination to satisfy the cocycle condition.\footnote{The subscript $TM$ on $Spin(3)$ reminds us that these are transition functions associated with the tangent bundle.} A change $h_{ij}\to \zeta_{ij} h_{ij}$, $\zeta_{ij} = \pm 1$, can be compensated by modifying the $SU(2)_g$ bundle, $\hat{U}^g_{ij} \to \zeta_{ij} \hat{U}^g_{ij}$, so the theory does not depend on the spin structure. Further, $w_2(E_s) + w_2(E_g) + w_2(TM) = 0 \,\,(mod\,\, 2)$, where we are now thinking of $w_2$'s as concrete cochains representing $H^2(M, Z_2)$ (as already noted, as a cohomology class, $w_2(TM) = 0$).

We now discuss the regularization of Eq.~(\ref{eq:QCD3M}) - in principle, such regularization is provided by the lattice we started with. An equivalent continuum regularization is obtained by using Pauli-Villars (PV) regulators with opposite mass for the Majoranas in the two valleys $v = 1,2$. Recall that for a single $SO(n)$ vector Majorana fermion coupled to an $SO(n)$ gauge field $A$, the PV regulated partition function is given by,\cite{wittenreview}
\beq Z_{\chi;PV, \pm}(A) = |Z_{\chi}(A)| \exp\left(\mp \pi i \eta(iD_A)/4\right) \label{eq:Zeta}\eeq
where the sign in the exponent is determined by the sign of Pauli-Villars mass. Here, $\eta(i D_A)$ is the $\eta$-invariant of the Dirac operator $i D_A = i \gamma^{\mu} (\d_{\mu} + i \omega_{\mu} - i A_{\mu})$,\footnote{Here, $\omega_{\mu}$ is the spin connection, which we included for further generality to describe curved manifolds \cite{GilkeyReview}.}
\bea \eta &=& \eta(0) + N_0 \nn
\eta(s) &=& \sum_{\lambda \neq 0} {\rm sgn}(\lambda) |\lambda|^{-s} \label{eq:etadef}\eea
with $\lambda$ in the above sum - eigenvalues of $i D_A$, and $N_0$  - the number of zero modes of $i D_A$. So, when we use PV regulators of opposite mass for the two valleys $v = 1, 2$, we obtain after integrating the Majorana fermions out
\beq \exp\left(-S_{QCD_{3}}[A^{s,g}]\right) = |Z_{\chi_{v=1}}(A^{s,g})|^2 \label{eq:ZQCD}\eeq
{with  $Z_{\chi_{v=1}}(A^{s,g})$ --- the partition function of Majoranas in just a single valley $v = 1$.} The theory thus defined obviously preserves $SO(3)_s$ as a non-anomalous symmetry. 
Likewise, time-reversal symmetry (last line of (\ref{eq:symQCD})) is preserved and non-anomalous --- this must be the case, as it is an onsite symmetry of the initial lattice model.\footnote{Strictly speaking, to show that $T$ is non-anomalous we have to regulate the theory on a non-orientable manifold. However, this can be done as the mass term $\chi^T \tau^y \mu^z \chi$ preserves $T$ - we can introduce a PV regulator with this mass term.} The discrete lattice symmetries in (\ref{eq:symQCD}) are global symmetries of (\ref{eq:ZQCD}), however, they are (in a certain sense) anomalous: there is no contradiction here, since they are not realized by the original lattice model in an on-site manner.

For future reference, we note that we can obtain an equivalent theory (\ref{eq:ZQCD}) by regulating both valleys in the same way (with the same sign of PV mass) and supplementing the action by a Chern-Simons  term for $A^{s,g}$:
\begin{widetext}
\beq S_{QCD_3} = \int_M [\bar{\chi}_v \gamma^{\mu} (\d_{\mu} + i \omega_{\mu} - i A^{s,g}) \chi_v]_{PV,+} - i CS_{SO(4)}[A^{s,g}, Y_4] - 4 i CS_g[Y_4] \label{eq:QCDsg}\eeq
\end{widetext}
where the subscript $PV_+$ indicates that the PV mass is the same for both valleys, see Eq.~(\ref{eq:Zeta}). 
We will use the notation where $CS_{SO(n)}[A, Y_4]$ is the Chern-Simons action for $SO(n)$ gauge field $A$ at level $1$,\footnote{{Level $1$ is defined so that for an $SO(2)$ subgroup acting on the first two components of an $n$-vector, $\sigma_{xy} = 1$. In other words, level $1$ corresponds to the $SO(n)$ response of $n$ identical copies of a $p_x + i p_y$ superconductor.}} and $CS_g[Y_4]$ is the gravitational Chern-Simons action (corresponding to the gravitational response of a $p_x+ip_y$ superconductor). The significance of the parameter $Y_4$ is as follows. We recall that a technical trick to define a Chern-Simons term is to extend the three-manifold $M$ to a four-manifold $Y_4$, and also extend the gauge field $A$ from $M$ to $Y_4$:
\bea CS_{SO(n)}[A, Y_4] &=&  \frac{\pi }{2\cdot (2
\pi)^2}\int_{Y_4}  {\rm tr}_{SO(n)} F \wedge F \nn
 CS_g[Y_4] &=& \frac{\pi}{8 \cdot
24\pi^2} \int_{Y_4} {\rm tr} R\wedge R \label{eq:CSdef}\eea
 where $F = d A - i A \wedge A$ is the $SO(n)$ field strength, $R$ is the Riemann curvature tensor, and the trace ${\rm tr}_{SO(n)}$ is in the $n$-dimensional vector representation.
In order for a theory to be a well-defined strictly 2+1D theory, it must be independent of the choice of the four-manifold $Y_4$ and the particular extension of the gauge field to $Y_4$. For our theory (\ref{eq:QCDsg}) this is actually guaranteed by the Atiyah-Patodi-Singer (APS) theorem. 
Indeed, recall that by APS theorem, if our three manifold $M$ is endowed with an $(SO(n)_A \times Spin(3)_{TM})/Z_2$ bundle $E$ (where $n$ is even) and $M$ is the boundary of a four-manifold $Y_4$ such that $E$ extends to a $(SO(n)_A \times Spin(4)_{TY_4})/Z_2$ bundle over $Y_4$ then \cite{GilkeyReview}
\be \frac{\pi}{2} \eta(i D^{SO(n)}_{A,M}) = CS_{SO(n)}[A, Y_4] + n CS_g[Y_4] - 2 \pi {\cal J}[A,Y_4] \label{eq:APS}\ee
where $2 {\cal J}[A, Y_4]$ is the index of the Dirac operator $i D_A$ on $Y_4$ with APS boundary conditions.\footnote{For an $SO(n)$ gauge-field the index is always even as $[CK, i D_A] = 0$, and $(CK)^2 = -1$, with $C$ - the charge conjugation operator and $K$ - complex conjugation.}  Since the left-hand-side of (\ref{eq:APS}) only depends on the boundary data,  the sum $CS_{SO(n)}[A, Y_4] + n CS_g[Y_4]$ is independent of the extension chosen modulo $2\pi$.\footnote{In the case when $M$ is supplemented with a spin-structure, which is preserved by the extension, $CS_{SO(n)}[A, Y_4]$ and $CS_g[Y_4]$ are {\it separately} independent of the extension.} This means that  (\ref{eq:QCDsg}) is a well-defined strictly 2+1D theory. Furthermore, integrating the fermions in (\ref{eq:QCDsg}) out using Eq.~(\ref{eq:Zeta}) and applying (\ref{eq:APS}), we recover the original regularization (\ref{eq:ZQCD}).

 Given the definition (\ref{eq:CSdef})  of  $SO(4)$ Chern-Simons term via the extension to $Y_4$, we can re-write it in terms of field strength of $SU(2)_s$ and $SU(2)_g$ gauge fields $A^s$ and $a^g$, or alternatively their $SO(3)$ representations,\footnote{In this section, we will be slightly sloppy and often use $A$ to label both the bundle and the connection. Moreover, we will  use $A^s$ to label both the $SO(3)_s$ bundle and the $SU(2)_s$ lift. The meaning will be clear from the context. Similarly, $A^5$ will denote both the $SO(5)$ bundle and the $Sp(4)$ lift.}   obtaining,
 \begin{widetext}
\bea CS_{SO(4)}[A^{s,g}, Y_4] + 4 CS_g[Y_4]  &=& CS_{SU(2)}[A^s, Y_4] + CS_{SU(2)}[a^g, Y_4] + 4 CS_g[Y_4] \label{CSU2} \\
&=&\frac{1}{2}CS_{SO(3)}[A^s, Y_4] + \frac{1}{2} CS_{SO(3)}[a^g,
Y_4] + 4 CS_g[Y_4] \label{eq:SO3sg} \eea \end{widetext}where as usual for an $SU(2)$ gauge field $A$, \beq
CS_{SU(2)}[A, Y_4] = \frac{1}{4 \pi} \int_{Y_4} {\rm tr}_{SU(2)} F
\wedge F \eeq and the trace is in the spin-$1/2$ representation.
The half-integer-level $SO(3)$ Chern-Simons terms in 
(\ref{eq:SO3sg}) are not independent of $Y_4$ individually, but
the full sum is independent of $Y_4$ (the integer-level $SU(2)$
terms in (\ref{CSU2}) are also not individually independent of $Y_4$,
since transition functions for $SU(2)_s$ and $SU(2)_g$ don't
independently satisfy the cocycle condition). It is instructive to
check this statement without appealing to the APS theorem. To show
that (\ref{eq:SO3sg}) is independent of $Y_4$, it
suffices to check that it vanishes modulo $2\pi$ when $Y_4$ has no boundary. Recalling that for an $SO(n)$ gauge bundle on a closed
manifold $Y_4$ the first Pontryagin number is given by \cite{Nakahara}
 \beq p_1 =
\frac{1}{2\cdot (2 \pi)^2} \int_{Y_4} {\rm tr}_{SO(n)} F\wedge
F \label{eq:Pontryagin} \eeq and the signature of the manifold is \cite{Nakahara}
\beq \sigma = -\frac{1}{24\pi^2} \int_{Y_4}  {\rm tr} R\wedge R \label{eq:sigma} \eeq
we must show that
\beq p_1^{SO(3)}[A^s, Y_4] + p_1^{SO(3)}[a^g, Y_4] - \sigma[Y_4] = 0 \quad {\rm}(mod\,\,4) \label{eq:ppSO3}\eeq
for any closed $Y_4$. Now, for an $SO(n)$ gauge bundle,\cite{ahseta}
\beq p_1 = {\cal P}(w_2) + 2 w_4 \quad  {\rm}(mod\,\,4) \label{eq:Pont}\eeq
where $w_i$ are the Stiefel-Whitney classes of the bundle and ${\cal P}: H^2(Z_2) \to H^4(Z_4)$ is the Pontryagin square operation, which satisfies ${\cal P}(a + b) = {\cal P}(a) + {\cal P}(b) + 2 a \cup b\,\,(mod\,\,4)$ (see appendix \ref{mathapp} for details). Recalling that $w_2[A^s] + w_2[a^g] + w_2(TY_4) = 0$, and that for $SO(3)$ bundles $w_4  =0$, Eq. (\ref{eq:ppSO3}) reduces to
\begin{widetext}
\beq p_1[A^s, Y_4] + p_1[a^g, Y_4] - \sigma[Y_4] = \int_{Y_4} \left( 2  {\cal P}(w_2[A^s]) + 2 w_2[A^s] \cup w_2[TY_4] + {\cal P} (w_2[TY_4])\right)- \sigma[Y_4] \label{eq:Pw}\eeq
\end{widetext}
where all manipulations are modulo $4$. On an orientable four-manifold $Y_4$, for any $a \in H^2(Z_2)$, $a \cup a = a \cup w_2[TY_4]$ (see e.g. Ref.~\cite{MilnoreStasheff}, p.~132); furthermore, ${\cal P}(a) = a \cup a \, (mod \,\,2)$, so the first two terms on the RHS of Eq.~(\ref{eq:Pw}) add to $0$ mod $4$. The remaining statement $\int_{Y_4}{\cal P}(w_2[TY_4]) = \sigma\,\,(mod \,\,4)$ is also true.\footnote{{\blue There is a theorem that on }any closed four-manifold, $w_2(TY_4)$ can be lifted to an integer cohomology class $\hat{w}_2 \in H^2(Z)$. (This is the statement that every four manifold admits a Spin$_c$ structure.) For an integer cohomology class $a$, ${\cal P}(a) = a\cup a$. So we need to show $\int_{Y_4} \hat{w}_2 \cup \hat{w}_2 = \sigma\, (mod\,\,4)$. Letting the unimodular matrix $Q$ be the intersection pairing on ${\rm Free}(H^2(Y_4, Z))$, we must show $(\hat{w}_2)^T Q \hat{w}_2 = \sigma(Q) \,(mod\,\,4)$, where $\sigma(Q)$ is the signature of the matrix $Q$. Actually, a stronger statement  $(\hat{w}_2)^T Q \hat{w}_2 = \sigma(Q) \,(mod\,\,8)$ holds and can be derived purely from algebra (see Ref.~\cite{MilnorHusemoller}, p. 24). Actually, this statement is familiar to users of the APS theorem.  Consider a general $Spin_c$ connection with field strength $F$ on a  four manifold.  The Atiyah-Singer theorem \cite{Nakahara,GilkeyReview}  then tells us that $ \frac{1}{2} \int \frac{F}{2\pi} \wedge \frac{F}{2\pi} - \frac{\sigma}{8} \in Z$.
Now a $Spin_c$ connection has
$\int \frac{F}{2\pi} = \int \frac{\hat{w}_2}{2} ~(mod ~Z)$ for any oriented two-cycle. Choosing a special case where $\frac{F}{2\pi} = \frac{\hat{w}_2}{2}$, we obtain the needed result.}

\subsection{Gauging $SO(5)$}
\label{sec:SO5g}

We are now ready to discuss gauging of the full $SO(5)$ global
symmetry of $QCD_3$. Given an $SO(5)$ gauge bundle with connection
$A^5$ on our three-manifold $M$, we attempt to lift it to $Sp(4) =
Spin(5)$. The resulting transition functions may not satisfy the
cocycle condition: the defect is $w_2[A^5]$. As before, we choose
$SU(2)_g$ transition functions so that the combination of
$Sp(4)$, $SU(2)_g$ and $Spin(3)_{TM}$ transition functions
satisfies the cocycle condition, i.e. $w_2[A^5] + w_2[a^g] +
w_2[TM] = 0\,\, (mod\,\,2)$, with $w_2[a^g]$ being the second
Stiefel-Whitney class of the $SO(3)_g$ gauge bundle. Thinking of
$(Sp(4) \times SU(2)_g)/Z_2$ as a subgroup of $SO(8)$, the
(unregulated) action becomes \beq L_{QCD_3} = \bar{\chi}
\gamma^{\mu} (\d_{\mu} + i \omega_{\mu} - i A^{5,g}_{\mu}) \chi
\eeq with $A^{5,g}$ living in the $so(8)$ Lie algebra, i.e. acting
on spin-color $m$ and valley indices $v$ of $\chi$. We must now
specify how to regulate the above action. We can no longer use
PV regulators of opposite mass for the two valleys since this will break
$SO(5)$  symmetry. Instead, we use a common PV regulator for the
$SO(8)$ vector $\chi$ and supplement the action by an $SO(8)$ Chern-Simons
term, \bea S_{QCD_3} =&& \int_M (\bar{\chi} \gamma^{\mu} (\d_{\mu} +
i \omega_{\mu} - i A^{5,g}_{\mu}) \chi)_{PV,+} \nn && - \frac{i}{2}
CS_{SO(8)}[A^{5,g}, Y_4] - 4 i CS_g[Y_4] \label{eq:QCD5} \eea with
the Chern-Simons terms again defined by extending to a four-manifold $Y_4$,
as before. When the $SO(5)$ bundle reduces to an $SO(3)_s$ bundle,
Eq.~(\ref{eq:QCD5}) reduces to Eq.~(\ref{eq:QCDsg}) as needed. But
are the Chern-Simons terms in Eq.~(\ref{eq:QCD5}) independent of $Y_4$ for an
arbitrary $SO(5)$ bundle? We will show that the answer is no:
Eq.~(\ref{eq:QCD5}) is not well-defined as a purely $2+1$D theory.
However, we will be able to define it as a surface of an $SO(5)$
protected $3+1$D bosonic SPT phase.

First, we observe that if $A^{5,g}$ was an arbitary $SO(8)$ bundle, (\ref{eq:QCD5}) obviously would not define a purely $2+1$D theory, as the $SO(8)$ level is fractional. {In fact, physically Eq.~(\ref{eq:QCD5}) is just the action of eight identical copies of a $3+1$D topological superconductor living on the space $Y_4$ and coupled to an $SO(8)$ gauge field. The bulk of such a state has a non-trivial $SO(8)$ response.} Indeed, by the APS theorem (\ref{eq:APS}), the partition function of (\ref{eq:QCD5}) after integrating the fermions out becomes,
\beq \exp\left(-S_{QCD_{3}}[A^{5,g}]\right)= |Z_{\chi}[A^{5,g}]| (-1)^{{\cal J}[A^{5,g}, Y_4]} \eeq
with $|Z_{\chi}[A^{5,g}]|$ --- the absolute value of the partition function for our eight $2+1$D Majorana fermions coupled to $A^{5,g}$, and $2 {\cal J}$ --- the index of $i D_{A^{5,g}}$ on  $Y_4$. For closed $Y_4$ and a general $SO(8)$ gauge field, ${\cal J}$ is not necessarily even, so $S_{QCD_{3}}[A^{5,g}]$ depends on the extension to $Y_4$. However, our $A^{5,g}$ is not the most general $SO(8)$ gauge field, rather we are dealing with an $(Sp(4) \times SU(2)_g \times Spin(3)_{TM})/{(Z_2 \times Z_2)}$ bundle: when this bundle is extended to the four-manifold $Y_4$, is Eq.~(\ref{eq:QCD5}) independent of the extension?\footnote{There is also one caveat here: does an extension always exist? For this, one must calculate the 3D bordism group --- a task that we will not attempt here.} We rewrite,
\beq \frac{1}{2} CS_{SO(8)}[A^{5,g}, Y_4] = \frac{1}{2} CS_{SO(5)}[A^5, Y_4] + \frac{1}{2} CS_{SO(3)}[a^g, Y_4] \eeq
Thus, to check whether (\ref{eq:QCD5}) is well-defined as a 2+1D theory, we must determine whether for closed $Y_4$,
\beq p_1^{SO(5)}[A^5, Y_4] + p_1^{SO(3)}[a^g, Y_4] - \sigma[Y_4] \stackrel{?}{=} 0\,\, (mod\,\,4)\eeq
We again use the identity (\ref{eq:Pont}), and $w_2[A^5] + w_2[a^g] + w_2[TM] = 0$. Repeating the manipulations below (\ref{eq:ppSO3}), we obtain
\beq p_1[A^5, Y_4] + p_1[a^g, Y_4] - \sigma[Y_4] = 2 w_4[A^5, Y_4] \neq 0 \,\, (mod\,\,4) \label{eq:pw4}\eeq
Thus, Eq.~(\ref{eq:QCD5}) generally depends on the extension to $Y_4$: given two extensions to $Y_4$ and $\tilde{Y}_4$, we have,
\begin{widetext}
\beq \exp\left(-S_{QCD_{3}}[A^{5,g}, \tilde{Y}_4] + S_{QCD_{3}}[A^{5,g}, Y_4]) \right) = \exp\left(\pi i \int_{\tilde{Y}_4 \cup \bar{Y}_4} w_4[A^5] \right) \label{eq:YY} \eeq
\end{widetext}
where the last integral is over the manifold obtained by gluing together $\tilde{Y}_4$ and $Y_4$ with reversed orientation. Crucially, the variation (\ref{eq:YY}) only depends on the extension of the $SO(5)$ bundle $A^5$, but not on the $SU(2)_g$ bundle. Still, we cannot promote $SO(5)$ to an on-site symmetry of a strictly $2+1$D theory. However, we can think of the theory (\ref{eq:QCD5}) as the surface of an $SO(5)$ protected $3+1$D SPT, as follows. Let $X_4$ be the physical $3+1$D manifold that our SPT phase lives on. There is an $SO(5)$ gauge field $A^5$ on $X_4$. When $X_4$ has no boundary, we let the partition function be:
\beq Z_{3+1}[A^5, X_4] = \exp\left(\pi i \int_{X_4} w_4[A^5]\right) \label{eq:ZbulkSO5}\eeq
When $X_4$ has a boundary, $\d X_4 = M$, Eq.~(\ref{eq:ZbulkSO5}) is not well-defined (not gauge-invariant under using different representatives of the cohomology class $w_4[A^5] \in H^4(X_4, Z_2)$). However, we can combine the surface action (\ref{eq:QCD5}) with the bulk action (\ref{eq:ZbulkSO5}), to obtain
\bea S^{{\rm bulk} + {\rm bound}}_{QCD_3} &=& \int_M [\bar{\chi}_v \gamma^{\mu} (\d_{\mu} + i \omega_{\mu} - i A^{5,g}) \chi_v]_{PV,+} \nn
&-& \frac{i}{2} CS_{SO(5)}[A^5, Y_4] - \frac{i}{2} CS_{SO(3)}[a^g, Y_4]  \nn
&-& 4 i CS_g[Y_4]+ \pi i \int_{X_4 \cup \bar{Y}_4} w_4[A^5] \label{eq:QCDbb} \eea
Again, the Chern-Simons terms in the second line of (\ref{eq:QCDbb}) are defined with the help of an extension to an auxiliary four-dimensional manifold $Y_4$. The term in the last line of Eq.~(\ref{eq:QCDbb}) involves an integral over a manifold obtained by gluing $X_4$ and $Y_4$ with its orientation reversed. Every line in Eq.~(\ref{eq:QCDbb}) is well-defined, however, the second and third lines individually depend on the extension to $Y_4$. However, the $Y_4$ dependence of the second and third lines cancels, due to Eq.~(\ref{eq:YY}). Therefore, Eq.~(\ref{eq:QCDbb}) is overall independent of $Y_4$. It does, however, depend on $A^5$ on the physical $3+1$D manifold $X_4$ and reduces to Eq.~(\ref{eq:ZbulkSO5}) when $X_4$ has no boundary. {Therefore, Eq.~(\ref{eq:QCDbb}) is a perfectly well defined action of an $SO(5)$ protected SPT phase on the $3+1$D manifold $X_4$ with boundary $M$: only the background $SO(5)$ gauge-field $A^5$ lives in the bulk; the fields $a^g$, $\chi$ live on the surface. The independence of the variation (\ref{eq:YY}) of $a^g$ was crucial for such an SPT interpretation to be possible.}

Let us obtain some intuition for the bulk topological term
(\ref{eq:ZbulkSO5}). First, consider gauging just the $SO(3)_{s}
\times SO(2)_{VBS}$ subgroup of $SO(5)$, i.e. the $SO(5)$ bundle is a direct sum of
$SO(3)_s$ and $SO(2)_{VBS}$ bundles. Label the corresponding gauge
bundles as $A^s$ and $A^{VBS}$. We have
\bea &&2 w^{SO(5)}_4[A^5] = p_1[A^5] - {\cal P}(w_2[A^5]) \nn &&= p^{SO(3)}_1[A^s] + p^{SO(2)}_1[A^{VBS}] - {\cal P}(w_2[A^s] +w_2[A^{VBS}]) \nn
&&= - 2 w_2[A^s] \cup w_2[A^{VBS}] \quad (mod\,\,4) 
\eea
where we've repeatedly used Eq.~(\ref{eq:Pont}), together with the fact that $w_2$ and $p_1$ are additive under $SO(n)$ bundle addition, and that $w_4 = 0$ for $SO(n)$ with $n \leq 3$. So,
\beq w_4[A^5] = w_2[A^s] \cup w_2[A^{VBS}]  = w_2[A^s] \cup \frac{F^{VBS}}{2\pi}\label{eq:w23}\eeq
where we've used the fact that for an $SO(2)$ gauge field, $w_2$ and the first Chern class $\frac{F}{2\pi}$ coincide mod $2$: $w_2 = \frac{F}{2\pi} \,\, (mod \,\,2)$.\footnote{Below, we will be somewhat cavalier using $\frac{F}{2\pi}$ to denote both the Chern class (in $H^2(Z)$) and the field-strength $F = dA$ of a $U(1)$ bundle.}  What is the physical interpretation of Eq.~(\ref{eq:w23})? Imagine we take our bulk manifold to be $S^2 \times \Sigma$, with $\Sigma$ an arbitrary two-dimensional surface. Place flux $2\pi$ of $F^{VBS}$ through $S^2$. Then the partition function (\ref{eq:ZbulkSO5}) is $Z = \exp(\pi i \int_{\Sigma} w_2[A^s])$. This is precisely the partition function 
 of $SO(3)_s$ protected $1+1$D Haldane phase on $\Sigma$.\footnote{One way to obtain this result is the following. Recall that the Haldane chain
is described by the $O(3)$  non-linear sigma model at  $\theta = 2\pi$.   In a $CP^1$ description the Lagrangian is \cite{WittenCP1}
${\cal L} = |(\partial_\mu - i a_\mu) z|^2 +  i \theta f/2\pi$,
where $f$ is the field strength associated with $a_\mu = -i z^\dagger \partial_\mu z$.
A background $SO(3)_s$ gauge field translates into a background $SU(2)_s$ gauge field for the $z$-field.  If, however, the $SO(3)_s$ gauge bundle has $w_2 \neq 0$, we cannot then
naively lift  it to $SU(2)_s$. However, $z$ is also coupled to the $U(1)$ gauge field $a$.  Any defect in the cocycle condition of the $SO(3)_s$ bundle can be compensated by a $\pi$ flux of the $U(1)$ gauge field. Thus, a configuration of the $SO(3)_s$ gauge field with a non-zero $w_2$ necessarily induces a background $a_\mu$ configuration where $\int_{\Sigma} \frac{f}{2\pi} = \int_{\Sigma} \frac{w_2[A^s]}{2}  ~(mod ~Z)$.
  For the Haldane chain, from the $CP^1$ Lagrangian with the $\theta$-term at {\blue $\theta = 2\pi$ we immediately see that  the partition function has an extra phase
$e^{i\pi \int_{\Sigma} w_2[A^s]}$}. Clearly on a closed manifold this is invariant under `gauge  transformations', $w_2 \rightarrow w_2 + dn$. 
However, in the presence of a boundary this is  no longer true. Gauge invariance can be mended by including a Wilson line $W[A^s]$ in the spin-$1/2$ representation along the boundary, {\em i.e}, the partition function now becomes $W[A^s] (-1)^{\int_{\Sigma} w_2}$. This boundary Wilson line is precisely the well known dangling spin-$1/2$ moment at the boundary of the Haldane chain.  } We know that the Haldane phase on a spatial interval $I = [0, 1]$ has dangling spin $1/2$s at the boundary. So, we can guess that if we consider the theory on the spatial manifold $S^2 \times I$, then there are dangling spin $1/2$s at the two ends of $I$ - i.e. at the locations of $SO(2)_{VBS}$ monopoles. We, therefore, conclude that the topological term (\ref{eq:w23}) makes monopoles of $SO(2)_{VBS}$ transform in the spin-$1/2$ representation of $SO(3)_s$. {This is precisely the conclusion reached by less formal methods in section \ref{sec:QCD3SPT}.} 

We can further focus on just the easy-plane subgroup $SO(2)_s$  of $SO(3)_s$, then Eq.~(\ref{eq:w23}) reduces to
\beq S = \pi i \int_{X_4} w_4[A^5] = \pi i \int_{X_4} \frac{F^s}{2\pi} \cup \frac{F^{VBS}}{2\pi} \eeq
This is precisely the mutual $\theta = \pi$ term for $U(1)_s \times U(1)_{VBS}$. Note that to protect this $\theta$ value from shifting, one needs to rely on some discrete symmetry, such as spin-flip symmetry or time-reversal.

What if we restrict ourselves to an $SO(4)$ subgroup of $SO(5)$? Let us write $SO(4) = (SU(2)_L \times SU(2)_R)/Z_2$. Label associated $SO(3)_L$ and $SO(3)_R$ gauge fields as $A^L$ and $A^R$. We have $w_2[A^L] = w_2[A^R] = w_2[A^5]$. Now,
\bea 2 w_4[A^5] &=& p_1[A^5] - {\cal P}(w_2[A^5]) \nn &=& \frac{1}{2} p^{SO(3)}_1[A^L] + \frac{1}{2} p^{SO(3)}_1[A^R] - {\cal P}(w_2[A^R])\nn &=&  \frac{1}{2} p_1[A^L] - \frac{1}{2} p_1[A^R]
\quad (mod\,\,4)\eea
Therefore, Eq.~(\ref{eq:ZbulkSO5}) reduces to
\bea S &=& \frac{1}{4} CS_{SO(3)}[A^L, X_4] - \frac{1}{4} CS_{SO(3)}[A^R, X_4]\nn &=& \frac{1}{2}  CS_{SU(2)}[A^L, X_4] - \frac{1}{2} CS_{SU(2)}[A^R, X_4]\eea
In $SU(2)$ terminology (\ref{eq:thetaSU2}), this corresponds to opposite $\theta$ angles for $SU(2)_L$ and $SU(2)_R$: $\theta_L = -\theta_R = \pi$. Again,  discrete symmetries, e.g. $Z_2 = O(4)/SO(4)$, which maps $R\leftrightarrow L$, and time-reversal, are required to fix these $\theta$ angles from flowing.

\subsection{Bulk parton construction for boson SPT with $SO(5) \times Z_2^T$ symmetry: Formal derivation}

{In this section, we reconsider from a more formal standpoint the parton construction of the $3+1$D $SO(5)\times Z^T_2$ boson SPT presented at the end of section \ref{sec:QCD3SPT}. We will show that this construction precisely recovers the bulk SPT ``discrete  $\theta$-angle" response in Eq.~(\ref{eq:ZbulkSO5}). This provides a more physical motivation for our $3+1$D bulk ``completion" of QCD$_3$ in Eq.~(\ref{eq:QCDbb}). }

As in section \ref{sec:QCD3SPT}, we begin by considering a 3+1D ``topological superconductor'' of fermions with $SO(8) \times Z_2^T$ symmetry.  We will then gauge an $SU(2)$ subgroup and show that the result is precisely the boson SPT of interest. 
As explained in section \ref{sec:QCD3SPT}, we represent the $SO(8) \times Z_2^T$ symmetric topological superconductor by eight massive Majorana fermions with an inverted mass, Eq.~(\ref{eq:8Maj}). 
Let us consider the partition function on a manifold $X_4$ in the presence of a background  $SO(8)$ gauge bundle $A^8$.  We will restrict ourselves to closed oriented manifolds. 
 The partition function will take the form (see appendix \ref{app:APS})
 \begin{equation}
 \label{zblktsc}
 Z_{TSC}[A^8] = | Z_{TSC}| e^{i\pi \left(\frac{p_1[A^8]}{2} - \frac{\sigma}{2} \right) }
 \end{equation}
 with the $p_1$, $\sigma$ given by Eqs.~(\ref{eq:Pontryagin}), (\ref{eq:sigma}). Further by the Atiyah-Singer index theorem (\ref{eq:APS}), we have 
 \begin{equation}
 \frac{p_1[A^8]}{2} - \frac{\sigma}{2} = {\cal J} \in Z.
 \end{equation}
 Here $2{\cal J}$ is the index of the $3+1$D massless Dirac operator 
 on $X_4$.  This implies that the partition function in Eq. (\ref{zblktsc}) is real (but not necessarily positive) as required for a time-reversal invariant SPT state. 
  
Note that on a spin manifold  $\frac{p_1[A^8]}{2}$ is itself an  integer (and so is $\frac{\sigma}{2}$). This follows because by (\ref{eq:Pont}),
$p_1[A^8] = w_2[A^8] \cup w_2[A^8] ~(mod ~2)$ = $w_2[A^8] \cup w_2[TX_4]  = 0 ~(mod ~2)$. However, as we eventually will want to describe a boson SPT it is important to be able to formulate the theory on a non-spin manifold. As usual, since the fermions see transition functions in $(SO(8)\times Spin(4)_{TX_4})/Z_2$, we only require these transition functions to satisfy the cocycle condition. Then though $\frac{p_1[A^8]}{2}$ and $\frac{\sigma}{2}$ are not separately integers, their sum is. 

Now, to construct the boson SPT phase, we gauge the $SU(2)_g$ subgroup of $SO(8)$, as in Eq.~(\ref{eq:QCD4}). We further couple the system to a background $SO(5)$ gauge field $A^5$. As we discussed in section \ref{sec:QCD3SPT}, the partons $\chi$ transform as $Sp(4)$ spinors under $SO(5)$, so they see a combined $(Sp(4)\times SU(2)_g)/Z_2$ gauge field. Together with the $Spin(4)_{TX_4}$ transition functions, this yields overall transition functions in $(Sp(4)\times SU(2)\times Spin(4)_{TX_4})/(Z_2 \times Z_2)$. This gives the by now familiar condition:
\begin{equation}
\label{w2a5agtm}
w_2[A^5] + w_2[a_g] + w_2[TX_4] = 0 ~(mod ~2) 
\end{equation}
For such an $SO(8)$ bundle we have 
 \begin{equation}
 p_1[A^8] = p_1[A^5] + p_1[a_g]
 \end{equation}
 Thus, applying Eq.~(\ref{eq:pw4}), we obtain:
  \begin{equation}
 \label{zblktsca5ag}
 Z_{TSC}[A^5,a_g] = | Z_{TSC}| e^{i\pi  \int_{X_4} w_4[A^5] }
 \end{equation}
 
 Thus far we treated $a_g$ as a background gauge field. Now we make it dynamical, {\em i.e} we integrate the partition function over $a_g$. There is no topological term for $a_g$ and its dynamics will be governed by  the usual Yang-Mills action, which is expected to confine all fields charged under $a_g$ leaving behind a {\blue trivial} gapped vacuum. The resulting theory has a partition function with a phase given precisely by $e^{i\pi  \int_{X_4} w_4[A^5] }$.  Further it can be formulated on a non-spin manifold {\blue with only a background $SO(5)$ gauge field. The local operators are bosonic and transform non-projectively under $SO(5)$.} Thus, we have constructed the desired boson SPT with $SO(5) \times Z_2^T$ symmetry, and the partition function matches Eq.~(\ref{eq:ZbulkSO5}) proposed in section \ref{sec:SO5g} based on consistency arguments.

\subsection{Chiral spin-liquid}
\label{sec:chiralsl}
We note that if we break time-reversal in $QCD_3$ with an $Sp(4)$ preserving mass term
\beq \delta L = m \bar{\chi} \chi\eeq
and make the mass $m$ large enough (compared say to the gauge coupling), we will drive the system into a topologically ordered phase. Integrating the gapped $\chi$s in Eq.~(\ref{eq:QCDbb}) out:
\begin{widetext}
\be S^{{\rm bulk} + {\rm bound}}_{QCD_3} =  {\rm sgn}(m)\left[\frac{i}{2} CS_{SO(5)}[A^5, Y_4] + i CS_{SU(2)}[a^g, Y_4] + 4 i CS_g[Y_4]\right] + \pi i \int_{X_4 \cup \bar{Y}_4} w_4[A^5]  \label{eq:QCDbbc} \ee \end{widetext}
Without loss of generality, choose $m > 0$. By looking at the action for $a_g$, we see that we get a $2+1$D $SU(2)_1$ topological order, which is just a semion state $\{1,s\}$. The semion $s$ is just the Majorana $\chi$. The chiral central charge $c = 2 - 1$, with $2 = 4 \times 1/2$ coming from the gravitational Chern-Simons term and $-1$ from integrating $a_g$ out. The semion $\chi$ transforms projectively under the $SO(5)$ symmetry --- as an $Sp(4)$ spinor. In particular, it carries spin-$1/2$ under $SO(3)_s$. Further, the $SO(5)$ response is given by a Chern-Simons term with level $k  = 1/2$; so the level of Chern-Simons response to $SO(3)_s$ gauge field is also $1/2$. We see that this state has all the properties of a chiral spin liquid \cite{wenbook}.

Now, it {would be} a little surprising if the chiral spin-liquid was in the vicinity of the deconfined quantum critical point. If we consider the $NCCP^1$ formulation (\ref{nccp1su2}),
then one operator with the same quantum numbers as $\bar{\chi} \chi$ is $\epsilon_{\mu \nu \lambda}  \d_{\rho} f^b_{\rho \mu}  f^b_{\nu \lambda}$
with $f^b_{\mu \nu} = \d_{\mu} b_{\nu} - \d_{\nu} b_{\mu}$. Given the number of derivatives, one would naively expect this term to be irrelevant in the $NCCP^1$ model. {If $NCCP^1$ and QCD$_3$ indeed share the same fixed point, this} would then imply that $\bar{\chi} \chi$ is irrelevant at the QCD$_3$ fixed point --- an unexpected, but not impossible scenario. It would be interesting to determine the scaling dimension of this operator numerically at the deconfined critical point. In the lattice magnet, it corresponds to the imaginary part of the plaquette ring exchange:
\begin{widetext}
\bea \bar{\chi} \chi &\sim& \frac{-i}{2}(P_{i+\hat{y}, i+\hat{x}+\hat{y}} P_{i + \hat{x}+\hat{y},i+\hat{x}}P_{i + \hat{x}, i} - h.c.) \nn
&\sim& \vec{S}_{i+\hat{x}+\hat{y}} \cdot(\vec{S}_i \times \vec{S}_{i +\hat{x}}) + \vec{S}_{i+\hat{y}}\cdot(\vec{S}_i\times \vec{S}_{i + \hat{x}}) +\vec{S}_{i+\hat{y}}\cdot(\vec{S}_{i+\hat{x}} \times \vec{S}_{i + \hat{x} + \hat{y}})+\vec{S}_{i+\hat{y}}\cdot(\vec{S}_i\times \vec{S}_{i + \hat{x}+\hat{y}}) \nonumber \eea
\end{widetext}
where $P_{ij} = 2 \vec{S}_i \cdot\vec{S}_j + 1/2$ is the exchange operator.

\section{Discussion and Implications of the dualities}
\label{implications}

The most fundamental question about both the ${\mathrm{SU}(2)}$ invariant and
the easy-plane $NCCP^1$ theories is whether they describe CFTs in
the IR.  We have not tackled this question head on in this paper.
We first discuss what follows if the dualities are assumed to hold
in their strong forms in the IR. As mentioned in the introduction,
the {various} theories could fail to flow to nontrivial fixed
points. In this scenario the dualities may still be relevant to
the `quasiuniversal' physics up to a large lengthscale. We discuss
these issues, and what is known about the IR fate of the
deconfined critical transitions, in Sec.~\ref{pseudocritical}.

\subsection{For deconfined criticality}


Many consequences of the emergent ${\mathrm{SO}(5)}$ symmetry have been explored numerically in Ref.~\cite{emergentso5}. In section \ref{PhaseD} we discussed an additional consequence for the phase diagram in the presence of perturbations that break ${\mathrm{SO}(5)}$. This will be interesting to explore in future numerical work.

We argued that the proposed duality web provides an explanation of this emergent ${\mathrm{SO}(5)}$ symmetry, {despite the fact that  the ${\mathrm{SO}(5)}$ symmetry is not manifest in any single member of the duality web.} In particular the proposed self-duality of the ${\mathrm{SU}(2)}$ invariant $NCCP^1$ model immediately implies emergent ${\mathrm{SO}(5)}$ symmetry in the IR.  We will discuss other numerical tests of the fermionic versions of this theory separately below. 

We also discussed the continuum $N_f = 2$ $QCD_3$ theory which has manifest ${\mathrm{SO}(5)}$ symmetry and which shares the same anomaly as the putative deconfined critical point, and may possibly flow to it in the IR. The IR fate of the $QCD_3$  theory is not currently known and is a good target for future numerical work. It will be particularly interesting to see if it shares the (quasi)universal power law correlations seen in other models equivalent to the ${\mathrm{SU}(2)}$-invariant $NCCP^1$.

For the easy plane model direct numerical simulations of quantum magnets find a first order transition. As we have emphasized several times,  the nature of the transition in this model is worth revisiting.
We have seen that this  model is dual to a version of fermionic $N_f = 2$ $QED_3$ with $U(1) \times U(1)$ symmetry.  For fermionic $QED_3$ with ${\mathrm{SU}(2)}$ flavor symmetry, there is some recent evidence that the theory is conformal in the IR.\cite{qedcft}  Further the results do not seem to be sensitive to whether the lattice regulator employed actually preserves full ${\mathrm{SU}(2)}$ flavor symmetry or whether it only has $U(1)$ flavor symmetry. In light of all this, more numerical studies of the models in the easy plane duality web are clearly called for.

The strongest form of the duality web of these theories asserts that all these theories flow in the IR to the same $O(4) \times Z_2^T$--invariant CFT.  Below we will describe some
implications of the enhanced symmetry expected in such a putative
critical theory.

If an $O(4) \times Z_2^T$ symmetric fixed point does exist then,
for the easy plane $NCCP^1$ to flow to it in the IR it must be
that perturbations that break  the symmetry to
$(U(1) \times (U(1) \rtimes Z_2)) \times Z_{2T}$ are irrelevant. {As discussed in Sec.~\ref{allowed1}, the simplest such perturbation is a N\'eel/VBS anisotropy which lies in the $(2,2)$ representation of $SO(4)$ (the quadrupled monopole operator $\Phi_2^4 + \Phi_2^{*4}$ pertinent to the lattice magnet lies in the same representation). Thus for enlarged $O(4)$ symmetry we need the scaling dimension $\Delta_{(2,2)} > 3$ at the $O(4)$ fixed point.}

If  an $O(4) \times Z_2^T$ symmetric fixed point exists, and the strongest form of the duality web holds, then a square lattice spin-$1/2$ quantum magnet with $XY$ symmetry can show a direct continuous Neel-VBS transition with enlarged $O(4)$ symmetry.

 An alternate possibility is that the $O(4) \times Z_2^T$ CFT exists and that ${\mathrm{SU}(2)}$ flavor symmetric $QED_3$ flows to it, but the $U(1) \times U(1)$ theories (easy plane $NCCP^1$ and $QED_3$ with the same symmetry) do not flow to that fixed point. This scenario can be tested by numerical simulations of the ${\mathrm{SU}(2)}$ symmetric $QED_3$ theory. We detail below how to test for emergent $O(4)$ symmetry. Should such a fixed point be found, it will be interesting to calculate the scaling dimension of operators transforming under the $(2,2)$  representation to test for relevance.
 
 Finally it is possible that an $O(4) \times Z_2^T$ fixed point of the kind we have described does not exist in the first place.  Therefore we next turn to the fermionic theories where this question is best addressed numerically. 
 
 {Useful analytical insights will also come from conformal bootstrap \cite{solvingIsing, bootstrappingO(N), SimmonsDuffinSO(5), Nakayama, ChesterPufuBootstrappingQED}.  Note also that the  duality webs open up the possibility of analytical results for deconfined critical points using large $N$ in the fermionic language \cite{GRACEY1993415, DyerMonopoleTaxonomy, ChesterPufuScalarOperators}.}

\subsection{For QED$_3$ and QED--GN}

The strong self-duality  for ${\mathrm{SU}(2)}$ flavor symmetric QED$_3$ implies an emergent $O(4)$ symmetry which leads to
  simple testable predictions.

{
The fermion bilinears
$\bar{\psi} \vec \sigma \psi$  are expected to be scaling fields, with dimension smaller than their engineering dimension of two,
transforming in the $(1,1)$ representation of $SO(4)$.
$O(4)$ symmetry relates them to 
strength-$2$  monopole operators in the $QED_3$ theory, so 
calculating correlations of monopole operators will allow interesting tests of the emergent symmetry. Presumably this requires some modifications of existing
numerical calculations of correlators in the QED$_3$ theory. We therefore also describe several tests using more ordinary correlators.}

It should be  fruitful to focus on correlations of the
conserved  $SO(4)$ currents. The operator $\bar{\psi} \gamma^0
\vec \sigma \psi$  (the time component of one of the $SU(2)$ currents) was already studied in Ref.~\cite{qedcft} and shown to have the expected scaling
dimension $2$.  One of the three currents of the other ${\mathrm{SU}(2)}$ is
a simple operator in $QED_3$: this is the gauge
flux $\epsilon_{\mu\nu\lambda} \partial_\nu a_\lambda$.  Therefore its
time component, the magnetic flux, is related by symmetry to
 $\bar{\psi} \gamma^0 \vec \sigma \psi$. Right at the critical
point these operators should have scaling dimension $2$; this follows from their conservation and is not a test of the
symmetry rotating them. A simple consequence of emergent $O(4)$ is that the universal amplitudes of the two
point functions should also be the same for these different currents. Specifically if we compare the correlators of the ${\mathrm{SU}(2)}$ currents with the correlators of $\frac{1}{2\pi} \epsilon_{\mu\nu\lambda} \partial_\nu a_\lambda$, they should have the same universal amplitude in addition to the same scaling dimension.

A more dramatic consequence arises if  we perturb the critical point by
turning on either a non-zero temperature $T$ or a fermion mass $m$, both of which preserve $SO(4)$. Then the current correlations  will involve a non-trivial universal scaling function
 \be
|k| F\left(\frac{m}{T}, \frac{\omega}{|k|},
\frac{|k|}{m}\right).
 \ee
Now $SO(4)$ symmetry predicts that this scaling function is
identical for the ${\mathrm{SU}(2)}$ current and for the $3$-flux of the gauge field. It will be very interesting to
test this.  For instance the ${\mathrm{SU}(2)}$ spin susceptibiltiy should be
described by the same crossover function as the diamagnetic
susceptibility of the gauge field, and likewise the ${\mathrm{SU}(2)}$ phase
stiffness should be described by the same crossover function as
the Meissner stiffness of the gauge field.

Finally, a representative of the important $(2,2)$ operator will be given by, {\it e.g},  ${2(\bar{\psi} \sigma^z \psi)^2  - (\bar{\psi} \sigma^x \psi)^2 - (\bar{\psi} \sigma^y \psi)^2}$ {(we are assuming that the other $SO(4)$ representations contributing to this operator are less relevant, as expected from the discussion in Sec.~\ref{allowed1})}. If $O(4)$ symmetry is established numerically, then the irrelevance of this $O(4)$--breaking perturbation can be tested.

For the QED$_3$--GN model, the first issue that should be addressed
numerically is whether the transition is second order at all (the duality with the $NCCP^1$ model suggests there should be critical behaviour at least up to a large length scale). Should such a second order transition be found, a number of its
properties can be predicted using our results.

First, if we measure $\phi$ correlations at this fixed point, we
are measuring correlations of the ${\mathrm{SO}(5)}$ vector. They can therefore
be compared with the N\'eel and VBS correlation functions known
from $NCCP^1$ simulations. Second, the  $\phi^2$  operator takes
us to the QED$_3$/easy plane $NCCP^1$ fixed point.  We know that
with ${\mathrm{SO}(5)}$ this is in the same representation as the operator
that tunes through the transition in the $NCCP^1$ theory (a component of $X^{(2)}$ in the notation of
Sec.~\ref{allowed2}). Hence the $\phi^2$
scaling dimension can be compared with results for $\nu$ at the
${\mathrm{SU}(2)}$--symmetric deconfined critical point.

More interestingly, the fermion bilinear
$\bar{\psi} \sigma^z \psi$ also corresponds to an element of $X^{(2)}$. Thus the vector
$\overline{\psi}  \vec \sigma \psi$ should have the same correlations
as  $\phi^2$ at the QED$_3$-GN fixed point (modulo subleading contributions) if there is full
${\mathrm{SO}(5)}$ symmetry. This last statement is particularly interesting as it does not involve comparing with a different theory --- both quantities are
calculated in the same simulation.

\subsection{Comparison between the $N=2$ QED$_3$, bilayer honeycomb lattice model, and easy plane spin models}

For the putative $O(4)$ fixed point, there are (at least) three
lattice model realizations that can be (and have been) studied
numerically: $N=2$ lattice QED$_3$, spin models
that realize the easy-plane deconfined transition ({if a model with a second order transition exists}), and the
bilayer honeycomb lattice interacting fermion model, studied in
Ref.~\cite{kevinQSH,so4qsh}, that realizes the transition between
a trivial and SPT boson insulator with explicit $SO(4)$ symmetry.
The critical exponents measured in different models should be
related to each other, which we discuss below.

The $N=2$ QED$_3$ was treated as a stable CFT in Ref.~\cite{qedcft},
so there is no correlation length critical exponent. But there is
still the anomalous dimensions associated with the
mass operators {$M_z = \bar{\psi}_1\psi_1 - \bar{\psi}_2 \psi_2$, $M_0 = \bar{\psi}_1\psi_1 + \bar{\psi}_2 \psi_2$,
\be \langle M_z(0) \ M_z(r) \rangle \sim
\frac{1}{r^{1+\eta_{\bar{\psi} \sigma^z \psi}}}, \quad  \langle M_0(0) \ M_0(r) \rangle \sim
\frac{1}{r^{1+\eta_{\bar{\psi}  \psi}}} \ee According to Ref.~\cite{qedcft},
$\eta_{\bar{\psi} \sigma^z \psi}\sim 1.0$. To our knowledge, a careful  study of $\eta_{\bar{\psi}  \psi}$ has not been performed in numerical simulations of QED$_3$ - we hope that future simulations will also address  this exponent.}

The bilayer honeycomb lattice model describes a bosonic
transition, which may {potentially} also be described by the $N=2$ QED$_3$.\footnote{{Since the microscopic model has an exact $SO(4)$ symmetry here, for QED$_3$ to describe it, the $SO(4)$ symmetry must necessarily emerge at the critical point.}} The
tuning parameter for this transition corresponds to the fermion
mass $m (\bar{\psi}_1 \psi_1 + \bar{\psi}_2 \psi_2)$ in the field
theory. There is a correlation length exponent $\nu_{bh}$ defined
as \beqn \xi \sim m^{-\nu_{bh}} \sim (J - J_c)^{- \nu_{bh}}, \eeqn
where  $J$ is the interaction on the lattice that is tuned to
the critical point. The O(4) order parameter ${n}_a$ has an
anomalous dimension $\eta_{bh}$: \beqn \langle {n}_a(0)
{n}_a(r) \rangle \sim \frac{1}{r^{1+\eta_{bh}}}. \eeqn
The easy-plane spin models have three different exponents: $\eta_{xy}$
(the same as $\eta_{vbs}$), $\eta_{z}$ and $\nu_{jq}$: 
\bea 
\xi&\sim& (Q - Q_c)^{-\nu_{jq}}, \nn 
 \langle S^x(0) \ S^x(r)\rangle &\sim& \frac{(-1)^r}{r^{1 + \eta_{xy}}}, \nn
  \langle S^z(0) \ S^z(r) \rangle &\sim& \frac{(-1)^r}{r^{1 + \eta_{z}}}.
\eeqn
Where $Q$ is a tuning parameter for the transition. {If the strong duality holds, we have the following relations:}
\bea 
3 -\frac{1}{\nu_{jq}} &=& \frac{1 + \eta_{\bar{\psi} \sigma^z \psi}}{2}, \nn
\eta_z &=& \eta_{\bar{\psi}\psi}, \nn
 \eta_{xy} &=& \eta_{bh}, \nn
  3 - \frac{1}{\nu_{bh}} &=& \frac{1 +\eta_z}{2}. \eea

\section{Critical and pseudocritical points}
\label{pseudocritical}

 It is not yet certain whether the $\mathrm{SU}(2)$--symmetric
$\nccp^1$ model has a true critical point, or whether it instead
shows `pseudocritical' behaviour with a very large but finite
correlation length. Here we review what is currently known from
simulations, and clarify what the latter possibility would mean
for the dualities presented here. We also briefly discuss the easy
plane--case.

Various lattice models that show a phase transition `in the
$\nccp^1$ universality class' have been studied numerically
\cite{SandvikJQ,melkokaulfan,lousandvikkawashima,Banerjeeetal,Sandviklogs,Kawashimadeconfinedcriticality,Jiangetal,deconfinedcriticalityflowJQ,DCPscalingviolations,emergentso5,MotrunichVishwanath2,kuklovetalDCPSU(2),Bartosch,CharrierAletPujol, Chenetal,Aletextendeddimer,powellmonopole}.\footnote{These include the J--Q model\cite{SandvikJQ, melkokaulfan, lousandvikkawashima, Banerjeeetal, Sandviklogs, Kawashimadeconfinedcriticality, Jiangetal, deconfinedcriticalityflowJQ} and a related loop model \cite{DCPscalingviolations,emergentso5}, lattice
\cite{MotrunichVishwanath2, kuklovetalDCPSU(2)} and
continuum  \cite{Bartosch} field theories, and the classical 3D
dimer model \cite{CharrierAletPujol, Chenetal, Aletextendeddimer, powellmonopole}, which has different microscopic
symmetries to the JQ model but has been argued to have the same
continuum description. See also numerical work on generalizations to
$\mathrm{SU}(n)$ with $n>2$ \cite{beach2009n,BanerjeeetalSU(3),KaulSU(3)SU(4),kaulsandviklargen}.} The  basic feature of these simulations is that the correlation length $\xi$ appears to
diverge as the critical point is approached, certainly becoming
larger than numerically accessible system sizes (up to 640 lattice
spacings in the model of Ref.~\cite{DCPscalingviolations}). At
these lengthscales  the standard signs of first-order behaviour,
e.g. double-peaked probability distributions, are absent. The
qualitative features of the transition are as expected from the
theory of deconfined criticality,\footnote{For example, in models
for the N\'eel-VBS transition there is a single direct transition
between the two phases,  the expected $U(1)$ symmetry for the VBS
order parameter indeed emerges at the critical point, and the N\'eel
vector has a large anomalous dimension, which was predicted as a
signature of deconfinement.} and finite size estimates of critical
exponents are roughly consistent between
different lattice models.

These features are consistent with a continuous transition (which
much recent work assumes). However, it was noted some time ago
that various naively `universal' quantities instead drift with
system size, leading to controversy about whether the transition was ultimately continuous or first order \cite{Jiangetal, deconfinedcriticalityflowJQ,
Kawashimadeconfinedcriticality, Sandviklogs, Banerjeeetal, kuklovetalDCPSU(2)}. Ref.~\cite{DCPscalingviolations} argued that these drifts are not merely
conventional  finite-size corrections to CFT scaling behaviour, since making this assumption leads to unphysical negative values for the
anomalous dimensions at large sizes, and suggested two possible
scenarios for reconciling the various numerical results. One
scenario is that the $\nccp^1$ model shows a continuous
transition, but with unconventional finite-size scaling behaviour
due to a dangerously irrelevant variable\footnote{Some features of the numerical
results are suggestive of this. See e.g. the discussion
of correlation functions in \cite{DCPscalingviolations}, and the
fits in \cite{sandvik2parameter}.} (see also \cite{KaulSU(3)SU(4), sandvik2parameter}).  The second scenario is that
$\nccp^1$ shows a  first order transition \cite{Jiangetal,
deconfinedcriticalityflowJQ, kuklovetalDCPSU(2)} which is
rendered anomalously weak by a quasi-universal mechanism \cite{DCPscalingviolations} which we discuss below.

Further complicating the issue, it was found numerically that
critical fluctuations at the deconfined transition are
$\mathrm{SO}(5)$ symmetric to a high level of precision
\cite{emergentso5}. Level degeneracies
found in the JQ model \cite{SandvikSpectrum} also support this enhanced symmetry   (the approximate equality of N\'eel and VBS scaling dimensions had been noticed earlier by Sandvik  \cite{sandvikVBS}).  At first sight $\mathrm{SO}(5)$ symmetry seems
to be strong evidence that the critical $\nccp^1$ model flows to
an $\mathrm{SO}(5)$--invariant CFT. However, subsequent
investigations \cite{SimmonsDuffinSO(5), Nakayama} of $\mathrm{SO}(5)$--symmetric CFTs using the
conformal bootstrap \cite{solvingIsing, bootstrappingO(N)} did
\textit{not} find a sufficiently stable\footnote{To describe the
(generic) deconfined critical point, the $\mathrm{SO}(5)$ CFT
should have no relevant operator that is invariant under all
symmetries, see Sec.~\ref{allowed2}. The bootstrap result constrains the
$\mathrm{SO}(5)$ vector's anomalous dimension, under the
assumption that there is no relevant symmetry-trivial operator.}
CFT in the expected region of parameter space. The
bootstrap shows that any sufficiently stable $\mathrm{SO}(5)$--invariant CFT  must have a
larger anomalous dimension for the $\mathrm{SO}(5)$ vector than is
expected from simulations of deconfined criticality
\cite{SimmonsDuffinSO(5), Nakayama}. In view of this, it makes
sense to revisit the weakly-first-order scenario with
$\mathrm{SO}(5)$ symmetry in mind.

At first sight a first order transition with $\xi \gg 1$ is
implausible because of a fine-tuning problem. If a theory has no
nontrivial stable fixed point, the obvious way to get a large
$\xi$ is to fine-tune it close to an unstable fixed
point.\footnote{Take for example the $q$-state Potts model in 3D,
with a large value of $q$. The transition is generically first
order due to  a cubic invariant in the Landau-Ginsburg action. One
could fine-tune the couplings to be close to the free fixed point,
giving a very weak first order transition. This mechanism is
non-generic, and is not the mechanism discussed in the text.}
Since this mechanism relies on fine--tuning, it is unlikely to be
the explanation for the apparent critical behaviour at the DCP,
which seems to be generic. However there is an alternative generic
mechanism for `pseudocritical' behaviour with very large $\xi$
\cite{nienhuispotts,nauenbergscalapino,cardynauenbergscalapino}. In this scenario,
the large $\xi$ can be understood in terms of a fixed point which
exists slightly outside the physical parameter space of the model
--- for example at slightly smaller spatial dimension $d_c$. The
structure of the RG flows close to $d_c$ implies an exponentially
large correlation length for $d\gtrsim d_c$. This mechanism
depends on an accident in the universal structure of the RG flows,
but it does not require fine-tuning of a given microscopic
Hamiltonian. Additionally, this scenario is plausible for the
$\nccp^1$ model (and indeed $\nccp^{n-1}$ for nearby values of
$n$), given what is known about the $d$--dimensional $\nccp^{n-1}$
model in various limits \cite{DCPscalingviolations}.

The basic mechanism is the annihilation of a stable and an
unstable fixed point as a parameter $\tau$ is varied. Here $\tau$
is a quantity which does not flow under RG, such as the spatial
dimension (in the case of $\nccp^1$) or the rank of a symmetry
group. Quite generally, close to $\tau_c$ the RG equation for the
coupling $\lambda$ which is becoming marginal looks like
\be\label{pseudocriticality rg equation} \f{\dd \lambda}{\dd \ln
L} = a\,(\tau_c - \tau) - \lambda^2, \ee where $a$ and $\tau_c$
are universal constants and $a>0$. For $\tau<\tau_c$ both fixed
points exist and for $\tau>\tau_c$ neither do. But for $\tau
\gtrsim \tau_c$ the RG flows become very slow close to
$\lambda=0$: the long RG time required to traverse the
`pseudocritical' region corresponds to a large lengthscale $\xi
\sim \ell_0 \exp{(\pi/\sqrt{a(\tau-\tau_c)})}$, where $\ell_0$ is
nonuniversal. The large amount of RG time spent near
$\lambda=0$ implies that the properties of the pseudocritical
regime are quasiuniversal in the limit of small ${\tau-\tau_c}$.  

In more detail, this is because (in the formal limit of small ${\tau-\tau_c}$) the \textit{subleading} RG couplings, $g_i$, have time to flow to well-defined pseudocritical values, independent of their bare values in a given microscopic model. (The relevant coupling which drives the transition is zero since we consider the theory in the critical plane.)  The RG flow is attracted to a quasiuniversal trajectory through coupling constant space, given by setting $g_i = 0$ up to corrections that are {exponentially} small in ${1/\sqrt{\tau-\tau_c}}$. A key point is that `quasiuniversality' holds to exponentially good precision in ${1/\sqrt{\tau-\tau_c}}$, despite the fact that the flow of $\lambda$ during a stretch of RG time of order $\ln \xi$ is larger than this.\footnote{We may see this in more detail using the logic of \cite{Wegner1976, cardynauenbergscalapino}. Starting with the RG equations for $\lambda$ and for the leading {irrelevant} coupling $g$ at  ${\tau=\tau_c}$, expand in ${\Delta^2 = \tau-\tau_c}$ assuming analyticity in $\Delta^2$ and the couplings. Using the freedom to make analytic redefinitions of the couplings gives (\ref{pseudocriticality rg equation}) and ${\dd g/ \dd t  = - (y + m \lambda) g}$, neglecting subleading corrections ($t$ is RG time). The zero of the latter equation at $g=0$ is preserved to all orders in $\Delta^2$. We have ${\lambda(t)=-\Delta \tan  \Delta [t-t_*]}$ with $t_* \simeq \pi/2\Delta - 1/\lambda(0)$. The correlation length is determined by setting $\lambda\sim -1$, giving ${\ln \xi = \pi / \Delta + O(1)}$. The subleading coupling behaves as $g(t) = g_0 e^{-yt} \left( \f{\Delta^2+\lambda(t)^2}{\Delta^2 + \lambda(0)^2}\right)^{m/2}$. Once the RG time is of order $1/\Delta$, $g(t)$ has become exponentially small in $1/\Delta$. On RG timescales of this order the typical variation in $\lambda(t)$, and therefore the typical `quasiuniversal' drift in large scale properties, is $O(\Delta)$.} This flow of $\lambda$ will lead to quasiuinversal drifts in e.g. effective exponents.

The $Q$-state Potts model in 2D provides an example of this
phenomenon with $\tau=Q$  \cite{baxterpotts,nienhuispotts,nauenbergscalapino,cardynauenbergscalapino,LeeKosterlitzPotts,Klumper, Buffenoir} (in this context $\lambda$ was
originally thought of as a fugacity for Potts vacancies
\cite{nienhuispotts}). For $Q<4$ both a critical and a
tricritical point exist, and they merge at $Q=4$. For $Q\gtrsim
4$, the Potts transition is very weakly first order. A priori the
above picture applies only for $(Q-4)\ll 1$, but empirically it is
found that the transition remains weakly first order at least for
$Q=5,\, 6,\, 7$, where $\xi \simeq 2512, \, 159,\, 48$
respectively on the square lattice \cite{Buffenoir}. This
mechanism for generating a small mass scale has also been
discussed in the context of 4D QCD \cite{Gies, kaplan,gukov2017rg}, with $\tau
=-N_f/N_c$.  (Fixed point annihilation phenomena have also been discussed in QED$_3$ \cite{giombi2016conformal, herbut2016chiral, gukov2017rg}, and in a Landau Ginsburg theory obtained from $NCCP^{n-1}$ by condensing the monopole \cite{PhaseTransitionsCPNSigmaModel}.\footnote{A fixed point annihilation phenomenon also occurs in the so called `compact' $CP^{n-1}$ model (an $SU(n)$--symmetric Landau-Ginsburg theory obtained from $NCCP^{n-1}$ by condensing the strength-1 monopole). However there the critical and tricritical fixed points annihilate when $n$ is \textit{increased} (with $n_c\sim 3$) rather than when $n$ is \textit{decreased} as is the case for $NCCP^{n-1}$ \cite{PhaseTransitionsCPNSigmaModel}.}) In the $\nccp^{n-1}$ model, it is plausible that there
is a range of $n$ where the transition is weakly first order but
can be rendered continuous by slightly decreasing the spatial
dimension.

It should be noted that the choice of deformation parameter $\tau$
is not unique; for example in the weakly first order regime of the
Potts model the transition can be made continuous by reducing
either $Q$ or $d$ (and in $\nccp^{n-1}$ we can certainly render
the transition continuous by a large enough increase in $n$).
Alternately, one may consider the theory only at the physical
value of $\tau$, and attribute the pseudocritical behaviour to
proximity to the nonunitary fixed points at $\lambda =  \pm
i\sqrt{a(\tau-\tau_c)}$.

A possible explanation for
the various numerical results for the deconfined transition is that there is a pseudocritical
regime within the $\mathrm{SO}(5)$--symmetric subspace of theory
space. If the effective scaling dimensions of allowed
$\mathrm{SO}(5)$ breaking perturbations --- specifically,  the
symmetric tensor $X^{(4)}_{abcd}$ discussed in Sec.~\ref{allowed2} --- are greater
than three, the $\nccp^1$ model and QED--Gross-Neveu can lie in
the basin of attraction of this regime, and will also show
pseudocritical behaviour. In this scenario the dualities we have
discussed apply to the physics at lengthscales up to $\xi$ (and somewhat beyond, see below).

To make the above possibility more concrete we may think of
$\nccp^1$ as a perturbation of an exactly
$\mathrm{SO}(5)$--invariant theory whose RG behaviour could in
principle be pinned down. The nonlinear sigma model is one
possibility: in the above scenario we would expect the  sigma
model at strong coupling to exhibit pseudocritical behaviour.
However the language of the sigma model does not give us an
obvious candidate for the deformation parameter $\tau$.
Tentatively, a renormalizable alternative may be the $N_c=2$,
$N_f=2$ QCD$_3$ discussed in Sec.~\ref{manifest SO(5)}. This theory has an
$\mathrm{SO}(5)$ symmetry which becomes explicit when it is
written in terms of Majorana fermions. The theory may
spontaneously break $\mathrm{SO}(5)$ on the scale set by the
coupling, in which case it is not very interesting. But another
possibility is that it generates a long lengthscale by the above
mechanism. If so, $N_f$ and $d$ are candidates for
deformation parameters $\tau$ which could produce a true fixed
point.

Another way to think about the possibility of an
$\mathrm{SO}(5)$-invariant pseudocritical regime is
to hypothesize an exact $\mathrm{SO}(5)$ symmetry for the nearby
nonunitary fixed points at imaginary $\lambda$; we thank S. Pufu for suggesting this. We would then view the $\nccp^1$
model and QED--Gross-Neveu on lengthscales $\lesssim \xi$ as
perturbations  away from this fixed point.

As an aside, note that in the present scenario $\mathrm{SO}(5)$ symmetry survives to lengthscales
even larger than $\xi$. The simplest possibility is that on scales
larger than $\xi$ the system flows to the ordered phase of the
$\mathrm{SO}(5)$ sigma model, representing a first order
transition for $\nccp^1$ --- although in principle it is possible
that the theory could flow to a new nontrivial fixed point. This
ordered sigma model is subject to anisotropies due to the
$\mathrm{SO}(5)$-breaking perturbations that are allowed in the
microscopic model. However these anisotropies are small since the
effective RG eigenvalue $y_4$ of the 4-index symmetric tensor is
negative in the pseudocritical regime: they do not become apparent
until a lengthscale $L_* \sim \xi^{1+|y_4|/3}\times (\text{bare
coupling})$. Simulations in the range $\xi \lesssim  L \lesssim
L_*$  would find four apparent Goldstone modes.\footnote{It is
clear from the drift in effective critical exponents that current
simulations are not in this regime (which would show simple
exponent values).}

A similar conjecture could in principle apply to the dualities
between $\mathrm{O}(4)$ invariant theories, although numerical
results for the easy-plane N\'eel-VBS transition suggest that a
first order transition with a rather shorter correlation length {than  in the $SU(2)$ case may be generic there} \cite{kukloveasyplane, kragseteasyplane, kauleasyplane1, kauleasyplane2,
Chenetal}. As noted in Sec.~\ref{allowed1}, it is also conceivable that
the fermionic theories flow to an $\mathrm{O}(4)$ symmetric fixed
point (or pseudocritical point), but that the easy plane model
does not, as a result of an additional perturbation allowed by
microscopic symmetry. Further numerical studies of both the easy plane model and of QED$_3$ along the lines described in Sec. \ref{implications} are clearly called for.

For quantum phase transitions, the pseudocriticality scenario implies that the system will show quantum critical behavior above a parametrically low temperature scale ${T^*\sim  J  \exp \left( -\pi/\sqrt{a(\tau-\tau_c)} \right)}$, where $J$ is a microscopic energy scale, with critical exponents drifting as the temperature changes. Criticality eventually disappears below $T^*$, and the system possibly crosses over to a first order transition. But for sufficiently low $T^*$, a quantum critical regime (the famous `critical fan') should be observable above $T^*$. Pseudocritical systems thus present interesting possibilities for phenomenology near quantum phase transitions.

\subsection*{Acknowledgments}

We thank 
F. Alet, 
J. Chalker,
Y. C. He, 
R. Kaul, 
A. C. Potter,
S. Powell, 
S. Pufu, 
S. Rychkov, 
A. Sandvik,  
N. Seiberg,
P. Serna,
D. Simmons-Duffin, 
A. Vishwanath, and 
E. Witten 
for helpful discussions.  We thank D. Simmons-Duffin for sharing unpublished bootstrap calculations. CW is supported by the Harvard
Society of Fellows. AN was supported by the Gordon and Betty Moore Foundation under the EPiQS initiative (grant No.~GBMF4303) and by the EPSRC Grant No.~EP/N028678/1. CX is funded by the David and Lucile Packard Foundation and NSF Grant No. DMR-1151208.
T.S. is supported by a US Department of Energy grant DE-SC0008739, and in part  
by a Simons Investigator award from the Simons
Foundation to Senthil Todadri.
CW, MM and CX thank
the hospitality of the Kavli Institute for Theoretical Physics,
which is supported in part by the National Science Foundation
under Grant No. NSF PHY-1125915.  Part of this work was performed by TS at the Aspen Center for Physics, which is supported by National Science Foundation grant PHY-1066293. Research at Perimeter Institute for Theoretical
Physics (MM) is supported by the Government of
Canada through the Department of Innovation, Science
and Economic Development and by the Province of Ontario
through the Ministry of Research and Innovation.

\appendix

\section{More precise Lagrangians}
\label{PreciseL}
In this Appendix we present the easy-plane dualities in a more precise notation. We define a Dirac fermion through a Pauli-Villars regulator, such that its partition function (under a general $U(1)$ gauge field $a$ and metric $g$) is given by $ Z_\psi =|Z_\psi| e^{-i\pi \eta[a,g]/2}$. In the usual notation (used in the main text), this would corresponds to a Dirac Lagrangian supplemented with a Chern-Simons term at level $k=-1/2$ for both $a$ and $g$. In general it is more precise to use the $\eta$-invariant instead of the less well-defined $k=-1/2$ Chern-Simons terms (for more details, see, for example, Ref.~\cite{wittenreview} for a review).

We also enforce the proper quantization of all the $U(1)$ gauge field (including dynamical and background ones) from the beginning, by working directly with $B_{1,2}$ (instead of $B,B'$) in the main text. The duality now reads:
\begin{widetext}
\bea
&&|D_bz_1|^2+|D_{b-B_1}z_2|^2-|z_1|^4-|z_2|^4+\frac{1}{2\pi}bdB_2    \nn
 \Longleftrightarrow \hspace{20pt} &&\bar{\psi}_1i\slashed{D}_a\psi_1+\bar{\psi}_2i\slashed{D}_{a+B_2-B_1}\psi_2+\frac{1}{4\pi}ada+\frac{1}{2\pi}adB_2+\frac{1}{4\pi}B_2dB_2+2CS_g,
\eea
\end{widetext}
where $CS_g$ is a gravitational Chern-Simons term, normalized such that in the absence of any other matter field, it leads to a thermal hall conductance $\kappa_{xy}=\frac{1}{2}\frac{\pi^2k_B^2T}{3h}$.

\section{Some other dualities}
\label{altdl} Here we describe an alternate set of dualities
between theories with global ${\mathrm{SU}(2)}$, $U(1)$ and ${\cal T}$
symmetries. We begin with a duality between $N = 2$ species of a
free massless Dirac fermion\footnote{In this and the next Appendix
we use the precise definition of the  Dirac partition function in
terms of the $\eta$-invariant. We will also include a coupling to
a background space-time metric $g$. } and a bosonic theory.

 To be concrete, we define the partition function of the massless Dirac fermion in terms of the $\eta$-invariant. To maintain ${\mathrm{SU}(2)}$ symmetry between the two species we must choose the same regularization for both species. Therefore we write the partition function of the $N = 2$ free massless Dirac fermion as
 \begin{equation}
 Z_\psi = |Z_\psi|   e^{-i\pi \eta[A,g]}
 \end{equation}
 Here $A, g$ are the background gauge field (strictly speaking a spin$_c$ connection) and metric respectively. We know that $e^{-i\pi \eta[A, g]} = e^{-i \int \frac{1}{4\pi} AdA + 2 CS[g]}$.
 Therefore the theory can be made time reversal invariant (while keeping ${\mathrm{SU}(2)}$ and $U(1)$) by adding $\int \frac{1}{4\pi} AdA + 2 CS[g]$ . Thus we consider\footnote{We have not shown explicitly background ${\mathrm{SU}(2)}$ gauge fields - they can be incorporated if needed.}
 \begin{equation}
 \label{free2f}
 {\cal L}_{0f} = i\overline{\psi}_\alpha D_A \psi_\alpha +  \frac{1}{4\pi} AdA + 2 CS[g]
 \end{equation}

We claim this   has a dual bosonic
description:
 \begin{equation}
 \label{2dualb0}
{\cal L}_{0b} = {\cal L}[Z_\alpha, b] + \frac{1}{2\pi} bdA -
\frac{1}{4\pi} bdb,
 \end{equation}
where $b$ is an ordinary $U(1)$ gauge field, and $Z_{\alpha}$ is a
spin-$1/2$ (under the global internal ${\mathrm{SU}(2)}$)  boson. As a check on this proposal, consider giving the
fermions a mass that preserves ${\mathrm{SU}(2)} \times U(1)$. If $m < 0$
(with our definition of the fermion determinant) we get $ \int
\frac{1}{4\pi} AdA + 2 CS[g]$ (which corresponds to a gapped phase
with $\sigma_{xy} = 1, \kappa_{xy} = 1$). If $m >0$, then we get
the same but with opposite sign ($\sigma_{xy} = -1, \kappa_{xy}/\kappa_0 =
-1$). The massless Dirac fermion sits right at the transition
between these two phases.

To match these from the boson side, if $Z_\alpha$ is in a trivial
insulator, we integrate it out first, and then integrate out $b$.
This gives $ \int \frac{1}{4\pi} AdA + 2 CS[g]$ which exactly
matches the fermion side with $m < 0$. The other phase is obtained
by putting $Z_\alpha$ in a boson integer quantum Hall state
(bIQH). Then integrating out $z_\alpha$ gives a term
$\frac{2}{4\pi}bdb$. Combining with the other terms already in the
boson action, and integrating out $b$, we get $- ( \int
\frac{1}{4\pi} AdA + 2 CS[g])$ which also exactly matches the
answer from the fermion side with $m > 0$.

Therefore what in the boson theory is the transition between a
trivial insulator and the bIQH state is dual to the free massless
Dirac fermion. This should not be surprising to us. In fact, we
already know two different forms of actions that describe bIQH
transitions: $QED_3$ with $N_f=2$ and the easy-plane $NCCP^1$. If
we use the $QED_3$ Lagrangian in Eq.~\eqref{2dualb0}, we simply
get back the free Dirac fermion theory. If we use the easy-plane
$NCCP^1$ theory instead, we obtain a ``composite boson" dual of
two free Dirac fermions discussed in Ref.~\cite{potter}. However,
in this representation the spin ${\mathrm{SU}(2)}$ symmetry is non-manifest.
Thus in the absence of a manifestly ${\mathrm{SU}(2)}$-invariant bosonic
representation of the bIQH transition, we will just keep the form
of Lagrangian in Eq.~\eqref{2dualb0} implicit.

We can also check monopole operators. A $2\pi$ monopole ${\cal M}_b$  of
$b$ carries $U_A(1)$ charge $1$, and $U_b(1)$ charge $-1$. Thus
$Z_\alpha {\cal M}_b$ has no $U_b(1)$ charge, has charge-$1$ under
$U_A(1)$, is an ${\mathrm{SU}(2)}$ doublet and is a fermion. We should
identify it with $\psi_\alpha$.


Now we follow the usual logic to generate other dualities. First
we let $A \to a$ and treat $a$ as a dynamical gauge field (spin$_c$
connection), and couple $\frac{1}{2\pi} adB + \frac{1}{4\pi} B dB$
where $B$ is an ordinary $U(1)$ gauge field. The fermion side
becomes
\begin{equation}
\label{qed2f0} {\cal L}_{1f} = i\overline{\psi}_\alpha
\slashed{D}_a \psi_\alpha  + \frac{1}{2\pi}
adB+ \frac{1}{4\pi} B dB +  2 CS[g]
 \end{equation}
In the boson side the $a$ dependence arises solely through the
term $\frac{1}{2\pi} (b + B) da$. Integrating out $a$ we set $b =
- B$ to get
\begin{equation}
 \label{2dualb1}
 {\cal L}_{1b} = {\cal L}[Z_\alpha, - B]
 \end{equation}
The two theories Eq.~\eqref{qed2f0} and \ref{2dualb1} are dual to
each other in the following sense. The phase transition in Eqn
\ref{2dualb1} between the trivial gapped phase and the bIQH state
is described by Eq.~\eqref{qed2f0} when the Dirac fermions are
massless.  Indeed we used this identification in various parts of the paper.

Now we make $B$ dynamical $B \rightarrow b$, and couple
$\frac{1}{2\pi} bdC$ where $C$ is an ordinary $U(1)$ gauge field.
In the fermion side the $b$ dependence occurs through the term
$\frac{1}{2\pi} (a+ C) db + \frac{1}{4\pi} b db$. Integrating out
$b$ leads to
  \begin{equation}
 \label{qed2f}
{\cal L}_{f} = i\overline{\psi}_\alpha D_a \psi_\alpha
-\frac{1}{4\pi}ada + \frac{1}{2\pi} adC-  \frac{1}{4\pi} C dC
 \end{equation}
The boson side becomes
 \begin{equation}
  \label{2dualb}
 {\cal L}_{b} = {\cal L}[Z_\alpha, - b]    + \frac{1}{2\pi} Cdb
 \end{equation}

This almost looks like the $NCCP^1$ model but we should remember
the precise sense in which Eqns. \ref{qed2f} and \ref{2dualb} are
dual. As in all the previous examples, the trivial to bIQH
transition of $z_\alpha$ is dual to the massless Dirac theory. Let
us interpret this phase transition in the boson side more clearly.
When $z_\alpha$ is in a trivial gapped phase, $C$ is Higgsed, and
we have broken $U_C(1)$ symmetry (a ``superfluid"). When
$z_\alpha$ is in a bIQH state, we have a $U(1)_2$ theory, and this
is really an ${\mathrm{SU}(2)}$ symmetric  chiral spin liquid where the
semion is a spin-$1/2$ spinon.

Thus the transition between a superfluid (that breaks $U(1)$ but
preserves ${\mathrm{SU}(2)}$)  and this chiral spin liquid with ${\mathrm{SU}(2)}$
symmetry is described by Eq.~\eqref{qed2f}.  It is easy to check
that this is reproduced by thinking directly about the fermions.

The duality between the theories in Eqns.
\ref{2dualb} (interpreted as above) and \ref{qed2f} should be
contrasted with the duality of the ${\mathrm{SU}(2)}$ invariant $NCCP^1$
model to the QED$_3$-GN model.   The two sets of dualities
describe two distinct phase transitions of the same underlying
spin system.  Further though both Eqn. \ref{2dualb} has the same
$SO(3) \times U(1)$ symmetry as $NCCP^1$, it - unlike $NCCP^1$ -
is not time reversal invariant.





\section{A different view of the dualities and emergent symmetries}
\label{sandt}

 Here we discuss the dualities and emergent symmetries from a point of view familiar in the high energy literature.
Caution is however needed as we describe below.

For  any $2+1$D CFT with a global $U(1)$ symmetry,
there is a formal  operation  on the path integral, denoted $S$, which is defined as follows:
\begin{equation}
\label{Sdef}
Z_S[B] = \int {\cal D}A ~ Z_{\mathrm{CFT}_1}[A] e^{\frac{i}{2\pi} \int d^3x AdB}.
\end{equation}
Here $Z_{\mathrm{CFT}_1}$[A] is the partition function of the $2+1$D CFT
in the presence of a {\em background} $U(1)$ gauge field $A$. The
operation  $S$  converts this  background gauge field  into a
dynamical one, without including a kinetic term for the field $A$. A new background $U(1)$ gauge field $B$, coupling to $dA/2\pi$ (which is conserved) is also introduced. This operation was defined and used by
Kapustin and Strassler \cite{kapstr}, and by Witten \cite{witten03}.
A different operation, $T$, was also introduced by Witten: this
simply shifts the level of the Chern-Simons term for the background
gauge field by $1$.

If the path-integral on the right hand side of Eqn. \ref{Sdef} is
well-defined, then $Z_S[B]$ is the partition function of a new
theory with a new global $U(1)$ symmetry ($B$ couples to the current of this symmetry). Further,  the theory $Z_S[B]$ is  conformally invariant, at least at the formal level --- it is to ensure
this that no kinetic term for $A$ is introduced in the definition
of $S$ --- and defines a new conformal field theory
which we denote CFT$_2$.

Schematically we write the $S$ operation  as  $S[\mathrm{CFT}_1] = \mathrm{CFT}_2$
where both CFTs have a global $U(1)$ symmetry.  The combination of
$S$ and $T$ then leads to a remarkable $SL(2,Z)$ action on the set of
$2+1$D CFTs with a global $U(1)$ symmetry \cite{witten03}. (See Refs~\cite{BurgessDolan,CARDY19821,CARDY198217,FradkinKivelsonSL2Z} for other appearances of this mathematical structure in related contexts.)
That is, $S$ and $T$ can be  shown formally to satisfy the
defining relations\footnote{Here the operation $S^2 = -1$ corresponds to changing the sign of the gauge coupling.} $S^2 = -
1$ and $\left(S T \right)^3 = 1$ of $SL(2, Z)$.

Let us think a bit more about $S$. A priori it is not evident that the path integral in Eq.~\ref{Sdef} is well-defined. To obtain some intuition, consider a modified operation which is certainly well-defined:
\begin{widetext}
\begin{equation}
\label{modS}
\widetilde{Z}_S[B; e^2] = \int {\cal D}A ~ Z_{\mathrm{CFT}_1}[A] \exp\left({-  \int d^3x \left( \frac{1}{2e^2} \left(dA\right)^2  - \frac{i}{2\pi}AdB \right)}\right).
\end{equation}
\end{widetext}
We have introduced a Maxwell term for the gauge field $A$ with coupling constant $e^2$. Formally, the original $S$ operation may be written
\begin{equation}
\label{Slim}
Z_S[B] = \lim_{e^2 \rightarrow \infty} \widetilde{Z}_S[B; e^2].
\end{equation}
However Eq.~\ref{Slim} has a more intuitive interpretation. Consider the theory at a fixed value of $e^2$. The gauge coupling introduces
a length scale $l_e \sim {1}/{e^2}$. At distances much smaller
than $l_e$ the physics is that of CFT$_1$ plus a decoupled free
photon, so this is a
`weakly' gauged version of CFT$_1$.  But the coupling between CFT$_1$ and the photon is relevant, so the physics  on distances much larger
than $l_e$ will be different. The limit $e^2 \rightarrow \infty$ is
 equivalent to describing the deep IR limit of  the $\widetilde{Z}_S$ theory, i.e. distances much greater than $l_e$.

At the formal level this deep IR limit is a new conformally invariant theory, CFT$_2$, described by $Z_S$. Formally, the relationship $S^2=-1$  (see below) also requires CFT$_2$ to be nontrivial if CFT$_1$ is. But  it is not obvious that the conclusions of these formal arguments will always hold in reality, at least for the non-supersymmetric, finite `$N$' theories of interest in this paper. For instance  if we take $CFT_1$ to be the theory of $N_f$ massless
$2$-component Dirac fermions ($N_f$ even) then we obtain  QED$_3$ for $\widetilde{Z}_S$. Whether or not this flows to a $CFT$ for general $N_f$, not necessarily large, is a long-standing issue which has  not yet been settled. For another example relevant to this paper, take CFT$_1$ to describe a pair of boson fields, each separately at the $U(1)$ Wilson Fisher fixed point, and  take $S$ to act on the diagonal $U(1)$ symmetry.  $\widetilde Z_S$ then describes the easy plane $NCCP^1$ model in a particular limit.\footnote{This limit is where the gauge coupling is much weaker than the quartic terms (so that the RG flow comes very close to the ungauged Wilson-Fisher fixed point before the gauge coupling drives it away). Also, in this example the bare quartic terms do not couple the two fields.}  As discussed in the main text, it is hardly clear that this flows to a CFT in the
IR. Similarly, if we start with the $O(4)$ Wilson-Fisher theory and use $S$ to gauge an appropriate $U(1)$ subgroup, we obtain the $SU(2)$--symmetric $NCCP^1$ model (in a similar limit). Whether or not this flows to a CFT is again a nontrivial question.\footnote{Note though that if we \textit{assume} the reliability of the $S$ operation then we can infer the existence of a CFT with the same symmetries, etc. as $NCCP^1$. It should be noted however that the formal construction  says nothing about the stability (number of relevant perturbations) of this CFT.} (Even when there is a flow to a CFT, we might expect to have to tune the coupling of any relevant symmetry-allowed operators of CFT$_1$ in order to be on this flow line, contrary to the expectation from the formal limit.)

What about the crucial relationship $S^2 = -1$? We apply the limiting procedure twice to  give the partition function
\be
\label{S2lim}
Z_{S^2}[C] = \,\lim_{e'^2 \rightarrow \infty}\,\lim_{e^2 \rightarrow \infty}\, \widetilde{Z}_{S^2}[C; e^2, e'^2],
\ee
with
\begin{widetext}
\begin{equation}
\label{modS2}
\widetilde{Z}_{S^2}[C; e^2, e'^2] = \int {\cal D}B \int {\cal D}A ~ Z_{\mathrm{CFT}_1}[A] \exp \left( 
{-  \int d^3x  \left( \frac{\left(dA\right)^2}{2e^2} + \frac{\left(dB \right)^2}{2e'^2} -\frac{i}{2\pi}   B d \left( C + A \right)   \right)  }  
\right).
\end{equation}
\end{widetext}
The proof in Ref.~\cite{witten03} of $S^2 = -1$ evaluates the path integral above in the absence of the Maxwell terms (as appropriate to the formal definition of $S$ in Eqn. \ref{Sdef}). The $B$ integral then acts as a delta function enforcing $C = -A$, and the right hand side becomes $Z_{\mathrm{CFT}_1}[-C]$.   

Even in the case where $l_e\gg l_{e'}$, 
\begin{equation}\label{other_order_of_limits}
 \lim_{e^2 \rightarrow \infty}\,\lim_{e'^2 \rightarrow \infty} \, \widetilde{Z}_{S^2}[C; e^2, e'^2],
\end{equation}
we may worry that this procedure will fail to give back the original CFT, due to  relevant terms generated by integrating out $B$ and then $A$.  For Eq.~\ref{S2lim}, where the order of limits is the opposite, it is even less clear that we will obtain $S^2=-1$. In general this is a nontrivial question about the structure of the RG flows.  

If the gauged CFT$_1$  does indeed  flow to a nontrivial CFT$_2$, then  characteristic lengthscale for this crossover is $l_e \sim {1}/{e^2}$. Gauging CFT$_2$ then introduces a new lengthscale  $l_{e'} \sim {1}/{e'^2}$, and the regime of interest is $l_{e'} \gg l_e$. In order for $S^2=-1$ to hold, the ultimate flow on scales $\gg l_{e'}$ must be to a copy of the CFT$_1$ fixed point. 

If the gauged CFT$_1$ instead flows to a trivial theory, representing for example a massive or symmetry-broken fixed point, then it is hard to see how $S^2 =-1$ can  ever be satisfied. {Here we are assuming that CFT$_1$ is nontrivial; it is certainly possible to have an example where CFT$_1$ and CFT$_2$ are both trivial, and $S^2=-1$}.\footnote{An instructive example is to take CFT$_1$ to be a trivial theory with partition function $Z_{\mathrm{CFT}_1} = 1$.  In a condensed matter context this is the fixed point theory for a bosonic Mott insulator. If we now couple the bosons to a dynamical gauge field to obtain $\widetilde{Z}_S$ we see that the new theory  is in  a phase where the new global $U(1)$ symmetry associated with conservation of $dA$ is spontaneously broken (a superfluid). If we  use the formal definition in Eq. \ref{Sdef} we get $Z_S[B] = \delta(B)$ as expected for a superfluid. In a formal sense this partition function is conformally invariant --- we may think of it as the fixed point description of a superfluid phase obtained by taking the phase stiffness of the Goldstone mode to $\infty$. Of course $Z_S$ is physically lacking as it misses the Goldstone physics.}

This discussion is intended to provide intuition for dangers which may arise in the formal use of the $S$ operation. It is of course conceivable that in practise they do not arise. We should also emphasise that the limiting procedure discussed above is not the only way to interpret the formal definition of $Z_S$: it is possible that when the above limits fail to give a nontrivial CFT, the $S$ operation can be rescued by an alternative implementation of the definition.

If we ignore all the caveats and {\em assume}  (as
is normally done in the literature) that there is a well-defined
$SL(2,Z)$ action on $2+1$D CFTs then we can make some powerful
statements. First it tells us that there is a CFT
which looks like easy plane $NCCP^1$, defined by the partition
function
\be\label{formal_easy_plane}
\int \mathcal{D}b  \, \left( Z_\mathrm{WF}[b] \right)^2 e^{i\int \frac{1}{2\pi} b dB}
\ee
where $Z_\mathrm{WF}[b]$ is the partition function for the Wilson Fisher fixed point of a single complex boson, with background gauge field $b$. A priori we do not know whether the standard easy plane $NCCP^1$ action (defined  with, e.g., an additional  Maxwell term) flows to
this CFT.

Second, if we assume that the formal $S$ operation gives a well defined action on CFTs, then it is natural to expect that basic boson-fermion duality, relating  a
single massless Dirac fermion to a Wilson-Fisher boson coupled to
a $U(1)_1$ gauge field, can be taken as an exact statement about path integrals:
\be Z_\mathrm{D}[A] = \int \mathcal{D}b \, Z_\mathrm{WF}[b] \, e^{i\int d^3x \frac{1}{4\pi} bdb
+ \frac{1}{2\pi} bdA} \ee
where $Z_D$ is the partition function of a free massless Dirac
fermion. We write this as (recall that $T$ shifts the level of the Chern Simons term)
\be
\mathrm{D} = ST[\mathrm{WF}].
\ee
The other boson-fermion duality then is
\be
\mathrm{D} = T^{-1} S^{-1} T^{-1}[\mathrm{WF}].
\ee
Multiplying the partition functions on both sides, shifting an
$AdA/4\pi$  to the left, and finally making $A$ dynamical we get
the duality of QED$_3$ to the easy plane CFT defined in (\ref{formal_easy_plane}).

The fermion side is manifestly $[{{\mathrm{SU}(2)} \times U(1)}]/{Z_2}$ invariant.
Further it is easy to see that it is exactly self-dual and the
dual side has the other $[{{\mathrm{SU}(2)} \times U(1)}]/{Z_2}$ as a manifest
symmetry. Altogether this implies the $O(4)$ symmetry. 

Within the framework of the present assumptions, these are all exact statements, regardless of  the relevance or irrelevance of operators
that break $O(4)$ to  ${\mathrm{SU}(2)} \times U(1)$ or $U(1) \times U(1)$. As mentioned above, it is possible that the fixed point we are describing is highly fine-tuned. But if  we  now make the natural \textit{further} assumption that the weak coupling limits
of all these gauge theories flow to the IR CFTs defined formally
but exactly by the path integrals above, and that there is no  fine tuning hidden in this flow, we indeed conclude
that the various symmetry-allowed perturbations are irrelevant. However we emphasize again that this view on the dualities and emergent symmetries is predicated on the reliability of the formal $SL(2,Z)$ action on $2+1$D CFTs which  as far as we are aware still remains conjectural.

\section{Some useful mathematical concepts}
\label{mathapp}

{
 
 It is convenient to consider a ``triangulation" of the   space-time manifold $M$ (we will be mostly interested in manifolds of dimension $D = 4$) : we represent points in space-time by a discrete lattice where each elementary unit is a $D$-simplex. Pick a local ordering of the vertices of the lattice. 
 A $k$-cochain lives on $k$-simplices, i.e. it is a function that depends on $(k + 1) $ vertices and takes values in some abelian group $G$. We will only need to consider the cases $Z$, $Z_2$ and $Z_4$. The corresponding cochain is then said to be an element of $C^k(M, G)$.   For instance a $2$-cochain in $C^2(M, Z_2)$ is a function $a_{ijk} = 0, 1$ while for a 2-cochain in $C^2(M, Z)$, the function $a_{ijk} \in Z$.  Here $(ijk)$ are the vertices of a triangular plaquette of the simplex.

 We can define a discrete derivative (known as a ``coboundary") operation $d$  that maps $k$-cochains to $(k+1)$-cochains:
 \begin{equation}
 (da)_ {(i_0, i_1, i_2, ....i_{k+1})} = \sum_{p = 0}^{k+1}  (-1)^p a_{i_0, i_1,......\hat{i_p},.....i_{k+1}}
 \end{equation}
 where the variable  $\hat{i_p}$ is omitted.  It is understood that the addition on the right side is performed in $G$ ({\em e.g}, mod $2$ addition for $G = Z_2$). It is readily checked that $d^2a = 0$.  The set of all $k$-cochains $a$ that satisfy $da = 0$  form a group under addition known as  the cocycle group $Z^k(M, G)$.  The set of all $a \in C^k(M, G)$ that may be written $a = db$ for some $b \in C^{(k-1)}(M, G)$ form a different group known as the coboundary group $B^k(M, G)$.  Clearly $B^k(M, G) \subset Z^k(M, G)$.  The cohomology group $H^k(M, G) = \frac{Z^k(M, G)}{B^k(M, G)}$.

 For two  2-cochains $a \in C^k(M,G)$ and $b \in C^l(M, G)$  , we define the cup product
 \begin{equation}
 (a \cup b)_{i_0, ......,i_{k+l}} = a_{i_0i_1i_2....i_{k })}b_{i_{k}i_{k+1}....i_{k+l}}
 \end{equation}
 where $i_0, \ldots i_{k+l}$ are assumed to be ordered.   The cup product satisfies
 \begin{equation}
 d(a \cup b) = da \cup b + (-1)^k a \cup db.
 \end{equation}
Clearly if $da = 0 = db$ then $d(a\cup b) = 0$. Thus the cup
product defines a product of cohomology classes. The cup product
is a generalization of the familiar wedge product of differential
forms.

The Poyntryagin square of $w \in H^2(M, Z_2)$ plays an important
role in our discussion (please see Ref.~\cite{akpsq} for
more discussion and references). We now specialize to 4-manifolds
$M = Y_4$. It is easiest to define  if $w$ can be lifted to   an
element  $\hat{w} \in H^2(Y_4, Z)$, i.e. 
$w = \hat{w}   ~(mod~ 2)$ and $d \hat{w} = 0$. In this case,  ${\cal P}(w) = \hat{w} \cup \hat{w}
(~mod~ 4)$.   If $w$ does not admit a lift to an integral cohomology class
then ${\cal P}$ is still a $mod~4$ quantity. It is defined to be
\begin{equation}
{\cal P}(w) = w \cup w + w \cup_1 dw~(mod~4)
\end{equation}
The new product $\cup_1$ is defined (for a $2$-cochain $a$ and a $3$-cochain $b$) as
\begin{equation}
(a \cup_1 b)_{01234} =  a_{034} b_{0123} + a_{014}b_{1234}
\end{equation}
It is readily seen that  ${\cal P}(w)$ transforms by a co-boundary under $w \to w + 2 n$, $w \to w + d m$, so it is well-defined on $H^2(Y_4, Z_2)$.  Note that as $w \in H^2(Y_4, Z_2)$, $d w = 0 ~(mod~2)$.  Thus we have
${\cal P}(w) = w \cup w ~(mod ~2)$.  It can be shown that
\begin{equation}
{\cal P}(w + w') = {\cal P}(w) + {\cal P}(w') + 2 w \cup w' ~(mod ~4)
\end{equation}
We will use this repeatedly.

}
  
\section{Topological superconductors and the APS theorem}
\label{app:APS}
In this appendix we review the field-theoretic description of topological superconductors in $2+1$ and $3+1$D. We follow Ref.~\cite{wittenreview} here {and adapt it to the SO(n) symmetric systems of interest in this paper}.

Let's begin with $2+1$D. A $p_x + i p_y$ superconductor can be represented by a massive 2-component Majorana fermion $\chi$:
\beq L = \bar{\chi} (D + m) \chi  \label{eq:LMajapp} \eeq
with $\bar{\chi} = \chi^T C^*$, $D = \gamma^{\mu} (\d_{\mu} + i \omega_{\mu})$,  and $C = \sigma^y$ - the charge-conjugation matrix.  $m > 0$ corresponds to the trivial superconductor and $m < 0$ to the $p_x + i p_y$ superconductor. The point $m  =0$ corresponds to the transition between these two phases. Now, the formal partition function of (\ref{eq:LMajapp}) on a closed manifold $M$ is 
\bea Z(m) &=& {\rm Pf}( C^{\dagger} (D + m)) = \pm \det(D + m)^{1/2} \nn &=&  \pm \prod_{\langle \lambda\rangle} (-i \lambda + m)^{N(\lambda)/2}\eea
where the product is over  eigenvalues $\lambda$ of the Dirac operator $i D$ (without repetitions) and $N(\lambda)$ is the multiplicity of the eigenvalue. Since $[C K, i D] = 0$ and $(CK)^2 = -1$, all eigenvalues of $i D$ are doubly degenerate. The above expression clearly requires regularization. We note that the partition function of the trivial superconductor at long wavelength (or equivalently in the $m \to \infty$ limit) is expected to be analytic in the curvature of the manifold and to have no topological terms, so it can be effectively set to $1$. It is then convenient to normalize other partitions functions by it. This can be understood as the physical justification of Pauli-Villars regularization. Then,
\beq Z(m)_{PV,+} = \lim_{M \to \infty} \frac{Z_{m}}{Z_{M}} =  \prod_{\langle \lambda\rangle} \frac{(-i \lambda + m)^{N(\lambda)/2}}{(-i \lambda + |M|)^{N(\lambda)/2}} \label{eq:Zreg} \eeq
Note that there is no sign ambiguity in Eq.~(\ref{eq:Zreg}). Indeed, we can reach any value of $m$ starting with the trivial insulator at $m  = \infty$. The requirement that the partition function during this process be analytic in $m$ removes the sign ambiguity. Now, when $m = 0$, we can write the partition function (\ref{eq:Zreg}) as $Z = |Z| e^{i \varphi}$,  
where the phase
\beq \varphi = -\frac12 \sum_{\langle \lambda\neq 0 \rangle} N(\lambda) {\rm sgn}(\lambda) \tan^{-1}{\frac{|M|}{|\lambda|}} \to  -\frac{\pi}{4} \sum_{\lambda \neq 0} {\rm sgn}(\lambda)\eeq
The sum in the last term is over all eigenvalues of $i D$ (repeated eigenvalues included) and we've taken the $M \to \infty$ limit naively. While the resulting final sum is formal, it can be equivalently regulated with the $\zeta$ function method, giving 
\beq Z(m=0)_{PV,+} = |Z(m = 0)| \exp(-\pi i \eta(i D)/4) \label{eq:Zetaapp}  \eeq 
with $\eta$ defined via Eq.~(\ref{eq:etadef}).\footnote{Note that $\eta$ in (\ref{eq:etadef}) includes a contribution $N_0$ from the zero modes  $\lambda = 0$. If there are any zero modes then the absolute value of the partition function, $|Z(m = 0)|$, vanishes and the phase is irrelevant.}  Deep in the $p_x + i p_y$ phase, we may set $m = - |M| \to -\infty$ in Eq.~(\ref{eq:Zreg}) and obtain a pure phase,
\beq Z_{p_x+i p_y} = \exp(-\pi i \eta(i D)/2) \label{eq:Zppip}\eeq

The APS theorem (\ref{eq:APS}) allows us to rewrite the partition function of a $p_x + i p_y$ superconductor, Eq.~(\ref{eq:Zppip}), as
\beq Z_{p_x + ip_y} = \exp(-i CS_g[Y_4])\eeq
where, as explained in the section \ref{sec:mathy}, the gravitational Chern-Simons term $CS_g$ (\ref{eq:CSdef}) is defined via a continuation of $M$ to an auxiliary four-manifold $Y_4$. The APS theorem guarantees that the result is independent of the continuation. The gravitational Chern-Simons term encodes precisely the thermal-Hall response of a $p_x + i p_y$ superconductor: $\frac{\kappa_{xy}}{T} = \frac12$. Note that our Majorana fermions $\chi$ require a spin structure, and the continuation of $M$ to $Y_4$ must preserve this spin structure. Thus, $CS_g[Y_4]$ is secretly spin-structure dependent (as, less surprisingly, is the $\eta$ invariant).

We can easily generalize the above discussion to $n$ identical Majorana fermions. Now we may couple the system to an $SO(n)$ gauge field $A$, so that the Dirac operator reads $D_A = \gamma^{\mu} (\d_{\mu} + i \omega_{\mu} - i A_{\mu})$.\footnote{$i A_{\mu}$ is real, so crucially, $[i D_A, CK] = 0$ still holds.} The partition function for $m \to -\infty$ (i.e. deep in the topological phase) again is (\ref{eq:Zppip}) with $\eta(iD)$ now referring to the full Dirac operator $D_A$. This can be rewritten using the APS theorem (\ref{eq:APS}) as,
\beq Z_{(p_x+ip_y)^n} = \exp\left(-i (CS_{SO(n)}[A,Y_4] + n CS_g[Y_4])\right)\eeq
with the $SO(n)$ Chern-Simons term, given by Eq.~(\ref{eq:CSdef}), again defined via the continuation to $Y_4$. Thus, $n$-copies of a $p_x + i p_y$ superconductor have $SO(n)$ response characterized by a Chern-Simons term at level $1$. For $n=2$ this simply corresponds to $\sigma_{xy} = 1$. There is also the expected thermal Hall response $\frac{\kappa_{xy}}{T} = n\times \frac12$ encoded in the gravitational Chern-Simons term.  As emphasized in section \ref{sec:mathy}, for even $n$ the Majorana fermions really see an $(SO(n) \times Spin(3)_{TM})/Z_2$ bundle, i.e. we don't need to separately specify the spin structure, but only the combination of $SO(n)$ and $Spin(3)_{TM}$ transition functions. Likewise, on $Y_4$ we again need to continue just the $(SO(n) \times Spin(3)_{TM})/Z_2$ bundle. 
Finally, the partition function for $m = 0$ (the transition point between a trivial phase and $n$ copies of a $p_x + i p_y$ superconductor) is again given by Eq.~(\ref{eq:Zetaapp}). 

Now, let us proceed to $3+1$D. A topological superconductor in class DIII can be represented by a massive (four-component) Majorana fermion. For generality, we work from the start with $n$ identical copies of a topological superconductor and couple the system to an $SO(n)$ gauge field $A$, so the continuum bulk action  is:
\beq L = \bar{\chi} (D_A + m) \chi\eeq
where again $D_{A} = \gamma^{\mu} (\d_{\mu} + i \omega - i A_{\mu})$, $\bar{\chi} = \chi^T C^*$ and $C$ is the charge-conjugation matrix. The phase with $m > 0$ may be taken (by convention) to represent the trivial superconductor and $m < 0$ - the topological superconductor. As before, we may set the partition function of the trivial phase, $m \to \infty$, to $1$, so the bulk partition function of the topological phase, $m \to -\infty$, on a closed four-manifold $X_4$ is:
\beq Z_{TSc} =  \prod_{\langle \lambda\rangle} \frac{(-i \lambda - |M|)^{N(\lambda)/2}}{(-i \lambda + |M|)^{N(\lambda)/2}}\eeq
with $\lambda$ - eigenvalues of $i D_A$. Crucially, again $[C K, iD_A] = 0$ and $(CK)^2 = -1$, so all eigenvalues are doubly degenerate. What is different compared to the previously discussed $p_x + ip_y$ case, is that  $\{\gamma^5, i D_A\} = 0$, so all non-zero eigenvalues of $i D_A$ come in pairs $\pm \lambda$; therefore, their contribution to the partition function cancels  and 
\beq Z_{TSc} = (-1)^{N_0/2}\eeq
where $N_0$ is the number of zero modes of $i D_A$. The zero modes can be chosen to be simultaneous eigenstates of $\gamma^5$. Suppose there are $N_\pm$ eigenstates with $\gamma^5 = \pm 1$. We note that $N_+$ and $N_-$ are separately even as $[CK, \gamma^5] = 0$. Therefore, we may rewrite $Z_{TSc} = (-1)^{(N_+ - N_-)/2}$. The difference $N_+ - N_- = 2 {\cal J}$ is known as the index of $i D_A$ (we include a prefactor of $2$ to emphasize that in the present situation it is even), and we may write:
\beq Z_{TSc} = (-1)^{\cal J}\eeq 
We see that the partition function is real, as it should be for a time-reversal invariant system on an orientable manifold. The Atiyah-Singer theorem\cite{Nakahara,GilkeyReview}, tells us that
\bea 2 {\cal J} &=& N_+ - N_- =  \frac{1}{\pi} (CS_{SO(n)}[A, X_4] + n CS_g[X_4]) \nn &=& p_1[A,X_4] - \frac{n \sigma[X_4]}{8} \label{eq:2J} \eea
with the Pontryagin number $p_1$ and signature $\sigma$ given by Eqs.~(\ref{eq:Pontryagin}), (\ref{eq:sigma}), so that we may re-write 
\beq Z_{TSc} = \exp\left[\frac{i}{2}(CS_{SO(n)}[A, X_4] + n CS_g[X_4])\right] \label{eq:ZTSctheta}\eeq
 When $n = 1$ (or more generally for odd $n$), we must pick a spin-structure for our fermions $\chi$ (in particular, $X_4$ must admit a spin structure); we then learn from (\ref{eq:2J}) that on a spin manifold $\sigma$ is a multiple of $16$. Furthermore, if we fix a spin-structure, $A$ is a true $SO(n)$ gauge field (with transition functions satisfying the cocycle condition), from which we learn that $p_1[A]$ on a spin manifold is even. Now, for even $n$, we don't require $X_4$ to admit a spin structure: the fermions see transition functions in $(SO(n)\times Spin(4)_{TX_4})/Z_2$ group, so while $p_1$ and $\frac{n \sigma}{8}$ themselves need not be even (in fact, $\sigma$ is an integer, so $\frac{n \sigma}{8}$ is generally a fraction), the combination $p_1 - \frac{n \sigma}{8}$ is an even integer.

Finally, let us discuss the case when the topological superconductor lives on a space $X_4$ with a boundary $M$. We know that the boundary supports $n$ gapless Majorana cones. The bulk+boundary partition function now is\cite{wittenreview}:
\beq S^{{\rm bulk+bound}}_{TSc} = \int_M [\bar{\chi} D_A \chi]_{PV,+} - \frac{i}{2} (CS_{SO(n)}[A, X_4] + n CS_g[X_4] ) \label{eq:TScbb} \eeq
The Majorana action (\ref{eq:TScbb}) depends only on the boundary data; on the other hand, the second term in Eq.~(\ref{eq:TScbb}) depends on the bulk, and in-fact, reduces to our previous expression (\ref{eq:ZTSctheta}) for a closed manifold. While each term in Eq.~(\ref{eq:TScbb}) is separately well defined, the time-reversal symmetry of (\ref{eq:TScbb}) is not obvious. However, using (\ref{eq:Zetaapp}) and the APS theorem (\ref{eq:APS}), we obtain 
\beq Z^{\rm bulk + bound}_{TSc} = |Z^{\rm bound}(m=0)| (-1)^{{\cal J}[A,X_4]}\eeq
where  $|Z^{\rm bound}(m=0)|$ is the absolute value of boundary Majorana fermion partition function, and $2 {\cal J}$ is the index of the bulk Dirac operator $i D_A$ with APS boundary conditions. The time-reversal symmetry is now manifest.

To obtain further physical intuition for the action (\ref{eq:TScbb}), we may break time-reversal symmetry on the surface with a mass term $m \bar{\chi} \chi$. Integrating the Majorana fermions out, by our preceding discussion we then obtain at long wave-length
\beq S =  \mp  \frac{i {\rm sgn}(m)}{2} (CS_{SO(n)}[A, X_4] + n CS_g[X_4] )\eeq
i.e. the surface has $SO(n)$ Chern-Simons response at level $1/2$ and thermal-Hall response with $\kappa_{xy}/T = n\times\frac{1}{4}$. This is precisely what we expect for the $T$-broken surface state of $n$-copies of a topological superconductor.

\bibliography{SO(5)}

\begin{thebibliography}{115}%
\makeatletter
\providecommand \@ifxundefined [1]{%
 \@ifx{#1\undefined}
}%
\providecommand \@ifnum [1]{%
 \ifnum #1\expandafter \@firstoftwo
 \else \expandafter \@secondoftwo
 \fi
}%
\providecommand \@ifx [1]{%
 \ifx #1\expandafter \@firstoftwo
 \else \expandafter \@secondoftwo
 \fi
}%
\providecommand \natexlab [1]{#1}%
\providecommand \enquote  [1]{``#1''}%
\providecommand \bibnamefont  [1]{#1}%
\providecommand \bibfnamefont [1]{#1}%
\providecommand \citenamefont [1]{#1}%
\providecommand \href@noop [0]{\@secondoftwo}%
\providecommand \href [0]{\begingroup \@sanitize@url \@href}%
\providecommand \@href[1]{\@@startlink{#1}\@@href}%
\providecommand \@@href[1]{\endgroup#1\@@endlink}%
\providecommand \@sanitize@url [0]{\catcode `\\12\catcode `\$12\catcode
  `\&12\catcode `\#12\catcode `\^12\catcode `\_12\catcode `\%12\relax}%
\providecommand \@@startlink[1]{}%
\providecommand \@@endlink[0]{}%
\providecommand \url  [0]{\begingroup\@sanitize@url \@url }%
\providecommand \@url [1]{\endgroup\@href {#1}{\urlprefix }}%
\providecommand \urlprefix  [0]{URL }%
\providecommand \Eprint [0]{\href }%
\providecommand \doibase [0]{http://dx.doi.org/}%
\providecommand \selectlanguage [0]{\@gobble}%
\providecommand \bibinfo  [0]{\@secondoftwo}%
\providecommand \bibfield  [0]{\@secondoftwo}%
\providecommand \translation [1]{[#1]}%
\providecommand \BibitemOpen [0]{}%
\providecommand \bibitemStop [0]{}%
\providecommand \bibitemNoStop [0]{.\EOS\space}%
\providecommand \EOS [0]{\spacefactor3000\relax}%
\providecommand \BibitemShut  [1]{\csname bibitem#1\endcsname}%
\let\auto@bib@innerbib\@empty
\bibitem [{\citenamefont {Peskin}(1978)}]{Peskin}%
  \BibitemOpen
  \bibfield  {author} {\bibinfo {author} {\bibfnamefont {Michael~E}\
  \bibnamefont {Peskin}},\ }\bibfield  {title} {\enquote {\bibinfo {title}
  {Mandelstam-'t hooft duality in abelian lattice models},}\ }\href {\doibase
  http://dx.doi.org/10.1016/0003-4916(78)90252-X} {\bibfield  {journal}
  {\bibinfo  {journal} {Annals of Physics}\ }\textbf {\bibinfo {volume}
  {113}},\ \bibinfo {pages} {122 -- 152} (\bibinfo {year} {1978})}\BibitemShut
  {NoStop}%
\bibitem [{\citenamefont {Dasgupta}\ and\ \citenamefont
  {Halperin}(1981)}]{bosonvortexdh}%
  \BibitemOpen
  \bibfield  {author} {\bibinfo {author} {\bibfnamefont {C.}~\bibnamefont
  {Dasgupta}}\ and\ \bibinfo {author} {\bibfnamefont {B.~I.}\ \bibnamefont
  {Halperin}},\ }\bibfield  {title} {\enquote {\bibinfo {title} {Phase
  transition in a lattice model of superconductivity},}\ }\href {\doibase
  10.1103/PhysRevLett.47.1556} {\bibfield  {journal} {\bibinfo  {journal}
  {Phys. Rev. Lett.}\ }\textbf {\bibinfo {volume} {47}},\ \bibinfo {pages}
  {1556--1560} (\bibinfo {year} {1981})}\BibitemShut {NoStop}%
\bibitem [{\citenamefont {Fisher}\ and\ \citenamefont
  {Lee}(1989)}]{bosonvortexfl}%
  \BibitemOpen
  \bibfield  {author} {\bibinfo {author} {\bibfnamefont {Matthew P.~A.}\
  \bibnamefont {Fisher}}\ and\ \bibinfo {author} {\bibfnamefont {D.~H.}\
  \bibnamefont {Lee}},\ }\bibfield  {title} {\enquote {\bibinfo {title}
  {Correspondence between two-dimensional bosons and a bulk superconductor in a
  magnetic field},}\ }\href {\doibase 10.1103/PhysRevB.39.2756} {\bibfield
  {journal} {\bibinfo  {journal} {Phys. Rev. B}\ }\textbf {\bibinfo {volume}
  {39}},\ \bibinfo {pages} {2756--2759} (\bibinfo {year} {1989})}\BibitemShut
  {NoStop}%
\bibitem [{\citenamefont {Senthil}\ \emph
  {et~al.}(2004{\natexlab{a}})\citenamefont {Senthil}, \citenamefont
  {Vishwanath}, \citenamefont {Balents}, \citenamefont {Sachdev},\ and\
  \citenamefont {Fisher}}]{deccp}%
  \BibitemOpen
  \bibfield  {author} {\bibinfo {author} {\bibfnamefont {T.}~\bibnamefont
  {Senthil}}, \bibinfo {author} {\bibfnamefont {Ashvin}\ \bibnamefont
  {Vishwanath}}, \bibinfo {author} {\bibfnamefont {Leon}\ \bibnamefont
  {Balents}}, \bibinfo {author} {\bibfnamefont {Subir}\ \bibnamefont
  {Sachdev}}, \ and\ \bibinfo {author} {\bibfnamefont {Matthew P.~A.}\
  \bibnamefont {Fisher}},\ }\bibfield  {title} {\enquote {\bibinfo {title}
  {Deconfined quantum critical points},}\ }\href@noop {} {\bibfield  {journal}
  {\bibinfo  {journal} {Science}\ }\textbf {\bibinfo {volume} {303}},\ \bibinfo
  {pages} {1490} (\bibinfo {year} {2004}{\natexlab{a}})}\BibitemShut {NoStop}%
\bibitem [{\citenamefont {Senthil}\ \emph
  {et~al.}(2004{\natexlab{b}})\citenamefont {Senthil}, \citenamefont {Balents},
  \citenamefont {Sachdev}, \citenamefont {Vishwanath},\ and\ \citenamefont
  {Fisher}}]{deccplong}%
  \BibitemOpen
  \bibfield  {author} {\bibinfo {author} {\bibfnamefont {T.}~\bibnamefont
  {Senthil}}, \bibinfo {author} {\bibfnamefont {Leon}\ \bibnamefont {Balents}},
  \bibinfo {author} {\bibfnamefont {Subir}\ \bibnamefont {Sachdev}}, \bibinfo
  {author} {\bibfnamefont {Ashvin}\ \bibnamefont {Vishwanath}}, \ and\ \bibinfo
  {author} {\bibfnamefont {Matthew P.~A.}\ \bibnamefont {Fisher}},\ }\bibfield
  {title} {\enquote {\bibinfo {title} {Quantum criticality beyond the
  landau-ginzburg-wilson paradigm},}\ }\href {\doibase
  10.1103/PhysRevB.70.144407} {\bibfield  {journal} {\bibinfo  {journal} {Phys.
  Rev. B}\ }\textbf {\bibinfo {volume} {70}},\ \bibinfo {pages} {144407}
  (\bibinfo {year} {2004}{\natexlab{b}})}\BibitemShut {NoStop}%
\bibitem [{\citenamefont {Sandvik}(2007)}]{SandvikJQ}%
  \BibitemOpen
  \bibfield  {author} {\bibinfo {author} {\bibfnamefont {Anders~W.}\
  \bibnamefont {Sandvik}},\ }\bibfield  {title} {\enquote {\bibinfo {title}
  {Evidence for deconfined quantum criticality in a two-dimensional heisenberg
  model with four-spin interactions},}\ }\href {\doibase
  10.1103/PhysRevLett.98.227202} {\bibfield  {journal} {\bibinfo  {journal}
  {Phys. Rev. Lett.}\ }\textbf {\bibinfo {volume} {98}},\ \bibinfo {pages}
  {227202} (\bibinfo {year} {2007})}\BibitemShut {NoStop}%
\bibitem [{\citenamefont {Melko}\ and\ \citenamefont
  {Kaul}(2008)}]{melkokaulfan}%
  \BibitemOpen
  \bibfield  {author} {\bibinfo {author} {\bibfnamefont {Roger~G.}\
  \bibnamefont {Melko}}\ and\ \bibinfo {author} {\bibfnamefont {Ribhu~K.}\
  \bibnamefont {Kaul}},\ }\bibfield  {title} {\enquote {\bibinfo {title}
  {Scaling in the fan of an unconventional quantum critical point},}\ }\href
  {\doibase 10.1103/PhysRevLett.100.017203} {\bibfield  {journal} {\bibinfo
  {journal} {Phys. Rev. Lett.}\ }\textbf {\bibinfo {volume} {100}},\ \bibinfo
  {pages} {017203} (\bibinfo {year} {2008})}\BibitemShut {NoStop}%
\bibitem [{\citenamefont {Lou}\ \emph {et~al.}(2009)\citenamefont {Lou},
  \citenamefont {Sandvik},\ and\ \citenamefont
  {Kawashima}}]{lousandvikkawashima}%
  \BibitemOpen
  \bibfield  {author} {\bibinfo {author} {\bibfnamefont {Jie}\ \bibnamefont
  {Lou}}, \bibinfo {author} {\bibfnamefont {Anders~W.}\ \bibnamefont
  {Sandvik}}, \ and\ \bibinfo {author} {\bibfnamefont {Naoki}\ \bibnamefont
  {Kawashima}},\ }\bibfield  {title} {\enquote {\bibinfo {title}
  {Antiferromagnetic to valence-bond-solid transitions in two-dimensional
  $\text{SU}(n)$ heisenberg models with multispin interactions},}\ }\href
  {\doibase 10.1103/PhysRevB.80.180414} {\bibfield  {journal} {\bibinfo
  {journal} {Phys. Rev. B}\ }\textbf {\bibinfo {volume} {80}},\ \bibinfo
  {pages} {180414} (\bibinfo {year} {2009})}\BibitemShut {NoStop}%
\bibitem [{\citenamefont {Banerjee}\ \emph {et~al.}(2010)\citenamefont
  {Banerjee}, \citenamefont {Damle},\ and\ \citenamefont
  {Alet}}]{Banerjeeetal}%
  \BibitemOpen
  \bibfield  {author} {\bibinfo {author} {\bibfnamefont {Argha}\ \bibnamefont
  {Banerjee}}, \bibinfo {author} {\bibfnamefont {Kedar}\ \bibnamefont {Damle}},
  \ and\ \bibinfo {author} {\bibfnamefont {Fabien}\ \bibnamefont {Alet}},\
  }\bibfield  {title} {\enquote {\bibinfo {title} {Impurity spin texture at a
  deconfined quantum critical point},}\ }\href {\doibase
  10.1103/PhysRevB.82.155139} {\bibfield  {journal} {\bibinfo  {journal} {Phys.
  Rev. B}\ }\textbf {\bibinfo {volume} {82}},\ \bibinfo {pages} {155139}
  (\bibinfo {year} {2010})}\BibitemShut {NoStop}%
\bibitem [{\citenamefont {Sandvik}(2010)}]{Sandviklogs}%
  \BibitemOpen
  \bibfield  {author} {\bibinfo {author} {\bibfnamefont {Anders~W.}\
  \bibnamefont {Sandvik}},\ }\bibfield  {title} {\enquote {\bibinfo {title}
  {Continuous quantum phase transition between an antiferromagnet and a
  valence-bond solid in two dimensions: Evidence for logarithmic corrections to
  scaling},}\ }\href {\doibase 10.1103/PhysRevLett.104.177201} {\bibfield
  {journal} {\bibinfo  {journal} {Phys. Rev. Lett.}\ }\textbf {\bibinfo
  {volume} {104}},\ \bibinfo {pages} {177201} (\bibinfo {year}
  {2010})}\BibitemShut {NoStop}%
\bibitem [{\citenamefont {Harada}\ \emph {et~al.}(2013)\citenamefont {Harada},
  \citenamefont {Suzuki}, \citenamefont {Okubo}, \citenamefont {Matsuo},
  \citenamefont {Lou}, \citenamefont {Watanabe}, \citenamefont {Todo},\ and\
  \citenamefont {Kawashima}}]{Kawashimadeconfinedcriticality}%
  \BibitemOpen
  \bibfield  {author} {\bibinfo {author} {\bibfnamefont {Kenji}\ \bibnamefont
  {Harada}}, \bibinfo {author} {\bibfnamefont {Takafumi}\ \bibnamefont
  {Suzuki}}, \bibinfo {author} {\bibfnamefont {Tsuyoshi}\ \bibnamefont
  {Okubo}}, \bibinfo {author} {\bibfnamefont {Haruhiko}\ \bibnamefont
  {Matsuo}}, \bibinfo {author} {\bibfnamefont {Jie}\ \bibnamefont {Lou}},
  \bibinfo {author} {\bibfnamefont {Hiroshi}\ \bibnamefont {Watanabe}},
  \bibinfo {author} {\bibfnamefont {Synge}\ \bibnamefont {Todo}}, \ and\
  \bibinfo {author} {\bibfnamefont {Naoki}\ \bibnamefont {Kawashima}},\
  }\bibfield  {title} {\enquote {\bibinfo {title} {Possibility of deconfined
  criticality in su($n$) heisenberg models at small $n$},}\ }\href {\doibase
  10.1103/PhysRevB.88.220408} {\bibfield  {journal} {\bibinfo  {journal} {Phys.
  Rev. B}\ }\textbf {\bibinfo {volume} {88}},\ \bibinfo {pages} {220408}
  (\bibinfo {year} {2013})}\BibitemShut {NoStop}%
\bibitem [{\citenamefont {Jiang}\ \emph {et~al.}(2008)\citenamefont {Jiang},
  \citenamefont {Nyfeler}, \citenamefont {Chandrasekharan},\ and\ \citenamefont
  {Wiese}}]{Jiangetal}%
  \BibitemOpen
  \bibfield  {author} {\bibinfo {author} {\bibfnamefont {F-J}\ \bibnamefont
  {Jiang}}, \bibinfo {author} {\bibfnamefont {M}~\bibnamefont {Nyfeler}},
  \bibinfo {author} {\bibfnamefont {S}~\bibnamefont {Chandrasekharan}}, \ and\
  \bibinfo {author} {\bibfnamefont {U-J}\ \bibnamefont {Wiese}},\ }\bibfield
  {title} {\enquote {\bibinfo {title} {From an antiferromagnet to a valence
  bond solid: evidence for a first-order phase transition},}\ }\href
  {http://stacks.iop.org/1742-5468/2008/i=02/a=P02009} {\bibfield  {journal}
  {\bibinfo  {journal} {Journal of Statistical Mechanics: Theory and
  Experiment}\ }\textbf {\bibinfo {volume} {2008}},\ \bibinfo {pages} {P02009}
  (\bibinfo {year} {2008})}\BibitemShut {NoStop}%
\bibitem [{\citenamefont {Chen}\ \emph {et~al.}(2013)\citenamefont {Chen},
  \citenamefont {Huang}, \citenamefont {Deng}, \citenamefont {Kuklov},
  \citenamefont {Prokof'ev},\ and\ \citenamefont
  {Svistunov}}]{deconfinedcriticalityflowJQ}%
  \BibitemOpen
  \bibfield  {author} {\bibinfo {author} {\bibfnamefont {Kun}\ \bibnamefont
  {Chen}}, \bibinfo {author} {\bibfnamefont {Yuan}\ \bibnamefont {Huang}},
  \bibinfo {author} {\bibfnamefont {Youjin}\ \bibnamefont {Deng}}, \bibinfo
  {author} {\bibfnamefont {A.~B.}\ \bibnamefont {Kuklov}}, \bibinfo {author}
  {\bibfnamefont {N.~V.}\ \bibnamefont {Prokof'ev}}, \ and\ \bibinfo {author}
  {\bibfnamefont {B.~V.}\ \bibnamefont {Svistunov}},\ }\bibfield  {title}
  {\enquote {\bibinfo {title} {Deconfined criticality flow in the heisenberg
  model with ring-exchange interactions},}\ }\href {\doibase
  10.1103/PhysRevLett.110.185701} {\bibfield  {journal} {\bibinfo  {journal}
  {Phys. Rev. Lett.}\ }\textbf {\bibinfo {volume} {110}},\ \bibinfo {pages}
  {185701} (\bibinfo {year} {2013})}\BibitemShut {NoStop}%
\bibitem [{\citenamefont {Nahum}\ \emph
  {et~al.}(2015{\natexlab{a}})\citenamefont {Nahum}, \citenamefont {Chalker},
  \citenamefont {Serna}, \citenamefont {Ortu\~no},\ and\ \citenamefont
  {Somoza}}]{DCPscalingviolations}%
  \BibitemOpen
  \bibfield  {author} {\bibinfo {author} {\bibfnamefont {Adam}\ \bibnamefont
  {Nahum}}, \bibinfo {author} {\bibfnamefont {J.~T.}\ \bibnamefont {Chalker}},
  \bibinfo {author} {\bibfnamefont {P.}~\bibnamefont {Serna}}, \bibinfo
  {author} {\bibfnamefont {M.}~\bibnamefont {Ortu\~no}}, \ and\ \bibinfo
  {author} {\bibfnamefont {A.~M.}\ \bibnamefont {Somoza}},\ }\bibfield  {title}
  {\enquote {\bibinfo {title} {Deconfined quantum criticality, scaling
  violations, and classical loop models},}\ }\href {\doibase
  10.1103/PhysRevX.5.041048} {\bibfield  {journal} {\bibinfo  {journal} {Phys.
  Rev. X}\ }\textbf {\bibinfo {volume} {5}},\ \bibinfo {pages} {041048}
  (\bibinfo {year} {2015}{\natexlab{a}})}\BibitemShut {NoStop}%
\bibitem [{\citenamefont {Nahum}\ \emph
  {et~al.}(2015{\natexlab{b}})\citenamefont {Nahum}, \citenamefont {Serna},
  \citenamefont {Chalker}, \citenamefont {Ortu\~no},\ and\ \citenamefont
  {Somoza}}]{emergentso5}%
  \BibitemOpen
  \bibfield  {author} {\bibinfo {author} {\bibfnamefont {Adam}\ \bibnamefont
  {Nahum}}, \bibinfo {author} {\bibfnamefont {P.}~\bibnamefont {Serna}},
  \bibinfo {author} {\bibfnamefont {J.~T.}\ \bibnamefont {Chalker}}, \bibinfo
  {author} {\bibfnamefont {M.}~\bibnamefont {Ortu\~no}}, \ and\ \bibinfo
  {author} {\bibfnamefont {A.~M.}\ \bibnamefont {Somoza}},\ }\bibfield  {title}
  {\enquote {\bibinfo {title} {Emergent so(5) symmetry at the n\'eel to
  valence-bond-solid transition},}\ }\href {\doibase
  10.1103/PhysRevLett.115.267203} {\bibfield  {journal} {\bibinfo  {journal}
  {Phys. Rev. Lett.}\ }\textbf {\bibinfo {volume} {115}},\ \bibinfo {pages}
  {267203} (\bibinfo {year} {2015}{\natexlab{b}})}\BibitemShut {NoStop}%
\bibitem [{\citenamefont {{Motrunich}}\ and\ \citenamefont
  {{Vishwanath}}(2008)}]{MotrunichVishwanath2}%
  \BibitemOpen
  \bibfield  {author} {\bibinfo {author} {\bibfnamefont {O.~I.}\ \bibnamefont
  {{Motrunich}}}\ and\ \bibinfo {author} {\bibfnamefont {A.}~\bibnamefont
  {{Vishwanath}}},\ }\bibfield  {title} {\enquote {\bibinfo {title}
  {{Comparative study of Higgs transition in one-component and two-component
  lattice superconductor models}},}\ }\href@noop {} {\bibfield  {journal}
  {\bibinfo  {journal} {ArXiv e-prints}\ } (\bibinfo {year} {2008})},\ \Eprint
  {http://arxiv.org/abs/0805.1494} {arXiv:0805.1494 [cond-mat.stat-mech]}
  \BibitemShut {NoStop}%
\bibitem [{\citenamefont {Kuklov}\ \emph {et~al.}(2008)\citenamefont {Kuklov},
  \citenamefont {Matsumoto}, \citenamefont {Prokof'ev}, \citenamefont
  {Svistunov},\ and\ \citenamefont {Troyer}}]{kuklovetalDCPSU(2)}%
  \BibitemOpen
  \bibfield  {author} {\bibinfo {author} {\bibfnamefont {A.~B.}\ \bibnamefont
  {Kuklov}}, \bibinfo {author} {\bibfnamefont {M.}~\bibnamefont {Matsumoto}},
  \bibinfo {author} {\bibfnamefont {N.~V.}\ \bibnamefont {Prokof'ev}}, \bibinfo
  {author} {\bibfnamefont {B.~V.}\ \bibnamefont {Svistunov}}, \ and\ \bibinfo
  {author} {\bibfnamefont {M.}~\bibnamefont {Troyer}},\ }\bibfield  {title}
  {\enquote {\bibinfo {title} {Deconfined criticality: Generic first-order
  transition in the su(2) symmetry case},}\ }\href {\doibase
  10.1103/PhysRevLett.101.050405} {\bibfield  {journal} {\bibinfo  {journal}
  {Phys. Rev. Lett.}\ }\textbf {\bibinfo {volume} {101}},\ \bibinfo {pages}
  {050405} (\bibinfo {year} {2008})}\BibitemShut {NoStop}%
\bibitem [{\citenamefont {Bartosch}(2013)}]{Bartosch}%
  \BibitemOpen
  \bibfield  {author} {\bibinfo {author} {\bibfnamefont {Lorenz}\ \bibnamefont
  {Bartosch}},\ }\bibfield  {title} {\enquote {\bibinfo {title} {Corrections to
  scaling in the critical theory of deconfined criticality},}\ }\href {\doibase
  10.1103/PhysRevB.88.195140} {\bibfield  {journal} {\bibinfo  {journal} {Phys.
  Rev. B}\ }\textbf {\bibinfo {volume} {88}},\ \bibinfo {pages} {195140}
  (\bibinfo {year} {2013})}\BibitemShut {NoStop}%
\bibitem [{\citenamefont {Charrier}\ \emph {et~al.}(2008)\citenamefont
  {Charrier}, \citenamefont {Alet},\ and\ \citenamefont
  {Pujol}}]{CharrierAletPujol}%
  \BibitemOpen
  \bibfield  {author} {\bibinfo {author} {\bibfnamefont {D.}~\bibnamefont
  {Charrier}}, \bibinfo {author} {\bibfnamefont {F.}~\bibnamefont {Alet}}, \
  and\ \bibinfo {author} {\bibfnamefont {P.}~\bibnamefont {Pujol}},\ }\bibfield
   {title} {\enquote {\bibinfo {title} {Gauge theory picture of an ordering
  transition in a dimer model},}\ }\href {\doibase
  10.1103/PhysRevLett.101.167205} {\bibfield  {journal} {\bibinfo  {journal}
  {Phys. Rev. Lett.}\ }\textbf {\bibinfo {volume} {101}},\ \bibinfo {pages}
  {167205} (\bibinfo {year} {2008})}\BibitemShut {NoStop}%
\bibitem [{\citenamefont {Chen}\ \emph {et~al.}(2009)\citenamefont {Chen},
  \citenamefont {Gukelberger}, \citenamefont {Trebst}, \citenamefont {Alet},\
  and\ \citenamefont {Balents}}]{Chenetal}%
  \BibitemOpen
  \bibfield  {author} {\bibinfo {author} {\bibfnamefont {Gang}\ \bibnamefont
  {Chen}}, \bibinfo {author} {\bibfnamefont {Jan}\ \bibnamefont {Gukelberger}},
  \bibinfo {author} {\bibfnamefont {Simon}\ \bibnamefont {Trebst}}, \bibinfo
  {author} {\bibfnamefont {Fabien}\ \bibnamefont {Alet}}, \ and\ \bibinfo
  {author} {\bibfnamefont {Leon}\ \bibnamefont {Balents}},\ }\bibfield  {title}
  {\enquote {\bibinfo {title} {Coulomb gas transitions in three-dimensional
  classical dimer models},}\ }\href {\doibase 10.1103/PhysRevB.80.045112}
  {\bibfield  {journal} {\bibinfo  {journal} {Phys. Rev. B}\ }\textbf {\bibinfo
  {volume} {80}},\ \bibinfo {pages} {045112} (\bibinfo {year}
  {2009})}\BibitemShut {NoStop}%
\bibitem [{\citenamefont {Charrier}\ and\ \citenamefont
  {Alet}(2010)}]{Aletextendeddimer}%
  \BibitemOpen
  \bibfield  {author} {\bibinfo {author} {\bibfnamefont {D.}~\bibnamefont
  {Charrier}}\ and\ \bibinfo {author} {\bibfnamefont {F.}~\bibnamefont
  {Alet}},\ }\bibfield  {title} {\enquote {\bibinfo {title} {Phase diagram of
  an extended classical dimer model},}\ }\href {\doibase
  10.1103/PhysRevB.82.014429} {\bibfield  {journal} {\bibinfo  {journal} {Phys.
  Rev. B}\ }\textbf {\bibinfo {volume} {82}},\ \bibinfo {pages} {014429}
  (\bibinfo {year} {2010})}\BibitemShut {NoStop}%
\bibitem [{\citenamefont {Sreejith}\ and\ \citenamefont
  {Powell}(2015)}]{powellmonopole}%
  \BibitemOpen
  \bibfield  {author} {\bibinfo {author} {\bibfnamefont {G.~J.}\ \bibnamefont
  {Sreejith}}\ and\ \bibinfo {author} {\bibfnamefont {Stephen}\ \bibnamefont
  {Powell}},\ }\bibfield  {title} {\enquote {\bibinfo {title} {Scaling
  dimensions of higher-charge monopoles at deconfined critical points},}\
  }\href {\doibase 10.1103/PhysRevB.92.184413} {\bibfield  {journal} {\bibinfo
  {journal} {Phys. Rev. B}\ }\textbf {\bibinfo {volume} {92}},\ \bibinfo
  {pages} {184413} (\bibinfo {year} {2015})}\BibitemShut {NoStop}%
\bibitem [{\citenamefont {Shao}\ \emph {et~al.}(2016)\citenamefont {Shao},
  \citenamefont {Guo},\ and\ \citenamefont {Sandvik}}]{sandvik2parameter}%
  \BibitemOpen
  \bibfield  {author} {\bibinfo {author} {\bibfnamefont {Hui}\ \bibnamefont
  {Shao}}, \bibinfo {author} {\bibfnamefont {Wenan}\ \bibnamefont {Guo}}, \
  and\ \bibinfo {author} {\bibfnamefont {Anders~W.}\ \bibnamefont {Sandvik}},\
  }\bibfield  {title} {\enquote {\bibinfo {title} {Quantum criticality with two
  length scales},}\ }\href {\doibase 10.1126/science.aad5007} {\bibfield
  {journal} {\bibinfo  {journal} {Science}\ }\textbf {\bibinfo {volume}
  {352}},\ \bibinfo {pages} {213--216} (\bibinfo {year} {2016})},\ \Eprint
  {http://arxiv.org/abs/http://science.sciencemag.org/content/352/6282/213.full.pdf}
  {http://science.sciencemag.org/content/352/6282/213.full.pdf} \BibitemShut
  {NoStop}%
\bibitem [{\citenamefont {Simmons-Duffin}()}]{SimmonsDuffinSO(5)}%
  \BibitemOpen
  \bibfield  {author} {\bibinfo {author} {\bibfnamefont {D.}~\bibnamefont
  {Simmons-Duffin}},\ }\href@noop {} {}\bibinfo {note}
  {Unpublished}\BibitemShut {NoStop}%
\bibitem [{\citenamefont {Nakayama}\ and\ \citenamefont
  {Ohtsuki}(2016)}]{Nakayama}%
  \BibitemOpen
  \bibfield  {author} {\bibinfo {author} {\bibfnamefont {Yu}~\bibnamefont
  {Nakayama}}\ and\ \bibinfo {author} {\bibfnamefont {Tomoki}\ \bibnamefont
  {Ohtsuki}},\ }\bibfield  {title} {\enquote {\bibinfo {title} {Necessary
  condition for emergent symmetry from the conformal bootstrap},}\ }\href
  {\doibase 10.1103/PhysRevLett.117.131601} {\bibfield  {journal} {\bibinfo
  {journal} {Phys. Rev. Lett.}\ }\textbf {\bibinfo {volume} {117}},\ \bibinfo
  {pages} {131601} (\bibinfo {year} {2016})}\BibitemShut {NoStop}%
\bibitem [{\citenamefont {Tanaka}\ and\ \citenamefont {Hu}(2005)}]{tanakahu}%
  \BibitemOpen
  \bibfield  {author} {\bibinfo {author} {\bibfnamefont {Akihiro}\ \bibnamefont
  {Tanaka}}\ and\ \bibinfo {author} {\bibfnamefont {Xiao}\ \bibnamefont {Hu}},\
  }\bibfield  {title} {\enquote {\bibinfo {title} {Many-body spin berry phases
  emerging from the $\ensuremath{\pi}$-flux state: Competition between
  antiferromagnetism and the valence-bond-solid state},}\ }\href {\doibase
  10.1103/PhysRevLett.95.036402} {\bibfield  {journal} {\bibinfo  {journal}
  {Phys. Rev. Lett.}\ }\textbf {\bibinfo {volume} {95}},\ \bibinfo {pages}
  {036402} (\bibinfo {year} {2005})}\BibitemShut {NoStop}%
\bibitem [{\citenamefont {Senthil}\ and\ \citenamefont
  {Fisher}(2006)}]{tsmpaf06}%
  \BibitemOpen
  \bibfield  {author} {\bibinfo {author} {\bibfnamefont {T.}~\bibnamefont
  {Senthil}}\ and\ \bibinfo {author} {\bibfnamefont {Matthew P.~A.}\
  \bibnamefont {Fisher}},\ }\bibfield  {title} {\enquote {\bibinfo {title}
  {Competing orders, nonlinear sigma models, and topological terms in quantum
  magnets},}\ }\href {\doibase 10.1103/PhysRevB.74.064405} {\bibfield
  {journal} {\bibinfo  {journal} {Phys. Rev. B}\ }\textbf {\bibinfo {volume}
  {74}},\ \bibinfo {pages} {064405} (\bibinfo {year} {2006})}\BibitemShut
  {NoStop}%
\bibitem [{\citenamefont {Wang}\ and\ \citenamefont
  {Senthil}(2015)}]{wangsenthil15b}%
  \BibitemOpen
  \bibfield  {author} {\bibinfo {author} {\bibfnamefont {Chong}\ \bibnamefont
  {Wang}}\ and\ \bibinfo {author} {\bibfnamefont {T.}~\bibnamefont {Senthil}},\
  }\bibfield  {title} {\enquote {\bibinfo {title} {Dual dirac liquid on the
  surface of the electron topological insulator},}\ }\href {\doibase
  10.1103/PhysRevX.5.041031} {\bibfield  {journal} {\bibinfo  {journal} {Phys.
  Rev. X}\ }\textbf {\bibinfo {volume} {5}},\ \bibinfo {pages} {041031}
  (\bibinfo {year} {2015})}\BibitemShut {NoStop}%
\bibitem [{\citenamefont {{Metlitski}}\ and\ \citenamefont
  {{Vishwanath}}(2016)}]{MaxAshvin15}%
  \BibitemOpen
  \bibfield  {author} {\bibinfo {author} {\bibfnamefont {M.~A.}\ \bibnamefont
  {{Metlitski}}}\ and\ \bibinfo {author} {\bibfnamefont {A.}~\bibnamefont
  {{Vishwanath}}},\ }\bibfield  {title} {\enquote {\bibinfo {title}
  {{Particle-vortex duality of two-dimensional Dirac fermion from
  electric-magnetic duality of three-dimensional topological insulators}},}\
  }\href {\doibase 10.1103/PhysRevB.93.245151} {\bibfield  {journal} {\bibinfo
  {journal} {\prb}\ }\textbf {\bibinfo {volume} {93}},\ \bibinfo {eid} {245151}
  (\bibinfo {year} {2016})},\ \Eprint {http://arxiv.org/abs/1505.05142}
  {arXiv:1505.05142 [cond-mat.str-el]} \BibitemShut {NoStop}%
\bibitem [{\citenamefont {Mross}\ \emph {et~al.}(2016)\citenamefont {Mross},
  \citenamefont {Alicea},\ and\ \citenamefont {Motrunich}}]{dualdrMAM}%
  \BibitemOpen
  \bibfield  {author} {\bibinfo {author} {\bibfnamefont {David~F.}\
  \bibnamefont {Mross}}, \bibinfo {author} {\bibfnamefont {Jason}\ \bibnamefont
  {Alicea}}, \ and\ \bibinfo {author} {\bibfnamefont {Olexei~I.}\ \bibnamefont
  {Motrunich}},\ }\bibfield  {title} {\enquote {\bibinfo {title} {Explicit
  derivation of duality between a free dirac cone and quantum electrodynamics
  in ($2+1$) dimensions},}\ }\href {\doibase 10.1103/PhysRevLett.117.016802}
  {\bibfield  {journal} {\bibinfo  {journal} {Phys. Rev. Lett.}\ }\textbf
  {\bibinfo {volume} {117}},\ \bibinfo {pages} {016802} (\bibinfo {year}
  {2016})}\BibitemShut {NoStop}%
\bibitem [{\citenamefont {Xu}\ and\ \citenamefont {You}(2015)}]{qeddual}%
  \BibitemOpen
  \bibfield  {author} {\bibinfo {author} {\bibfnamefont {Cenke}\ \bibnamefont
  {Xu}}\ and\ \bibinfo {author} {\bibfnamefont {Yi-Zhuang}\ \bibnamefont
  {You}},\ }\bibfield  {title} {\enquote {\bibinfo {title} {Self-dual quantum
  electrodynamics as boundary state of the three-dimensional bosonic
  topological insulator},}\ }\href {\doibase 10.1103/PhysRevB.92.220416}
  {\bibfield  {journal} {\bibinfo  {journal} {Phys. Rev. B}\ }\textbf {\bibinfo
  {volume} {92}},\ \bibinfo {pages} {220416} (\bibinfo {year} {2015})},\
  \Eprint {http://arxiv.org/abs/cond-mat/0010440} {cond-mat/0010440}
  \BibitemShut {NoStop}%
\bibitem [{\citenamefont {Seiberg}\ \emph {et~al.}(2016)\citenamefont
  {Seiberg}, \citenamefont {Senthil}, \citenamefont {Wang},\ and\ \citenamefont
  {Witten}}]{seiberg1}%
  \BibitemOpen
  \bibfield  {author} {\bibinfo {author} {\bibfnamefont {Nathan}\ \bibnamefont
  {Seiberg}}, \bibinfo {author} {\bibfnamefont {T.}~\bibnamefont {Senthil}},
  \bibinfo {author} {\bibfnamefont {Chong}\ \bibnamefont {Wang}}, \ and\
  \bibinfo {author} {\bibfnamefont {Edward}\ \bibnamefont {Witten}},\
  }\bibfield  {title} {\enquote {\bibinfo {title} {A duality web in dimensions
  and condensed matter physics},}\ }\href {\doibase
  http://dx.doi.org/10.1016/j.aop.2016.08.007} {\bibfield  {journal} {\bibinfo
  {journal} {Annals of Physics}\ }\textbf {\bibinfo {volume} {374}},\ \bibinfo
  {pages} {395 -- 433} (\bibinfo {year} {2016})}\BibitemShut {NoStop}%
\bibitem [{\citenamefont {{Karch}}\ and\ \citenamefont
  {{Tong}}(2016)}]{karchtong}%
  \BibitemOpen
  \bibfield  {author} {\bibinfo {author} {\bibfnamefont {A.}~\bibnamefont
  {{Karch}}}\ and\ \bibinfo {author} {\bibfnamefont {D.}~\bibnamefont
  {{Tong}}},\ }\bibfield  {title} {\enquote {\bibinfo {title} {{Particle-Vortex
  Duality from 3D Bosonization}},}\ }\href {\doibase 10.1103/PhysRevX.6.031043}
  {\bibfield  {journal} {\bibinfo  {journal} {Physical Review X}\ }\textbf
  {\bibinfo {volume} {6}},\ \bibinfo {eid} {031043} (\bibinfo {year} {2016})},\
  \Eprint {http://arxiv.org/abs/1606.01893} {arXiv:1606.01893 [hep-th]}
  \BibitemShut {NoStop}%
\bibitem [{\citenamefont {{Murugan}}\ and\ \citenamefont
  {{Nastase}}(2016)}]{murugan}%
  \BibitemOpen
  \bibfield  {author} {\bibinfo {author} {\bibfnamefont {J.}~\bibnamefont
  {{Murugan}}}\ and\ \bibinfo {author} {\bibfnamefont {H.}~\bibnamefont
  {{Nastase}}},\ }\bibfield  {title} {\enquote {\bibinfo {title}
  {{Particle-vortex duality in topological insulators and superconductors}},}\
  }\href@noop {} {\bibfield  {journal} {\bibinfo  {journal} {ArXiv e-prints}\ }
  (\bibinfo {year} {2016})},\ \Eprint {http://arxiv.org/abs/1606.01912}
  {arXiv:1606.01912 [hep-th]} \BibitemShut {NoStop}%
\bibitem [{\citenamefont {Hsin}\ and\ \citenamefont
  {Seiberg}(2016)}]{seiberg2}%
  \BibitemOpen
  \bibfield  {author} {\bibinfo {author} {\bibfnamefont {Po-Shen}\ \bibnamefont
  {Hsin}}\ and\ \bibinfo {author} {\bibfnamefont {Nathan}\ \bibnamefont
  {Seiberg}},\ }\bibfield  {title} {\enquote {\bibinfo {title} {Level/rank
  duality and chern-simons-matter theories},}\ }\href {\doibase
  10.1007/JHEP09(2016)095} {\bibfield  {journal} {\bibinfo  {journal} {Journal
  of High Energy Physics}\ }\textbf {\bibinfo {volume} {2016}},\ \bibinfo
  {pages} {95} (\bibinfo {year} {2016})}\BibitemShut {NoStop}%
\bibitem [{\citenamefont {{Wang}}\ and\ \citenamefont
  {{Senthil}}(2016)}]{wangsenthil15a}%
  \BibitemOpen
  \bibfield  {author} {\bibinfo {author} {\bibfnamefont {C.}~\bibnamefont
  {{Wang}}}\ and\ \bibinfo {author} {\bibfnamefont {T.}~\bibnamefont
  {{Senthil}}},\ }\bibfield  {title} {\enquote {\bibinfo {title}
  {{Time-Reversal Symmetric U (1 ) Quantum Spin Liquids}},}\ }\href {\doibase
  10.1103/PhysRevX.6.011034} {\bibfield  {journal} {\bibinfo  {journal}
  {Physical Review X}\ }\textbf {\bibinfo {volume} {6}},\ \bibinfo {eid}
  {011034} (\bibinfo {year} {2016})},\ \Eprint
  {http://arxiv.org/abs/1505.03520} {arXiv:1505.03520 [cond-mat.str-el]}
  \BibitemShut {NoStop}%
\bibitem [{\citenamefont {{Metlitski}}(2015)}]{Max15}%
  \BibitemOpen
  \bibfield  {author} {\bibinfo {author} {\bibfnamefont {M.~A.}\ \bibnamefont
  {{Metlitski}}},\ }\bibfield  {title} {\enquote {\bibinfo {title}
  {{$S$-duality of $u(1)$ gauge theory with $\theta =\pi$ on non-orientable
  manifolds: Applications to topological insulators and superconductors}},}\
  }\href@noop {} {\bibfield  {journal} {\bibinfo  {journal} {ArXiv e-prints}\ }
  (\bibinfo {year} {2015})},\ \Eprint {http://arxiv.org/abs/1510.05663}
  {arXiv:1510.05663 [hep-th]} \BibitemShut {NoStop}%
\bibitem [{\citenamefont {Son}(2015)}]{sonphcfl}%
  \BibitemOpen
  \bibfield  {author} {\bibinfo {author} {\bibfnamefont {Dam~Thanh}\
  \bibnamefont {Son}},\ }\bibfield  {title} {\enquote {\bibinfo {title} {Is the
  composite fermion a dirac particle?}}\ }\href {\doibase
  10.1103/PhysRevX.5.031027} {\bibfield  {journal} {\bibinfo  {journal} {Phys.
  Rev. X}\ }\textbf {\bibinfo {volume} {5}},\ \bibinfo {pages} {031027}
  (\bibinfo {year} {2015})}\BibitemShut {NoStop}%
\bibitem [{\citenamefont {Wang}\ and\ \citenamefont
  {Senthil}(2016{\natexlab{a}})}]{wangsenthil15c}%
  \BibitemOpen
  \bibfield  {author} {\bibinfo {author} {\bibfnamefont {Chong}\ \bibnamefont
  {Wang}}\ and\ \bibinfo {author} {\bibfnamefont {T.}~\bibnamefont {Senthil}},\
  }\bibfield  {title} {\enquote {\bibinfo {title} {Half-filled landau level,
  topological insulator surfaces, and three-dimensional quantum spin
  liquids},}\ }\href {\doibase 10.1103/PhysRevB.93.085110} {\bibfield
  {journal} {\bibinfo  {journal} {Phys. Rev. B}\ }\textbf {\bibinfo {volume}
  {93}},\ \bibinfo {pages} {085110} (\bibinfo {year}
  {2016}{\natexlab{a}})}\BibitemShut {NoStop}%
\bibitem [{\citenamefont {Geraedts}\ \emph {et~al.}(2016)\citenamefont
  {Geraedts}, \citenamefont {Zaletel}, \citenamefont {Mong}, \citenamefont
  {Metlitski}, \citenamefont {Vishwanath},\ and\ \citenamefont
  {Motrunich}}]{geraedtsnum}%
  \BibitemOpen
  \bibfield  {author} {\bibinfo {author} {\bibfnamefont {Scott~D.}\
  \bibnamefont {Geraedts}}, \bibinfo {author} {\bibfnamefont {Michael~P.}\
  \bibnamefont {Zaletel}}, \bibinfo {author} {\bibfnamefont {Roger S.~K.}\
  \bibnamefont {Mong}}, \bibinfo {author} {\bibfnamefont {Max~A.}\ \bibnamefont
  {Metlitski}}, \bibinfo {author} {\bibfnamefont {Ashvin}\ \bibnamefont
  {Vishwanath}}, \ and\ \bibinfo {author} {\bibfnamefont {Olexei~I.}\
  \bibnamefont {Motrunich}},\ }\bibfield  {title} {\enquote {\bibinfo {title}
  {The half-filled landau level: The case for dirac composite fermions},}\
  }\href {\doibase 10.1126/science.aad4302} {\bibfield  {journal} {\bibinfo
  {journal} {Science}\ }\textbf {\bibinfo {volume} {352}},\ \bibinfo {pages}
  {197--201} (\bibinfo {year} {2016})},\ \Eprint
  {http://arxiv.org/abs/http://science.sciencemag.org/content/352/6282/197.full.pdf}
  {http://science.sciencemag.org/content/352/6282/197.full.pdf} \BibitemShut
  {NoStop}%
\bibitem [{\citenamefont {Aharony}(2016)}]{Aharony2016}%
  \BibitemOpen
  \bibfield  {author} {\bibinfo {author} {\bibfnamefont {Ofer}\ \bibnamefont
  {Aharony}},\ }\bibfield  {title} {\enquote {\bibinfo {title} {Baryons,
  monopoles and dualities in chern-simons-matter theories},}\ }\href {\doibase
  10.1007/JHEP02(2016)093} {\bibfield  {journal} {\bibinfo  {journal} {Journal
  of High Energy Physics}\ }\textbf {\bibinfo {volume} {2016}},\ \bibinfo
  {pages} {93} (\bibinfo {year} {2016})}\BibitemShut {NoStop}%
\bibitem [{\citenamefont {{Aharony}}\ \emph {et~al.}(2013)\citenamefont
  {{Aharony}}, \citenamefont {{Seiberg}},\ and\ \citenamefont
  {{Tachikawa}}}]{ahseta}%
  \BibitemOpen
  \bibfield  {author} {\bibinfo {author} {\bibfnamefont {O.}~\bibnamefont
  {{Aharony}}}, \bibinfo {author} {\bibfnamefont {N.}~\bibnamefont
  {{Seiberg}}}, \ and\ \bibinfo {author} {\bibfnamefont {Y.}~\bibnamefont
  {{Tachikawa}}},\ }\bibfield  {title} {\enquote {\bibinfo {title} {{Reading
  between the lines of four-dimensional gauge theories}},}\ }\href {\doibase
  10.1007/JHEP08(2013)115} {\bibfield  {journal} {\bibinfo  {journal} {Journal
  of High Energy Physics}\ }\textbf {\bibinfo {volume} {8}},\ \bibinfo {eid}
  {115} (\bibinfo {year} {2013})},\ \Eprint {http://arxiv.org/abs/1305.0318}
  {arXiv:1305.0318 [hep-th]} \BibitemShut {NoStop}%
\bibitem [{\citenamefont {{Karthik}}\ and\ \citenamefont
  {{Narayanan}}(2016)}]{qedcft}%
  \BibitemOpen
  \bibfield  {author} {\bibinfo {author} {\bibfnamefont {N.}~\bibnamefont
  {{Karthik}}}\ and\ \bibinfo {author} {\bibfnamefont {R.}~\bibnamefont
  {{Narayanan}}},\ }\bibfield  {title} {\enquote {\bibinfo {title} {{Scale
  invariance of parity-invariant three-dimensional QED}},}\ }\href {\doibase
  10.1103/PhysRevD.94.065026} {\bibfield  {journal} {\bibinfo  {journal}
  {\prd}\ }\textbf {\bibinfo {volume} {94}},\ \bibinfo {eid} {065026} (\bibinfo
  {year} {2016})},\ \Eprint {http://arxiv.org/abs/1606.04109} {arXiv:1606.04109
  [hep-th]} \BibitemShut {NoStop}%
\bibitem [{\citenamefont {{Slagle}}\ \emph {et~al.}(2015)\citenamefont
  {{Slagle}}, \citenamefont {{You}},\ and\ \citenamefont {{Xu}}}]{kevinQSH}%
  \BibitemOpen
  \bibfield  {author} {\bibinfo {author} {\bibfnamefont {K.}~\bibnamefont
  {{Slagle}}}, \bibinfo {author} {\bibfnamefont {Y.-Z.}\ \bibnamefont {{You}}},
  \ and\ \bibinfo {author} {\bibfnamefont {C.}~\bibnamefont {{Xu}}},\
  }\bibfield  {title} {\enquote {\bibinfo {title} {{Exotic quantum phase
  transitions of strongly interacting topological insulators}},}\ }\href
  {\doibase 10.1103/PhysRevB.91.115121} {\bibfield  {journal} {\bibinfo
  {journal} {\prb}\ }\textbf {\bibinfo {volume} {91}},\ \bibinfo {eid} {115121}
  (\bibinfo {year} {2015})},\ \Eprint {http://arxiv.org/abs/1409.7401}
  {arXiv:1409.7401 [cond-mat.str-el]} \BibitemShut {NoStop}%
\bibitem [{\citenamefont {{He}}\ \emph {et~al.}(2016)\citenamefont {{He}},
  \citenamefont {{Wu}}, \citenamefont {{You}}, \citenamefont {{Xu}},
  \citenamefont {{Meng}},\ and\ \citenamefont {{Lu}}}]{so4qsh}%
  \BibitemOpen
  \bibfield  {author} {\bibinfo {author} {\bibfnamefont {Y.-Y.}\ \bibnamefont
  {{He}}}, \bibinfo {author} {\bibfnamefont {H.-Q.}\ \bibnamefont {{Wu}}},
  \bibinfo {author} {\bibfnamefont {Y.-Z.}\ \bibnamefont {{You}}}, \bibinfo
  {author} {\bibfnamefont {C.}~\bibnamefont {{Xu}}}, \bibinfo {author}
  {\bibfnamefont {Z.~Y.}\ \bibnamefont {{Meng}}}, \ and\ \bibinfo {author}
  {\bibfnamefont {Z.-Y.}\ \bibnamefont {{Lu}}},\ }\bibfield  {title} {\enquote
  {\bibinfo {title} {{Bona fide interaction-driven topological phase transition
  in correlated symmetry-protected topological states}},}\ }\href {\doibase
  10.1103/PhysRevB.93.115150} {\bibfield  {journal} {\bibinfo  {journal}
  {\prb}\ }\textbf {\bibinfo {volume} {93}},\ \bibinfo {eid} {115150} (\bibinfo
  {year} {2016})},\ \Eprint {http://arxiv.org/abs/1508.06389} {arXiv:1508.06389
  [cond-mat.str-el]} \BibitemShut {NoStop}%
\bibitem [{\citenamefont {{Hands}}\ \emph {et~al.}(2004)\citenamefont
  {{Hands}}, \citenamefont {{Kogut}}, \citenamefont {{Scorzato}},\ and\
  \citenamefont {{Strouthos}}}]{kogut2}%
  \BibitemOpen
  \bibfield  {author} {\bibinfo {author} {\bibfnamefont {S.~J.}\ \bibnamefont
  {{Hands}}}, \bibinfo {author} {\bibfnamefont {J.~B.}\ \bibnamefont
  {{Kogut}}}, \bibinfo {author} {\bibfnamefont {L.}~\bibnamefont {{Scorzato}}},
  \ and\ \bibinfo {author} {\bibfnamefont {C.~G.}\ \bibnamefont
  {{Strouthos}}},\ }\bibfield  {title} {\enquote {\bibinfo {title} {{Noncompact
  three-dimensional quantum electrodynamics with N$_{f}$ =1 and N$_{f}$ =4}},}\
  }\href {\doibase 10.1103/PhysRevB.70.104501} {\bibfield  {journal} {\bibinfo
  {journal} {\prb}\ }\textbf {\bibinfo {volume} {70}},\ \bibinfo {eid} {104501}
  (\bibinfo {year} {2004})},\ \Eprint {http://arxiv.org/abs/hep-lat/0404013}
  {hep-lat/0404013} \BibitemShut {NoStop}%
\bibitem [{\citenamefont {Kapustin}\ and\ \citenamefont
  {Strassler}(1999)}]{kapstr}%
  \BibitemOpen
  \bibfield  {author} {\bibinfo {author} {\bibfnamefont {Anton}\ \bibnamefont
  {Kapustin}}\ and\ \bibinfo {author} {\bibfnamefont {Matthew~J.}\ \bibnamefont
  {Strassler}},\ }\bibfield  {title} {\enquote {\bibinfo {title} {On mirror
  symmetry in three dimensional abelian gauge theories},}\ }\href
  {http://stacks.iop.org/1126-6708/1999/i=04/a=021} {\bibfield  {journal}
  {\bibinfo  {journal} {Journal of High Energy Physics}\ }\textbf {\bibinfo
  {volume} {1999}},\ \bibinfo {pages} {021} (\bibinfo {year}
  {1999})}\BibitemShut {NoStop}%
\bibitem [{\citenamefont {Witten}(2003)}]{witten03}%
  \BibitemOpen
  \bibfield  {author} {\bibinfo {author} {\bibfnamefont {Edward}\ \bibnamefont
  {Witten}},\ }\bibfield  {title} {\enquote {\bibinfo {title} {{SL(2,Z) Action
  On Three-Dimensional Conformal Field Theories With Abelian Symmetry}},}\
  }\href@noop {} {\bibfield  {journal} {\bibinfo  {journal} {ArXiv e-prints}\ }
  (\bibinfo {year} {2003})},\ \Eprint {http://arxiv.org/abs/hep-th/0307041}
  {arXiv:hep-th/0307041} \BibitemShut {NoStop}%
\bibitem [{\citenamefont {Motrunich}\ and\ \citenamefont
  {Vishwanath}(2004)}]{lesikav04}%
  \BibitemOpen
  \bibfield  {author} {\bibinfo {author} {\bibfnamefont {Olexei~I.}\
  \bibnamefont {Motrunich}}\ and\ \bibinfo {author} {\bibfnamefont {Ashvin}\
  \bibnamefont {Vishwanath}},\ }\bibfield  {title} {\enquote {\bibinfo {title}
  {Emergent photons and transitions in the $\mathrm{O}(3)$ sigma model with
  hedgehog suppression},}\ }\href {\doibase 10.1103/PhysRevB.70.075104}
  {\bibfield  {journal} {\bibinfo  {journal} {Phys. Rev. B}\ }\textbf {\bibinfo
  {volume} {70}},\ \bibinfo {pages} {075104} (\bibinfo {year}
  {2004})}\BibitemShut {NoStop}%
\bibitem [{\citenamefont {Haldane}(1988)}]{HaldaneBerry}%
  \BibitemOpen
  \bibfield  {author} {\bibinfo {author} {\bibfnamefont {F.~D.~M.}\
  \bibnamefont {Haldane}},\ }\bibfield  {title} {\enquote {\bibinfo {title}
  {O(3) nonlinear $\ensuremath{\sigma}$ model and the topological distinction
  between integer- and half-integer-spin antiferromagnets in two dimensions},}\
  }\href {\doibase 10.1103/PhysRevLett.61.1029} {\bibfield  {journal} {\bibinfo
   {journal} {Phys. Rev. Lett.}\ }\textbf {\bibinfo {volume} {61}},\ \bibinfo
  {pages} {1029--1032} (\bibinfo {year} {1988})}\BibitemShut {NoStop}%
\bibitem [{\citenamefont {Read}\ and\ \citenamefont {Sachdev}(1990)}]{ReSaSUN}%
  \BibitemOpen
  \bibfield  {author} {\bibinfo {author} {\bibfnamefont {N.}~\bibnamefont
  {Read}}\ and\ \bibinfo {author} {\bibfnamefont {Subir}\ \bibnamefont
  {Sachdev}},\ }\bibfield  {title} {\enquote {\bibinfo {title} {Spin-peierls,
  valence-bond solid, and n\'eel ground states of low-dimensional quantum
  antiferromagnets},}\ }\href {\doibase 10.1103/PhysRevB.42.4568} {\bibfield
  {journal} {\bibinfo  {journal} {Phys. Rev. B}\ }\textbf {\bibinfo {volume}
  {42}},\ \bibinfo {pages} {4568--4589} (\bibinfo {year} {1990})}\BibitemShut
  {NoStop}%
\bibitem [{\citenamefont {{Levin}}\ and\ \citenamefont
  {{Senthil}}(2004)}]{mlts04}%
  \BibitemOpen
  \bibfield  {author} {\bibinfo {author} {\bibfnamefont {M.}~\bibnamefont
  {{Levin}}}\ and\ \bibinfo {author} {\bibfnamefont {T.}~\bibnamefont
  {{Senthil}}},\ }\bibfield  {title} {\enquote {\bibinfo {title} {{Deconfined
  quantum criticality and N{\'e}el order via dimer disorder}},}\ }\href
  {\doibase 10.1103/PhysRevB.70.220403} {\bibfield  {journal} {\bibinfo
  {journal} {\prb}\ }\textbf {\bibinfo {volume} {70}},\ \bibinfo {eid} {220403}
  (\bibinfo {year} {2004})},\ \Eprint {http://arxiv.org/abs/cond-mat/0405702}
  {cond-mat/0405702} \BibitemShut {NoStop}%
\bibitem [{\citenamefont {{Vishwanath}}\ and\ \citenamefont
  {{Senthil}}(2013)}]{ashvinsenthil}%
  \BibitemOpen
  \bibfield  {author} {\bibinfo {author} {\bibfnamefont {A.}~\bibnamefont
  {{Vishwanath}}}\ and\ \bibinfo {author} {\bibfnamefont {T.}~\bibnamefont
  {{Senthil}}},\ }\bibfield  {title} {\enquote {\bibinfo {title} {{Physics of
  Three-Dimensional Bosonic Topological Insulators: Surface-Deconfined
  Criticality and Quantized Magnetoelectric Effect}},}\ }\href {\doibase
  10.1103/PhysRevX.3.011016} {\bibfield  {journal} {\bibinfo  {journal}
  {Physical Review X}\ }\textbf {\bibinfo {volume} {3}},\ \bibinfo {eid}
  {011016} (\bibinfo {year} {2013})},\ \Eprint {http://arxiv.org/abs/1209.3058}
  {arXiv:1209.3058 [cond-mat.str-el]} \BibitemShut {NoStop}%
\bibitem [{\citenamefont {{Witten}}(2016)}]{wittenreview}%
  \BibitemOpen
  \bibfield  {author} {\bibinfo {author} {\bibfnamefont {E.}~\bibnamefont
  {{Witten}}},\ }\bibfield  {title} {\enquote {\bibinfo {title} {{Fermion path
  integrals and topological phases}},}\ }\href {\doibase
  10.1103/RevModPhys.88.035001} {\bibfield  {journal} {\bibinfo  {journal}
  {Reviews of Modern Physics}\ }\textbf {\bibinfo {volume} {88}},\ \bibinfo
  {eid} {035001} (\bibinfo {year} {2016})},\ \Eprint
  {http://arxiv.org/abs/1508.04715} {arXiv:1508.04715 [cond-mat.mes-hall]}
  \BibitemShut {NoStop}%
\bibitem [{\citenamefont {{Borokhov}}\ \emph {et~al.}(2002)\citenamefont
  {{Borokhov}}, \citenamefont {{Kapustin}},\ and\ \citenamefont
  {{Wu}}}]{kapustinqed}%
  \BibitemOpen
  \bibfield  {author} {\bibinfo {author} {\bibfnamefont {V.}~\bibnamefont
  {{Borokhov}}}, \bibinfo {author} {\bibfnamefont {A.}~\bibnamefont
  {{Kapustin}}}, \ and\ \bibinfo {author} {\bibfnamefont {X.}~\bibnamefont
  {{Wu}}},\ }\bibfield  {title} {\enquote {\bibinfo {title} {{Topological
  Disorder Operators in Three-Dimensional Conformal Field Theory}},}\ }\href
  {\doibase 10.1088/1126-6708/2002/11/049} {\bibfield  {journal} {\bibinfo
  {journal} {Journal of High Energy Physics}\ }\textbf {\bibinfo {volume}
  {11}},\ \bibinfo {eid} {049} (\bibinfo {year} {2002})},\ \Eprint
  {http://arxiv.org/abs/hep-th/0206054} {hep-th/0206054} \BibitemShut {NoStop}%
\bibitem [{\citenamefont {{Benini}}\ \emph {et~al.}(2017)\citenamefont
  {{Benini}}, \citenamefont {{Hsin}},\ and\ \citenamefont
  {{Seiberg}}}]{bhse2017}%
  \BibitemOpen
  \bibfield  {author} {\bibinfo {author} {\bibfnamefont {F.}~\bibnamefont
  {{Benini}}}, \bibinfo {author} {\bibfnamefont {P.-S.}\ \bibnamefont
  {{Hsin}}}, \ and\ \bibinfo {author} {\bibfnamefont {N.}~\bibnamefont
  {{Seiberg}}},\ }\bibfield  {title} {\enquote {\bibinfo {title} {{Comments on
  Global Symmetries, Anomalies, and Duality in (2+1)d}},}\ }\href@noop {}
  {\bibfield  {journal} {\bibinfo  {journal} {ArXiv e-prints}\ } (\bibinfo
  {year} {2017})},\ \Eprint {http://arxiv.org/abs/1702.07035} {arXiv:1702.07035
  [cond-mat.str-el]} \BibitemShut {NoStop}%
\bibitem [{\citenamefont {Kuklov}\ \emph {et~al.}(2006)\citenamefont {Kuklov},
  \citenamefont {Prokof'ev}, \citenamefont {Svistunov},\ and\ \citenamefont
  {Troyer}}]{kukloveasyplane}%
  \BibitemOpen
  \bibfield  {author} {\bibinfo {author} {\bibfnamefont {A.B.}\ \bibnamefont
  {Kuklov}}, \bibinfo {author} {\bibfnamefont {N.~V.}\ \bibnamefont
  {Prokof'ev}}, \bibinfo {author} {\bibfnamefont {B.V.}\ \bibnamefont
  {Svistunov}}, \ and\ \bibinfo {author} {\bibfnamefont {M.}~\bibnamefont
  {Troyer}},\ }\bibfield  {title} {\enquote {\bibinfo {title} {Deconfined
  criticality, runaway flow in the two-component scalar electrodynamics and
  weak first-order superfluid-solid transitions},}\ }\href {\doibase
  http://dx.doi.org/10.1016/j.aop.2006.04.007} {\bibfield  {journal} {\bibinfo
  {journal} {Annals of Physics}\ }\textbf {\bibinfo {volume} {321}},\ \bibinfo
  {pages} {1602 -- 1621} (\bibinfo {year} {2006})}\BibitemShut {NoStop}%
\bibitem [{\citenamefont {Kragset}\ \emph {et~al.}(2006)\citenamefont
  {Kragset}, \citenamefont {Sm\o{}rgrav}, \citenamefont {Hove}, \citenamefont
  {Nogueira},\ and\ \citenamefont {Sudb\o{}}}]{kragseteasyplane}%
  \BibitemOpen
  \bibfield  {author} {\bibinfo {author} {\bibfnamefont {S.}~\bibnamefont
  {Kragset}}, \bibinfo {author} {\bibfnamefont {E.}~\bibnamefont
  {Sm\o{}rgrav}}, \bibinfo {author} {\bibfnamefont {J.}~\bibnamefont {Hove}},
  \bibinfo {author} {\bibfnamefont {F.~S.}\ \bibnamefont {Nogueira}}, \ and\
  \bibinfo {author} {\bibfnamefont {A.}~\bibnamefont {Sudb\o{}}},\ }\bibfield
  {title} {\enquote {\bibinfo {title} {First-order phase transition in
  easy-plane quantum antiferromagnets},}\ }\href {\doibase
  10.1103/PhysRevLett.97.247201} {\bibfield  {journal} {\bibinfo  {journal}
  {Phys. Rev. Lett.}\ }\textbf {\bibinfo {volume} {97}},\ \bibinfo {pages}
  {247201} (\bibinfo {year} {2006})}\BibitemShut {NoStop}%
\bibitem [{\citenamefont {D'Emidio}\ and\ \citenamefont
  {Kaul}(2016)}]{kauleasyplane1}%
  \BibitemOpen
  \bibfield  {author} {\bibinfo {author} {\bibfnamefont {Jonathan}\
  \bibnamefont {D'Emidio}}\ and\ \bibinfo {author} {\bibfnamefont {Ribhu~K.}\
  \bibnamefont {Kaul}},\ }\bibfield  {title} {\enquote {\bibinfo {title}
  {First-order superfluid to valence-bond solid phase transitions in easy-plane
  $\mathrm{SU}(n)$ magnets for small $n$},}\ }\href {\doibase
  10.1103/PhysRevB.93.054406} {\bibfield  {journal} {\bibinfo  {journal} {Phys.
  Rev. B}\ }\textbf {\bibinfo {volume} {93}},\ \bibinfo {pages} {054406}
  (\bibinfo {year} {2016})}\BibitemShut {NoStop}%
\bibitem [{\citenamefont {{D'Emidio}}\ and\ \citenamefont
  {{Kaul}}(2016)}]{kauleasyplane2}%
  \BibitemOpen
  \bibfield  {author} {\bibinfo {author} {\bibfnamefont {J.}~\bibnamefont
  {{D'Emidio}}}\ and\ \bibinfo {author} {\bibfnamefont {R.~K.}\ \bibnamefont
  {{Kaul}}},\ }\bibfield  {title} {\enquote {\bibinfo {title} {{New easy-plane
  $\mathbb{CP}^{N-1}$ fixed points}},}\ }\href@noop {} {\bibfield  {journal}
  {\bibinfo  {journal} {ArXiv e-prints}\ } (\bibinfo {year} {2016})},\ \Eprint
  {http://arxiv.org/abs/1610.07702} {arXiv:1610.07702 [cond-mat.str-el]}
  \BibitemShut {NoStop}%
\bibitem [{\citenamefont {{Chen}}\ \emph {et~al.}(1993)\citenamefont {{Chen}},
  \citenamefont {{Fisher}},\ and\ \citenamefont {{Wu}}}]{ChenFisherWu}%
  \BibitemOpen
  \bibfield  {author} {\bibinfo {author} {\bibfnamefont {W.}~\bibnamefont
  {{Chen}}}, \bibinfo {author} {\bibfnamefont {M.~P.~A.}\ \bibnamefont
  {{Fisher}}}, \ and\ \bibinfo {author} {\bibfnamefont {Y.-S.}\ \bibnamefont
  {{Wu}}},\ }\bibfield  {title} {\enquote {\bibinfo {title} {{Mott transition
  in an anyon gas}},}\ }\href {\doibase 10.1103/PhysRevB.48.13749} {\bibfield
  {journal} {\bibinfo  {journal} {\prb}\ }\textbf {\bibinfo {volume} {48}},\
  \bibinfo {pages} {13749--13761} (\bibinfo {year} {1993})},\ \Eprint
  {http://arxiv.org/abs/cond-mat/9301037} {cond-mat/9301037} \BibitemShut
  {NoStop}%
\bibitem [{\citenamefont {{Barkeshli}}\ and\ \citenamefont
  {{McGreevy}}(2012)}]{BarkeshliMcGreevy}%
  \BibitemOpen
  \bibfield  {author} {\bibinfo {author} {\bibfnamefont {M.}~\bibnamefont
  {{Barkeshli}}}\ and\ \bibinfo {author} {\bibfnamefont {J.}~\bibnamefont
  {{McGreevy}}},\ }\bibfield  {title} {\enquote {\bibinfo {title} {{A
  continuous transition between fractional quantum Hall and superfluid
  states}},}\ }\href@noop {} {\bibfield  {journal} {\bibinfo  {journal} {ArXiv
  e-prints}\ } (\bibinfo {year} {2012})},\ \Eprint
  {http://arxiv.org/abs/1201.4393} {arXiv:1201.4393 [cond-mat.str-el]}
  \BibitemShut {NoStop}%
\bibitem [{\citenamefont {{Lu}}\ and\ \citenamefont
  {{Vishwanath}}(2012)}]{luav12}%
  \BibitemOpen
  \bibfield  {author} {\bibinfo {author} {\bibfnamefont {Y.-M.}\ \bibnamefont
  {{Lu}}}\ and\ \bibinfo {author} {\bibfnamefont {A.}~\bibnamefont
  {{Vishwanath}}},\ }\bibfield  {title} {\enquote {\bibinfo {title} {{Theory
  and classification of interacting integer topological phases in two
  dimensions: A Chern-Simons approach}},}\ }\href {\doibase
  10.1103/PhysRevB.86.125119} {\bibfield  {journal} {\bibinfo  {journal}
  {\prb}\ }\textbf {\bibinfo {volume} {86}},\ \bibinfo {eid} {125119} (\bibinfo
  {year} {2012})},\ \Eprint {http://arxiv.org/abs/1205.3156} {arXiv:1205.3156
  [cond-mat.str-el]} \BibitemShut {NoStop}%
\bibitem [{\citenamefont {{Senthil}}\ and\ \citenamefont
  {{Levin}}(2013)}]{tsml13}%
  \BibitemOpen
  \bibfield  {author} {\bibinfo {author} {\bibfnamefont {T.}~\bibnamefont
  {{Senthil}}}\ and\ \bibinfo {author} {\bibfnamefont {M.}~\bibnamefont
  {{Levin}}},\ }\bibfield  {title} {\enquote {\bibinfo {title} {{Integer
  Quantum Hall Effect for Bosons}},}\ }\href {\doibase
  10.1103/PhysRevLett.110.046801} {\bibfield  {journal} {\bibinfo  {journal}
  {Physical Review Letters}\ }\textbf {\bibinfo {volume} {110}},\ \bibinfo
  {eid} {046801} (\bibinfo {year} {2013})},\ \Eprint
  {http://arxiv.org/abs/1206.1604} {arXiv:1206.1604 [cond-mat.str-el]}
  \BibitemShut {NoStop}%
\bibitem [{\citenamefont {{Grover}}\ and\ \citenamefont
  {{Vishwanath}}(2013)}]{tarunashvin}%
  \BibitemOpen
  \bibfield  {author} {\bibinfo {author} {\bibfnamefont {T.}~\bibnamefont
  {{Grover}}}\ and\ \bibinfo {author} {\bibfnamefont {A.}~\bibnamefont
  {{Vishwanath}}},\ }\bibfield  {title} {\enquote {\bibinfo {title} {{Quantum
  phase transition between integer quantum Hall states of bosons}},}\ }\href
  {\doibase 10.1103/PhysRevB.87.045129} {\bibfield  {journal} {\bibinfo
  {journal} {\prb}\ }\textbf {\bibinfo {volume} {87}},\ \bibinfo {eid} {045129}
  (\bibinfo {year} {2013})},\ \Eprint {http://arxiv.org/abs/1210.0907}
  {arXiv:1210.0907 [cond-mat.str-el]} \BibitemShut {NoStop}%
\bibitem [{\citenamefont {Lu}\ and\ \citenamefont {Lee}(2014)}]{LuLee}%
  \BibitemOpen
  \bibfield  {author} {\bibinfo {author} {\bibfnamefont {Yuan-Ming}\
  \bibnamefont {Lu}}\ and\ \bibinfo {author} {\bibfnamefont {Dung-Hai}\
  \bibnamefont {Lee}},\ }\bibfield  {title} {\enquote {\bibinfo {title}
  {Quantum phase transitions between bosonic symmetry-protected topological
  phases in two dimensions: Emergent ${\mathrm{qed}}_{3}$ and anyon
  superfluid},}\ }\href {\doibase 10.1103/PhysRevB.89.195143} {\bibfield
  {journal} {\bibinfo  {journal} {Phys. Rev. B}\ }\textbf {\bibinfo {volume}
  {89}},\ \bibinfo {pages} {195143} (\bibinfo {year} {2014})}\BibitemShut
  {NoStop}%
\bibitem [{\citenamefont {{Fuji}}\ \emph {et~al.}(2016)\citenamefont {{Fuji}},
  \citenamefont {{He}}, \citenamefont {{Bhattacharjee}},\ and\ \citenamefont
  {{Pollmann}}}]{hewire}%
  \BibitemOpen
  \bibfield  {author} {\bibinfo {author} {\bibfnamefont {Y.}~\bibnamefont
  {{Fuji}}}, \bibinfo {author} {\bibfnamefont {Y.-C.}\ \bibnamefont {{He}}},
  \bibinfo {author} {\bibfnamefont {S.}~\bibnamefont {{Bhattacharjee}}}, \ and\
  \bibinfo {author} {\bibfnamefont {F.}~\bibnamefont {{Pollmann}}},\ }\bibfield
   {title} {\enquote {\bibinfo {title} {{Bridging coupled wires and lattice
  Hamiltonian for two-component bosonic quantum Hall states}},}\ }\href
  {\doibase 10.1103/PhysRevB.93.195143} {\bibfield  {journal} {\bibinfo
  {journal} {\prb}\ }\textbf {\bibinfo {volume} {93}},\ \bibinfo {eid} {195143}
  (\bibinfo {year} {2016})},\ \Eprint {http://arxiv.org/abs/1603.05109}
  {arXiv:1603.05109 [cond-mat.str-el]} \BibitemShut {NoStop}%
\bibitem [{\citenamefont {Cheng}\ and\ \citenamefont {Xu}(2016)}]{xucheng}%
  \BibitemOpen
  \bibfield  {author} {\bibinfo {author} {\bibfnamefont {Meng}\ \bibnamefont
  {Cheng}}\ and\ \bibinfo {author} {\bibfnamefont {Cenke}\ \bibnamefont {Xu}},\
  }\bibfield  {title} {\enquote {\bibinfo {title} {Series of (2+1)-dimensional
  stable self-dual interacting conformal field theories},}\ }\href {\doibase
  10.1103/PhysRevB.94.214415} {\bibfield  {journal} {\bibinfo  {journal} {Phys.
  Rev. B}\ }\textbf {\bibinfo {volume} {94}},\ \bibinfo {pages} {214415}
  (\bibinfo {year} {2016})}\BibitemShut {NoStop}%
\bibitem [{\citenamefont {{Bi}}\ \emph {et~al.}(2015)\citenamefont {{Bi}},
  \citenamefont {{Rasmussen}}, \citenamefont {{Slagle}},\ and\ \citenamefont
  {{Xu}}}]{BiNLSM}%
  \BibitemOpen
  \bibfield  {author} {\bibinfo {author} {\bibfnamefont {Z.}~\bibnamefont
  {{Bi}}}, \bibinfo {author} {\bibfnamefont {A.}~\bibnamefont {{Rasmussen}}},
  \bibinfo {author} {\bibfnamefont {K.}~\bibnamefont {{Slagle}}}, \ and\
  \bibinfo {author} {\bibfnamefont {C.}~\bibnamefont {{Xu}}},\ }\bibfield
  {title} {\enquote {\bibinfo {title} {{Classification and description of
  bosonic symmetry protected topological phases with semiclassical nonlinear
  sigma models}},}\ }\href {\doibase 10.1103/PhysRevB.91.134404} {\bibfield
  {journal} {\bibinfo  {journal} {\prb}\ }\textbf {\bibinfo {volume} {91}},\
  \bibinfo {eid} {134404} (\bibinfo {year} {2015})},\ \Eprint
  {http://arxiv.org/abs/1309.0515} {arXiv:1309.0515 [cond-mat.str-el]}
  \BibitemShut {NoStop}%
\bibitem [{\citenamefont {Wen}(2004)}]{wenbook}%
  \BibitemOpen
  \bibfield  {author} {\bibinfo {author} {\bibfnamefont {Xiao-Gang}\
  \bibnamefont {Wen}},\ }\href
  {http://libproxy.mit.edu/login?url=http://search.ebscohost.com/login.aspx?direct=true&db=nlebk&AN=186592&site=ehost-live}
  {\emph {\bibinfo {title} {Quantum Field Theory of Many-body Systems : From
  the Origin of Sound to an Origin of Light and Electrons.}}},\ Oxford Graduate
  Texts\ (\bibinfo  {publisher} {OUP Premium},\ \bibinfo {year}
  {2004})\BibitemShut {NoStop}%
\bibitem [{\citenamefont {{Abanov}}\ and\ \citenamefont
  {{Wiegmann}}(2000)}]{AbanovWiegmann}%
  \BibitemOpen
  \bibfield  {author} {\bibinfo {author} {\bibfnamefont {A.~G.}\ \bibnamefont
  {{Abanov}}}\ and\ \bibinfo {author} {\bibfnamefont {P.~B.}\ \bibnamefont
  {{Wiegmann}}},\ }\bibfield  {title} {\enquote {\bibinfo {title} {{Theta-terms
  in nonlinear sigma-models}},}\ }\href {\doibase
  10.1016/S0550-3213(99)00820-2} {\bibfield  {journal} {\bibinfo  {journal}
  {Nuclear Physics B}\ }\textbf {\bibinfo {volume} {570}},\ \bibinfo {pages}
  {685--698} (\bibinfo {year} {2000})},\ \Eprint
  {http://arxiv.org/abs/hep-th/9911025} {hep-th/9911025} \BibitemShut {NoStop}%
\bibitem [{\citenamefont {Metlitski}\ \emph {et~al.}(2013)\citenamefont
  {Metlitski}, \citenamefont {Kane},\ and\ \citenamefont {Fisher}}]{MKF2013}%
  \BibitemOpen
  \bibfield  {author} {\bibinfo {author} {\bibfnamefont {Max~A.}\ \bibnamefont
  {Metlitski}}, \bibinfo {author} {\bibfnamefont {C.~L.}\ \bibnamefont {Kane}},
  \ and\ \bibinfo {author} {\bibfnamefont {Matthew P.~A.}\ \bibnamefont
  {Fisher}},\ }\bibfield  {title} {\enquote {\bibinfo {title} {Bosonic
  topological insulator in three dimensions and the statistical witten
  effect},}\ }\href {\doibase 10.1103/PhysRevB.88.035131} {\bibfield  {journal}
  {\bibinfo  {journal} {Phys. Rev. B}\ }\textbf {\bibinfo {volume} {88}},\
  \bibinfo {pages} {035131} (\bibinfo {year} {2013})}\BibitemShut {NoStop}%
\bibitem [{\citenamefont {Wang}\ and\ \citenamefont {Senthil}(2013)}]{WS2013}%
  \BibitemOpen
  \bibfield  {author} {\bibinfo {author} {\bibfnamefont {Chong}\ \bibnamefont
  {Wang}}\ and\ \bibinfo {author} {\bibfnamefont {T.}~\bibnamefont {Senthil}},\
  }\bibfield  {title} {\enquote {\bibinfo {title} {Boson topological
  insulators: A window into highly entangled quantum phases},}\ }\href
  {\doibase 10.1103/PhysRevB.87.235122} {\bibfield  {journal} {\bibinfo
  {journal} {Phys. Rev. B}\ }\textbf {\bibinfo {volume} {87}},\ \bibinfo
  {pages} {235122} (\bibinfo {year} {2013})}\BibitemShut {NoStop}%
\bibitem [{\citenamefont {{Oon}}\ \emph {et~al.}(2013)\citenamefont {{Oon}},
  \citenamefont {{Cho}},\ and\ \citenamefont {{Xu}}}]{XuParton}%
  \BibitemOpen
  \bibfield  {author} {\bibinfo {author} {\bibfnamefont {J.}~\bibnamefont
  {{Oon}}}, \bibinfo {author} {\bibfnamefont {G.~Y.}\ \bibnamefont {{Cho}}}, \
  and\ \bibinfo {author} {\bibfnamefont {C.}~\bibnamefont {{Xu}}},\ }\bibfield
  {title} {\enquote {\bibinfo {title} {{Two-dimensional symmetry-protected
  topological phases with PSU(N) and time-reversal symmetry}},}\ }\href
  {\doibase 10.1103/PhysRevB.88.014425} {\bibfield  {journal} {\bibinfo
  {journal} {\prb}\ }\textbf {\bibinfo {volume} {88}},\ \bibinfo {eid} {014425}
  (\bibinfo {year} {2013})},\ \Eprint {http://arxiv.org/abs/1212.1726}
  {arXiv:1212.1726 [cond-mat.str-el]} \BibitemShut {NoStop}%
\bibitem [{\citenamefont {{Wang}}\ \emph {et~al.}(2015)\citenamefont {{Wang}},
  \citenamefont {{Nahum}},\ and\ \citenamefont {{Senthil}}}]{WangNahumSenthil}%
  \BibitemOpen
  \bibfield  {author} {\bibinfo {author} {\bibfnamefont {C.}~\bibnamefont
  {{Wang}}}, \bibinfo {author} {\bibfnamefont {A.}~\bibnamefont {{Nahum}}}, \
  and\ \bibinfo {author} {\bibfnamefont {T.}~\bibnamefont {{Senthil}}},\
  }\bibfield  {title} {\enquote {\bibinfo {title} {{Topological paramagnetism
  in frustrated spin-1 Mott insulators}},}\ }\href {\doibase
  10.1103/PhysRevB.91.195131} {\bibfield  {journal} {\bibinfo  {journal}
  {\prb}\ }\textbf {\bibinfo {volume} {91}},\ \bibinfo {eid} {195131} (\bibinfo
  {year} {2015})},\ \Eprint {http://arxiv.org/abs/1501.01047} {arXiv:1501.01047
  [cond-mat.str-el]} \BibitemShut {NoStop}%
\bibitem [{\citenamefont {Wang}\ and\ \citenamefont {Senthil}(2014)}]{3dfspt2}%
  \BibitemOpen
  \bibfield  {author} {\bibinfo {author} {\bibfnamefont {Chong}\ \bibnamefont
  {Wang}}\ and\ \bibinfo {author} {\bibfnamefont {T.}~\bibnamefont {Senthil}},\
  }\bibfield  {title} {\enquote {\bibinfo {title} {Interacting fermionic
  topological insulators/superconductors in three dimensions},}\ }\href
  {\doibase 10.1103/PhysRevB.89.195124} {\bibfield  {journal} {\bibinfo
  {journal} {Phys. Rev. B}\ }\textbf {\bibinfo {volume} {89}},\ \bibinfo
  {pages} {195124} (\bibinfo {year} {2014})}\BibitemShut {NoStop}%
\bibitem [{\citenamefont {Wang}\ and\ \citenamefont
  {Senthil}(2016{\natexlab{b}})}]{wangtscflll16}%
  \BibitemOpen
  \bibfield  {author} {\bibinfo {author} {\bibfnamefont {Chong}\ \bibnamefont
  {Wang}}\ and\ \bibinfo {author} {\bibfnamefont {T.}~\bibnamefont {Senthil}},\
  }\bibfield  {title} {\enquote {\bibinfo {title} {Composite fermi liquids in
  the lowest landau level},}\ }\href {\doibase 10.1103/PhysRevB.94.245107}
  {\bibfield  {journal} {\bibinfo  {journal} {Phys. Rev. B}\ }\textbf {\bibinfo
  {volume} {94}},\ \bibinfo {pages} {245107} (\bibinfo {year}
  {2016}{\natexlab{b}})}\BibitemShut {NoStop}%
\bibitem [{\citenamefont {{Sodemann}}\ \emph {et~al.}(2016)\citenamefont
  {{Sodemann}}, \citenamefont {{Kimchi}}, \citenamefont {{Wang}},\ and\
  \citenamefont {{Senthil}}}]{skws17}%
  \BibitemOpen
  \bibfield  {author} {\bibinfo {author} {\bibfnamefont {I.}~\bibnamefont
  {{Sodemann}}}, \bibinfo {author} {\bibfnamefont {I.}~\bibnamefont
  {{Kimchi}}}, \bibinfo {author} {\bibfnamefont {C.}~\bibnamefont {{Wang}}}, \
  and\ \bibinfo {author} {\bibfnamefont {T.}~\bibnamefont {{Senthil}}},\
  }\bibfield  {title} {\enquote {\bibinfo {title} {{Composite fermion duality
  for half-filled multicomponent Landau Levels}},}\ }\href@noop {} {\bibfield
  {journal} {\bibinfo  {journal} {ArXiv e-prints}\ } (\bibinfo {year}
  {2016})},\ \Eprint {http://arxiv.org/abs/1609.08616} {arXiv:1609.08616
  [cond-mat.str-el]} \BibitemShut {NoStop}%
\bibitem [{\citenamefont {Nakahara}(2003)}]{Nakahara}%
  \BibitemOpen
  \bibfield  {author} {\bibinfo {author} {\bibfnamefont {Mikio}\ \bibnamefont
  {Nakahara}},\ }\href@noop {} {\emph {\bibinfo {title} {Geometry, Topology and
  Physics}}},\ Graduate Student Series in Physics\ (\bibinfo  {publisher}
  {Institute of Physics Publishing},\ \bibinfo {year} {2003})\BibitemShut
  {NoStop}%
\bibitem [{\citenamefont {Eguchi}\ \emph {et~al.}(1980)\citenamefont {Eguchi},
  \citenamefont {Gilkey},\ and\ \citenamefont {Hanson}}]{GilkeyReview}%
  \BibitemOpen
  \bibfield  {author} {\bibinfo {author} {\bibfnamefont {Tohru}\ \bibnamefont
  {Eguchi}}, \bibinfo {author} {\bibfnamefont {Peter~B.}\ \bibnamefont
  {Gilkey}}, \ and\ \bibinfo {author} {\bibfnamefont {Andrew~J.}\ \bibnamefont
  {Hanson}},\ }\bibfield  {title} {\enquote {\bibinfo {title} {Gravitation,
  gauge theories and differential geometry},}\ }\href {\doibase
  http://dx.doi.org/10.1016/0370-1573(80)90130-1} {\bibfield  {journal}
  {\bibinfo  {journal} {Physics Reports}\ }\textbf {\bibinfo {volume} {66}},\
  \bibinfo {pages} {213 -- 393} (\bibinfo {year} {1980})}\BibitemShut {NoStop}%
\bibitem [{\citenamefont {Milnor}\ and\ \citenamefont
  {Stasheff}(1974)}]{MilnoreStasheff}%
  \BibitemOpen
  \bibfield  {author} {\bibinfo {author} {\bibfnamefont {John}\ \bibnamefont
  {Milnor}}\ and\ \bibinfo {author} {\bibfnamefont {James}\ \bibnamefont
  {Stasheff}},\ }\href@noop {} {\emph {\bibinfo {title} {Characteristic
  classes}}}\ (\bibinfo  {publisher} {Princeton University Press},\ \bibinfo
  {year} {1974})\BibitemShut {NoStop}%
\bibitem [{\citenamefont {Milnor}\ and\ \citenamefont
  {Husemoller}(1973)}]{MilnorHusemoller}%
  \BibitemOpen
  \bibfield  {author} {\bibinfo {author} {\bibfnamefont {J}~\bibnamefont
  {Milnor}}\ and\ \bibinfo {author} {\bibfnamefont {D}~\bibnamefont
  {Husemoller}},\ }\href@noop {} {\emph {\bibinfo {title} {Symmetric bilinear
  forms}}}\ (\bibinfo  {publisher} {Springer-Verlag Berlin New York},\ \bibinfo
  {year} {1973})\BibitemShut {NoStop}%
\bibitem [{\citenamefont {Witten}(1979)}]{WittenCP1}%
  \BibitemOpen
  \bibfield  {author} {\bibinfo {author} {\bibfnamefont {Edward}\ \bibnamefont
  {Witten}},\ }\bibfield  {title} {\enquote {\bibinfo {title} {Instatons, the
  quark model, and the 1/n expansion},}\ }\href {\doibase
  http://dx.doi.org/10.1016/0550-3213(79)90243-8} {\bibfield  {journal}
  {\bibinfo  {journal} {Nuclear Physics B}\ }\textbf {\bibinfo {volume}
  {149}},\ \bibinfo {pages} {285 -- 320} (\bibinfo {year} {1979})}\BibitemShut
  {NoStop}%
\bibitem [{\citenamefont {El-Showk}\ \emph {et~al.}(2012)\citenamefont
  {El-Showk}, \citenamefont {Paulos}, \citenamefont {Poland}, \citenamefont
  {Rychkov}, \citenamefont {Simmons-Duffin},\ and\ \citenamefont
  {Vichi}}]{solvingIsing}%
  \BibitemOpen
  \bibfield  {author} {\bibinfo {author} {\bibfnamefont {Sheer}\ \bibnamefont
  {El-Showk}}, \bibinfo {author} {\bibfnamefont {Miguel~F.}\ \bibnamefont
  {Paulos}}, \bibinfo {author} {\bibfnamefont {David}\ \bibnamefont {Poland}},
  \bibinfo {author} {\bibfnamefont {Slava}\ \bibnamefont {Rychkov}}, \bibinfo
  {author} {\bibfnamefont {David}\ \bibnamefont {Simmons-Duffin}}, \ and\
  \bibinfo {author} {\bibfnamefont {Alessandro}\ \bibnamefont {Vichi}},\
  }\bibfield  {title} {\enquote {\bibinfo {title} {Solving the 3d ising model
  with the conformal bootstrap},}\ }\href {\doibase 10.1103/PhysRevD.86.025022}
  {\bibfield  {journal} {\bibinfo  {journal} {Phys. Rev. D}\ }\textbf {\bibinfo
  {volume} {86}},\ \bibinfo {pages} {025022} (\bibinfo {year}
  {2012})}\BibitemShut {NoStop}%
\bibitem [{\citenamefont {{Kos}}\ \emph {et~al.}(2014)\citenamefont {{Kos}},
  \citenamefont {{Poland}},\ and\ \citenamefont
  {{Simmons-Duffin}}}]{bootstrappingO(N)}%
  \BibitemOpen
  \bibfield  {author} {\bibinfo {author} {\bibfnamefont {F.}~\bibnamefont
  {{Kos}}}, \bibinfo {author} {\bibfnamefont {D.}~\bibnamefont {{Poland}}}, \
  and\ \bibinfo {author} {\bibfnamefont {D.}~\bibnamefont {{Simmons-Duffin}}},\
  }\bibfield  {title} {\enquote {\bibinfo {title} {{Bootstrapping the O( N )
  vector models}},}\ }\href {\doibase 10.1007/JHEP06(2014)091} {\bibfield
  {journal} {\bibinfo  {journal} {Journal of High Energy Physics}\ }\textbf
  {\bibinfo {volume} {6}},\ \bibinfo {eid} {91} (\bibinfo {year} {2014})},\
  \Eprint {http://arxiv.org/abs/1307.6856} {arXiv:1307.6856 [hep-th]}
  \BibitemShut {NoStop}%
\bibitem [{\citenamefont {{Chester}}\ and\ \citenamefont
  {{Pufu}}(2016{\natexlab{a}})}]{ChesterPufuBootstrappingQED}%
  \BibitemOpen
  \bibfield  {author} {\bibinfo {author} {\bibfnamefont {S.~M.}\ \bibnamefont
  {{Chester}}}\ and\ \bibinfo {author} {\bibfnamefont {S.~S.}\ \bibnamefont
  {{Pufu}}},\ }\bibfield  {title} {\enquote {\bibinfo {title} {{Towards
  bootstrapping QED$_{3}$}},}\ }\href {\doibase 10.1007/JHEP08(2016)019}
  {\bibfield  {journal} {\bibinfo  {journal} {Journal of High Energy Physics}\
  }\textbf {\bibinfo {volume} {8}},\ \bibinfo {eid} {19} (\bibinfo {year}
  {2016}{\natexlab{a}})},\ \Eprint {http://arxiv.org/abs/1601.03476}
  {arXiv:1601.03476 [hep-th]} \BibitemShut {NoStop}%
\bibitem [{\citenamefont {Gracey}(1993)}]{GRACEY1993415}%
  \BibitemOpen
  \bibfield  {author} {\bibinfo {author} {\bibfnamefont {J.A.}\ \bibnamefont
  {Gracey}},\ }\bibfield  {title} {\enquote {\bibinfo {title} {Electron mass
  anomalous dimension at $o(1/n_f^2)$ in quantum electrodynamics},}\ }\href
  {\doibase http://dx.doi.org/10.1016/0370-2693(93)91017-H} {\bibfield
  {journal} {\bibinfo  {journal} {Physics Letters B}\ }\textbf {\bibinfo
  {volume} {317}},\ \bibinfo {pages} {415 -- 420} (\bibinfo {year}
  {1993})}\BibitemShut {NoStop}%
\bibitem [{\citenamefont {{Dyer}}\ \emph {et~al.}(2013)\citenamefont {{Dyer}},
  \citenamefont {{Mezei}},\ and\ \citenamefont
  {{Pufu}}}]{DyerMonopoleTaxonomy}%
  \BibitemOpen
  \bibfield  {author} {\bibinfo {author} {\bibfnamefont {E.}~\bibnamefont
  {{Dyer}}}, \bibinfo {author} {\bibfnamefont {M.}~\bibnamefont {{Mezei}}}, \
  and\ \bibinfo {author} {\bibfnamefont {S.~S.}\ \bibnamefont {{Pufu}}},\
  }\bibfield  {title} {\enquote {\bibinfo {title} {{Monopole Taxonomy in
  Three-Dimensional Conformal Field Theories}},}\ }\href@noop {} {\bibfield
  {journal} {\bibinfo  {journal} {ArXiv e-prints}\ } (\bibinfo {year}
  {2013})},\ \Eprint {http://arxiv.org/abs/1309.1160} {arXiv:1309.1160
  [hep-th]} \BibitemShut {NoStop}%
\bibitem [{\citenamefont {{Chester}}\ and\ \citenamefont
  {{Pufu}}(2016{\natexlab{b}})}]{ChesterPufuScalarOperators}%
  \BibitemOpen
  \bibfield  {author} {\bibinfo {author} {\bibfnamefont {S.~M.}\ \bibnamefont
  {{Chester}}}\ and\ \bibinfo {author} {\bibfnamefont {S.~S.}\ \bibnamefont
  {{Pufu}}},\ }\bibfield  {title} {\enquote {\bibinfo {title} {{Anomalous
  dimensions of scalar operators in QED$_{3}$}},}\ }\href {\doibase
  10.1007/JHEP08(2016)069} {\bibfield  {journal} {\bibinfo  {journal} {Journal
  of High Energy Physics}\ }\textbf {\bibinfo {volume} {8}},\ \bibinfo {eid}
  {69} (\bibinfo {year} {2016}{\natexlab{b}})},\ \Eprint
  {http://arxiv.org/abs/1603.05582} {arXiv:1603.05582 [hep-th]} \BibitemShut
  {NoStop}%
\bibitem [{\citenamefont {Beach}\ \emph {et~al.}(2009)\citenamefont {Beach},
  \citenamefont {Alet}, \citenamefont {Mambrini},\ and\ \citenamefont
  {Capponi}}]{beach2009n}%
  \BibitemOpen
  \bibfield  {author} {\bibinfo {author} {\bibfnamefont {KSD}\ \bibnamefont
  {Beach}}, \bibinfo {author} {\bibfnamefont {Fabien}\ \bibnamefont {Alet}},
  \bibinfo {author} {\bibfnamefont {Matthieu}\ \bibnamefont {Mambrini}}, \ and\
  \bibinfo {author} {\bibfnamefont {Sylvain}\ \bibnamefont {Capponi}},\
  }\bibfield  {title} {\enquote {\bibinfo {title} {Su (n) heisenberg model on
  the square lattice: A continuous-n quantum monte carlo study},}\ }\href@noop
  {} {\bibfield  {journal} {\bibinfo  {journal} {Physical Review B}\ }\textbf
  {\bibinfo {volume} {80}},\ \bibinfo {pages} {184401} (\bibinfo {year}
  {2009})}\BibitemShut {NoStop}%
\bibitem [{\citenamefont {Banerjee}\ \emph {et~al.}(2011)\citenamefont
  {Banerjee}, \citenamefont {Damle},\ and\ \citenamefont
  {Alet}}]{BanerjeeetalSU(3)}%
  \BibitemOpen
  \bibfield  {author} {\bibinfo {author} {\bibfnamefont {Argha}\ \bibnamefont
  {Banerjee}}, \bibinfo {author} {\bibfnamefont {Kedar}\ \bibnamefont {Damle}},
  \ and\ \bibinfo {author} {\bibfnamefont {Fabien}\ \bibnamefont {Alet}},\
  }\bibfield  {title} {\enquote {\bibinfo {title} {Impurity spin texture at the
  critical point between n\'eel-ordered and valence-bond-solid states in
  two-dimensional su(3) quantum antiferromagnets},}\ }\href {\doibase
  10.1103/PhysRevB.83.235111} {\bibfield  {journal} {\bibinfo  {journal} {Phys.
  Rev. B}\ }\textbf {\bibinfo {volume} {83}},\ \bibinfo {pages} {235111}
  (\bibinfo {year} {2011})}\BibitemShut {NoStop}%
\bibitem [{\citenamefont {Kaul}(2011)}]{KaulSU(3)SU(4)}%
  \BibitemOpen
  \bibfield  {author} {\bibinfo {author} {\bibfnamefont {Ribhu~K.}\
  \bibnamefont {Kaul}},\ }\bibfield  {title} {\enquote {\bibinfo {title}
  {Quantum criticality in su(3) and su(4) antiferromagnets},}\ }\href {\doibase
  10.1103/PhysRevB.84.054407} {\bibfield  {journal} {\bibinfo  {journal} {Phys.
  Rev. B}\ }\textbf {\bibinfo {volume} {84}},\ \bibinfo {pages} {054407}
  (\bibinfo {year} {2011})}\BibitemShut {NoStop}%
\bibitem [{\citenamefont {Kaul}\ and\ \citenamefont
  {Sandvik}(2012)}]{kaulsandviklargen}%
  \BibitemOpen
  \bibfield  {author} {\bibinfo {author} {\bibfnamefont {Ribhu~K.}\
  \bibnamefont {Kaul}}\ and\ \bibinfo {author} {\bibfnamefont {Anders~W.}\
  \bibnamefont {Sandvik}},\ }\bibfield  {title} {\enquote {\bibinfo {title}
  {Lattice model for the $\mathrm{SU}(n)$ n\'eel to valence-bond solid quantum
  phase transition at large $n$},}\ }\href {\doibase
  10.1103/PhysRevLett.108.137201} {\bibfield  {journal} {\bibinfo  {journal}
  {Phys. Rev. Lett.}\ }\textbf {\bibinfo {volume} {108}},\ \bibinfo {pages}
  {137201} (\bibinfo {year} {2012})}\BibitemShut {NoStop}%
\bibitem [{\citenamefont {Suwa}\ \emph {et~al.}(2016)\citenamefont {Suwa},
  \citenamefont {Sen},\ and\ \citenamefont {Sandvik}}]{SandvikSpectrum}%
  \BibitemOpen
  \bibfield  {author} {\bibinfo {author} {\bibfnamefont {Hidemaro}\
  \bibnamefont {Suwa}}, \bibinfo {author} {\bibfnamefont {Arnab}\ \bibnamefont
  {Sen}}, \ and\ \bibinfo {author} {\bibfnamefont {Anders~W.}\ \bibnamefont
  {Sandvik}},\ }\bibfield  {title} {\enquote {\bibinfo {title} {Level
  spectroscopy in a two-dimensional quantum magnet: Linearly dispersing spinons
  at the deconfined quantum critical point},}\ }\href {\doibase
  10.1103/PhysRevB.94.144416} {\bibfield  {journal} {\bibinfo  {journal} {Phys.
  Rev. B}\ }\textbf {\bibinfo {volume} {94}},\ \bibinfo {pages} {144416}
  (\bibinfo {year} {2016})}\BibitemShut {NoStop}%
\bibitem [{\citenamefont {Sandvik}(2012)}]{sandvikVBS}%
  \BibitemOpen
  \bibfield  {author} {\bibinfo {author} {\bibfnamefont {Anders~W.}\
  \bibnamefont {Sandvik}},\ }\bibfield  {title} {\enquote {\bibinfo {title}
  {Finite-size scaling and boundary effects in two-dimensional valence-bond
  solids},}\ }\href {\doibase 10.1103/PhysRevB.85.134407} {\bibfield  {journal}
  {\bibinfo  {journal} {Phys. Rev. B}\ }\textbf {\bibinfo {volume} {85}},\
  \bibinfo {pages} {134407} (\bibinfo {year} {2012})}\BibitemShut {NoStop}%
\bibitem [{\citenamefont {Nienhuis}\ \emph {et~al.}(1979)\citenamefont
  {Nienhuis}, \citenamefont {Berker}, \citenamefont {Riedel},\ and\
  \citenamefont {Schick}}]{nienhuispotts}%
  \BibitemOpen
  \bibfield  {author} {\bibinfo {author} {\bibfnamefont {B.}~\bibnamefont
  {Nienhuis}}, \bibinfo {author} {\bibfnamefont {A.~N.}\ \bibnamefont
  {Berker}}, \bibinfo {author} {\bibfnamefont {Eberhard~K.}\ \bibnamefont
  {Riedel}}, \ and\ \bibinfo {author} {\bibfnamefont {M.}~\bibnamefont
  {Schick}},\ }\bibfield  {title} {\enquote {\bibinfo {title} {First- and
  second-order phase transitions in potts models: Renormalization-group
  solution},}\ }\href {\doibase 10.1103/PhysRevLett.43.737} {\bibfield
  {journal} {\bibinfo  {journal} {Phys. Rev. Lett.}\ }\textbf {\bibinfo
  {volume} {43}},\ \bibinfo {pages} {737--740} (\bibinfo {year}
  {1979})}\BibitemShut {NoStop}%
\bibitem [{\citenamefont {Nauenberg}\ and\ \citenamefont
  {Scalapino}(1980)}]{nauenbergscalapino}%
  \BibitemOpen
  \bibfield  {author} {\bibinfo {author} {\bibfnamefont {M.}~\bibnamefont
  {Nauenberg}}\ and\ \bibinfo {author} {\bibfnamefont {D.~J.}\ \bibnamefont
  {Scalapino}},\ }\bibfield  {title} {\enquote {\bibinfo {title} {Singularities
  and scaling functions at the potts-model multicritical point},}\ }\href
  {\doibase 10.1103/PhysRevLett.44.837} {\bibfield  {journal} {\bibinfo
  {journal} {Phys. Rev. Lett.}\ }\textbf {\bibinfo {volume} {44}},\ \bibinfo
  {pages} {837--840} (\bibinfo {year} {1980})}\BibitemShut {NoStop}%
\bibitem [{\citenamefont {Cardy}\ \emph {et~al.}(1980)\citenamefont {Cardy},
  \citenamefont {Nauenberg},\ and\ \citenamefont
  {Scalapino}}]{cardynauenbergscalapino}%
  \BibitemOpen
  \bibfield  {author} {\bibinfo {author} {\bibfnamefont {John~L.}\ \bibnamefont
  {Cardy}}, \bibinfo {author} {\bibfnamefont {M.}~\bibnamefont {Nauenberg}}, \
  and\ \bibinfo {author} {\bibfnamefont {D.~J.}\ \bibnamefont {Scalapino}},\
  }\bibfield  {title} {\enquote {\bibinfo {title} {Scaling theory of the
  potts-model multicritical point},}\ }\href {\doibase
  10.1103/PhysRevB.22.2560} {\bibfield  {journal} {\bibinfo  {journal} {Phys.
  Rev. B}\ }\textbf {\bibinfo {volume} {22}},\ \bibinfo {pages} {2560--2568}
  (\bibinfo {year} {1980})}\BibitemShut {NoStop}%
\bibitem [{\citenamefont {Wegner}(1976)}]{Wegner1976}%
  \BibitemOpen
  \bibfield  {author} {\bibinfo {author} {\bibfnamefont {F.~J.}\ \bibnamefont
  {Wegner}},\ }\bibfield  {title} {\enquote {\bibinfo {title} {The critical
  state, general aspects},}\ }in\ \href@noop {} {\emph {\bibinfo {booktitle}
  {Phase transitions and critical phenomena}}},\ Vol.~\bibinfo {volume} {6},\
  \bibinfo {editor} {edited by\ \bibinfo {editor} {\bibfnamefont
  {C.}~\bibnamefont {Domb}}\ and\ \bibinfo {editor} {\bibfnamefont {M.~S.}\
  \bibnamefont {Green}}}\ (\bibinfo  {publisher} {Academic Press},\ \bibinfo
  {address} {London},\ \bibinfo {year} {1976})\ Chap.~\bibinfo {chapter}
  {2}\BibitemShut {NoStop}%
\bibitem [{\citenamefont {Baxter}(1973)}]{baxterpotts}%
  \BibitemOpen
  \bibfield  {author} {\bibinfo {author} {\bibfnamefont {R~J}\ \bibnamefont
  {Baxter}},\ }\bibfield  {title} {\enquote {\bibinfo {title} {Potts model at
  the critical temperature},}\ }\href
  {http://stacks.iop.org/0022-3719/6/i=23/a=005} {\bibfield  {journal}
  {\bibinfo  {journal} {Journal of Physics C: Solid State Physics}\ }\textbf
  {\bibinfo {volume} {6}},\ \bibinfo {pages} {L445} (\bibinfo {year}
  {1973})}\BibitemShut {NoStop}%
\bibitem [{\citenamefont {Lee}\ and\ \citenamefont
  {Kosterlitz}(1990)}]{LeeKosterlitzPotts}%
  \BibitemOpen
  \bibfield  {author} {\bibinfo {author} {\bibfnamefont {Jooyoung}\
  \bibnamefont {Lee}}\ and\ \bibinfo {author} {\bibfnamefont {J.~M.}\
  \bibnamefont {Kosterlitz}},\ }\bibfield  {title} {\enquote {\bibinfo {title}
  {New numerical method to study phase transitions},}\ }\href {\doibase
  10.1103/PhysRevLett.65.137} {\bibfield  {journal} {\bibinfo  {journal} {Phys.
  Rev. Lett.}\ }\textbf {\bibinfo {volume} {65}},\ \bibinfo {pages} {137--140}
  (\bibinfo {year} {1990})}\BibitemShut {NoStop}%
\bibitem [{\citenamefont {Kl{\"u}mper}\ \emph {et~al.}(1989)\citenamefont
  {Kl{\"u}mper}, \citenamefont {Schadschneider},\ and\ \citenamefont
  {Zittartz}}]{Klumper}%
  \BibitemOpen
  \bibfield  {author} {\bibinfo {author} {\bibfnamefont {A.}~\bibnamefont
  {Kl{\"u}mper}}, \bibinfo {author} {\bibfnamefont {A.}~\bibnamefont
  {Schadschneider}}, \ and\ \bibinfo {author} {\bibfnamefont {J.}~\bibnamefont
  {Zittartz}},\ }\bibfield  {title} {\enquote {\bibinfo {title} {Inversion
  relations, phase transitions and transfer matrix excitations for special spin
  models in two dimensions},}\ }\href {\doibase 10.1007/BF01312692} {\bibfield
  {journal} {\bibinfo  {journal} {Zeitschrift f{\"u}r Physik B Condensed
  Matter}\ }\textbf {\bibinfo {volume} {76}},\ \bibinfo {pages} {247--258}
  (\bibinfo {year} {1989})}\BibitemShut {NoStop}%
\bibitem [{\citenamefont {Buffenoir}\ and\ \citenamefont
  {Wallon}(1993)}]{Buffenoir}%
  \BibitemOpen
  \bibfield  {author} {\bibinfo {author} {\bibfnamefont {E}~\bibnamefont
  {Buffenoir}}\ and\ \bibinfo {author} {\bibfnamefont {S}~\bibnamefont
  {Wallon}},\ }\bibfield  {title} {\enquote {\bibinfo {title} {The correlation
  length of the potts model at the first-order transition point},}\ }\href
  {http://stacks.iop.org/0305-4470/26/i=13/a=009} {\bibfield  {journal}
  {\bibinfo  {journal} {Journal of Physics A: Mathematical and General}\
  }\textbf {\bibinfo {volume} {26}},\ \bibinfo {pages} {3045} (\bibinfo {year}
  {1993})}\BibitemShut {NoStop}%
\bibitem [{\citenamefont {Gies}\ and\ \citenamefont {Jaeckel}(2006)}]{Gies}%
  \BibitemOpen
  \bibfield  {author} {\bibinfo {author} {\bibfnamefont {H.}~\bibnamefont
  {Gies}}\ and\ \bibinfo {author} {\bibfnamefont {J.}~\bibnamefont {Jaeckel}},\
  }\bibfield  {title} {\enquote {\bibinfo {title} {Chiral phase structure of
  qcd with many flavors},}\ }\href {\doibase 10.1140/epjc/s2006-02475-0}
  {\bibfield  {journal} {\bibinfo  {journal} {The European Physical Journal C -
  Particles and Fields}\ }\textbf {\bibinfo {volume} {46}},\ \bibinfo {pages}
  {433--438} (\bibinfo {year} {2006})}\BibitemShut {NoStop}%
\bibitem [{\citenamefont {Kaplan}\ \emph {et~al.}(2009)\citenamefont {Kaplan},
  \citenamefont {Lee}, \citenamefont {Son},\ and\ \citenamefont
  {Stephanov}}]{kaplan}%
  \BibitemOpen
  \bibfield  {author} {\bibinfo {author} {\bibfnamefont {David~B.}\
  \bibnamefont {Kaplan}}, \bibinfo {author} {\bibfnamefont {Jong-Wan}\
  \bibnamefont {Lee}}, \bibinfo {author} {\bibfnamefont {Dam~T.}\ \bibnamefont
  {Son}}, \ and\ \bibinfo {author} {\bibfnamefont {Mikhail~A.}\ \bibnamefont
  {Stephanov}},\ }\bibfield  {title} {\enquote {\bibinfo {title} {Conformality
  lost},}\ }\href {\doibase 10.1103/PhysRevD.80.125005} {\bibfield  {journal}
  {\bibinfo  {journal} {Phys. Rev. D}\ }\textbf {\bibinfo {volume} {80}},\
  \bibinfo {pages} {125005} (\bibinfo {year} {2009})}\BibitemShut {NoStop}%
\bibitem [{\citenamefont {Gukov}(2017)}]{gukov2017rg}%
  \BibitemOpen
  \bibfield  {author} {\bibinfo {author} {\bibfnamefont {Sergei}\ \bibnamefont
  {Gukov}},\ }\bibfield  {title} {\enquote {\bibinfo {title} {Rg flows and
  bifurcations},}\ }\href@noop {} {\bibfield  {journal} {\bibinfo  {journal}
  {Nuclear Physics B}\ }\textbf {\bibinfo {volume} {919}},\ \bibinfo {pages}
  {583--638} (\bibinfo {year} {2017})}\BibitemShut {NoStop}%
\bibitem [{\citenamefont {Giombi}\ \emph {et~al.}(2016)\citenamefont {Giombi},
  \citenamefont {Klebanov},\ and\ \citenamefont
  {Tarnopolsky}}]{giombi2016conformal}%
  \BibitemOpen
  \bibfield  {author} {\bibinfo {author} {\bibfnamefont {Simone}\ \bibnamefont
  {Giombi}}, \bibinfo {author} {\bibfnamefont {Igor~R}\ \bibnamefont
  {Klebanov}}, \ and\ \bibinfo {author} {\bibfnamefont {Grigory}\ \bibnamefont
  {Tarnopolsky}},\ }\bibfield  {title} {\enquote {\bibinfo {title} {Conformal
  qed d, f-theorem and the $\epsilon$-expansion},}\ }\href@noop {} {\bibfield
  {journal} {\bibinfo  {journal} {Journal of Physics A: Mathematical and
  Theoretical}\ }\textbf {\bibinfo {volume} {49}},\ \bibinfo {pages} {135403}
  (\bibinfo {year} {2016})}\BibitemShut {NoStop}%
\bibitem [{\citenamefont {Herbut}(2016)}]{herbut2016chiral}%
  \BibitemOpen
  \bibfield  {author} {\bibinfo {author} {\bibfnamefont {Igor~F}\ \bibnamefont
  {Herbut}},\ }\bibfield  {title} {\enquote {\bibinfo {title} {Chiral symmetry
  breaking in three-dimensional quantum electrodynamics as fixed point
  annihilation},}\ }\href@noop {} {\bibfield  {journal} {\bibinfo  {journal}
  {Physical Review D}\ }\textbf {\bibinfo {volume} {94}},\ \bibinfo {pages}
  {025036} (\bibinfo {year} {2016})}\BibitemShut {NoStop}%
\bibitem [{\citenamefont {Nahum}\ \emph {et~al.}(2013)\citenamefont {Nahum},
  \citenamefont {Chalker}, \citenamefont {Serna}, \citenamefont {Ortu\~no},\
  and\ \citenamefont {Somoza}}]{PhaseTransitionsCPNSigmaModel}%
  \BibitemOpen
  \bibfield  {author} {\bibinfo {author} {\bibfnamefont {Adam}\ \bibnamefont
  {Nahum}}, \bibinfo {author} {\bibfnamefont {J.~T.}\ \bibnamefont {Chalker}},
  \bibinfo {author} {\bibfnamefont {P.}~\bibnamefont {Serna}}, \bibinfo
  {author} {\bibfnamefont {M.}~\bibnamefont {Ortu\~no}}, \ and\ \bibinfo
  {author} {\bibfnamefont {A.~M.}\ \bibnamefont {Somoza}},\ }\bibfield  {title}
  {\enquote {\bibinfo {title} {Phase transitions in three-dimensional loop
  models and the $c{P}^{n\ensuremath{-}1}$ sigma model},}\ }\href {\doibase
  10.1103/PhysRevB.88.134411} {\bibfield  {journal} {\bibinfo  {journal} {Phys.
  Rev. B}\ }\textbf {\bibinfo {volume} {88}},\ \bibinfo {pages} {134411}
  (\bibinfo {year} {2013})}\BibitemShut {NoStop}%
\bibitem [{\citenamefont {{Potter}}\ \emph {et~al.}(2016)\citenamefont
  {{Potter}}, \citenamefont {{Wang}}, \citenamefont {{Metlitski}},\ and\
  \citenamefont {{Vishwanath}}}]{potter}%
  \BibitemOpen
  \bibfield  {author} {\bibinfo {author} {\bibfnamefont {A.~C.}\ \bibnamefont
  {{Potter}}}, \bibinfo {author} {\bibfnamefont {C.}~\bibnamefont {{Wang}}},
  \bibinfo {author} {\bibfnamefont {M.~A.}\ \bibnamefont {{Metlitski}}}, \ and\
  \bibinfo {author} {\bibfnamefont {A.}~\bibnamefont {{Vishwanath}}},\
  }\bibfield  {title} {\enquote {\bibinfo {title} {{Realizing topological
  surface states in a lower-dimensional flat band}},}\ }\href@noop {}
  {\bibfield  {journal} {\bibinfo  {journal} {ArXiv e-prints}\ } (\bibinfo
  {year} {2016})},\ \Eprint {http://arxiv.org/abs/1609.08618} {arXiv:1609.08618
  [cond-mat.str-el]} \BibitemShut {NoStop}%
\bibitem [{\citenamefont {Burgess}\ and\ \citenamefont
  {Dolan}(2001)}]{BurgessDolan}%
  \BibitemOpen
  \bibfield  {author} {\bibinfo {author} {\bibfnamefont {C.~P.}\ \bibnamefont
  {Burgess}}\ and\ \bibinfo {author} {\bibfnamefont {Brian~P.}\ \bibnamefont
  {Dolan}},\ }\bibfield  {title} {\enquote {\bibinfo {title} {Particle--vortex
  duality and the modular group: Applications to the quantum hall effect and
  other two-dimensional systems},}\ }\href {\doibase
  10.1103/PhysRevB.63.155309} {\bibfield  {journal} {\bibinfo  {journal} {Phys.
  Rev. B}\ }\textbf {\bibinfo {volume} {63}},\ \bibinfo {pages} {155309}
  (\bibinfo {year} {2001})}\BibitemShut {NoStop}%
\bibitem [{\citenamefont {Cardy}\ and\ \citenamefont
  {Rabinovici}(1982)}]{CARDY19821}%
  \BibitemOpen
  \bibfield  {author} {\bibinfo {author} {\bibfnamefont {John~L.}\ \bibnamefont
  {Cardy}}\ and\ \bibinfo {author} {\bibfnamefont {Eliezer}\ \bibnamefont
  {Rabinovici}},\ }\bibfield  {title} {\enquote {\bibinfo {title} {Phase
  structure of $z_p$ models in the presence of a $\theta$ parameter},}\ }\href
  {\doibase http://dx.doi.org/10.1016/0550-3213(82)90463-1} {\bibfield
  {journal} {\bibinfo  {journal} {Nuclear Physics B}\ }\textbf {\bibinfo
  {volume} {205}},\ \bibinfo {pages} {1 -- 16} (\bibinfo {year}
  {1982})}\BibitemShut {NoStop}%
\bibitem [{\citenamefont {Cardy}(1982)}]{CARDY198217}%
  \BibitemOpen
  \bibfield  {author} {\bibinfo {author} {\bibfnamefont {John~L.}\ \bibnamefont
  {Cardy}},\ }\bibfield  {title} {\enquote {\bibinfo {title} {Duality and the
  $\theta$ parameter in abelian lattice models},}\ }\href {\doibase
  http://dx.doi.org/10.1016/0550-3213(82)90464-3} {\bibfield  {journal}
  {\bibinfo  {journal} {Nuclear Physics B}\ }\textbf {\bibinfo {volume}
  {205}},\ \bibinfo {pages} {17 -- 26} (\bibinfo {year} {1982})}\BibitemShut
  {NoStop}%
\bibitem [{\citenamefont {Fradkin}\ and\ \citenamefont
  {Kivelson}(1996)}]{FradkinKivelsonSL2Z}%
  \BibitemOpen
  \bibfield  {author} {\bibinfo {author} {\bibfnamefont {Eduardo}\ \bibnamefont
  {Fradkin}}\ and\ \bibinfo {author} {\bibfnamefont {Steven}\ \bibnamefont
  {Kivelson}},\ }\bibfield  {title} {\enquote {\bibinfo {title} {Modular
  invariance, self-duality and the phase transition between quantum hall
  plateaus},}\ }\href@noop {} {\bibfield  {journal} {\bibinfo  {journal}
  {Nuclear Physics B}\ }\textbf {\bibinfo {volume} {474}},\ \bibinfo {pages}
  {543 -- 574} (\bibinfo {year} {1996})}\BibitemShut {NoStop}%
\bibitem [{\citenamefont {Kapustin}\ and\ \citenamefont
  {Thorngren}(2013)}]{akpsq}%
  \BibitemOpen
  \bibfield  {author} {\bibinfo {author} {\bibfnamefont {Anton}\ \bibnamefont
  {Kapustin}}\ and\ \bibinfo {author} {\bibfnamefont {Ryan}\ \bibnamefont
  {Thorngren}},\ }\bibfield  {title} {\enquote {\bibinfo {title} {{Topological
  Field Theory on a Lattice, Discrete Theta-Angles and Confinement}},}\
  }\href@noop {} {\bibfield  {journal} {\bibinfo  {journal} {ArXiv e-prints}\ }
  (\bibinfo {year} {2013})},\ \Eprint {http://arxiv.org/abs/arXiv:1308.2926}
  {arXiv:arXiv:1308.2926 [hep-th]} \BibitemShut {NoStop}%
\end{thebibliography}%

\end{document}